\documentclass[fleqn,usenatbib]{mnras}

\usepackage[T1]{fontenc}
\usepackage{ae,aecompl}

\usepackage{appendix}

\usepackage{amsmath}
\usepackage{longtable}
\usepackage{color}
\usepackage{float}

\usepackage{setspace}
\usepackage{amsfonts}
\usepackage{times}
\usepackage{graphicx}
\usepackage{framed}

\def\be{\begin{equation}}
\def\ee{\end{equation}}
\def\ba{\begin{eqnarray}}
\def\ea{\end{eqnarry}}
\def\bal#1\eal{\begin{align}#1\end{align}}

\newcommand{\eqnref}[1]{Equation~(\ref{#1})}

\title[A line-binned treatment of opacities]{A line-binned treatment of opacities for the spectra and\\
light curves from neutron star mergers}

\author[C.~J.~Fontes et al.]{
C.~J.~Fontes$^{1}$\thanks{E-mail: cjf@lanl.gov},
C.~L.~Fryer$^{1,2,3}$,
A.~L.~Hungerford$^{1}$,
R.~T.~Wollaeger$^{1}$, and
O.~Korobkin$^{1}$
\\
$^1$Los Alamos National Laboratory, Los Alamos, NM 87545, USA\\
$^2$Physics Department, University of Arizona, Tucson, AZ 85721, USA\\
$^3$Physics and Astronomy Department, University of New Mexico, Albuquerque,
NM 87131, USA
}

\date{Accepted XXX. Received YYY; in original form ZZZ}

\pubyear{2019}

\hypersetup{
        colorlinks=true,
        linkcolor=blue,
        citecolor=blue,
        urlcolor=cyan
}
\begin{document}
\label{firstpage}
\pagerange{\pageref{firstpage}--\pageref{lastpage}}
\maketitle



\begin{abstract}

The electromagnetic observations of GW170817 were able to dramatically
increase our understanding of neutron star mergers beyond what we learned
from gravitational waves alone. These observations provided insight on all
aspects of the merger from the nature of the gamma-ray burst to the
characteristics of the ejected material.
The ejecta of neutron star mergers are expected to produce such
electromagnetic transients, called kilonovae or macronovae.
Characteristics of the
ejecta include large velocity gradients,
relative to supernovae,
and the presence of heavy
$r$-process elements, which pose significant challenges to the
accurate calculation of radiative opacities and radiation
transport. For example, these opacities include a dense forest of
bound-bound features arising from near-neutral lanthanide and actinide
elements. Here we investigate the use of fine-structure,
line-binned opacities that preserve the integral of the opacity over frequency.
Advantages of this area-preserving approach over the
traditional expansion-opacity formalism include
the ability to pre-calculate opacity tables that are independent of
the type of hydrodynamic expansion and that eliminate the computational expense
of calculating opacities within radiation-transport simulations.
Tabular opacities are generated for all 14 lanthanides as well as a
representative actinide element, uranium.
We demonstrate that spectral simulations produced with the line-binned
opacities agree well with results produced with the more accurate
continuous Monte Carlo Sobolev approach, as well as with the
commonly used expansion-opacity formalism.
Additional investigations illustrate the convergence of opacity
with respect to the number of included lines, and elucidate
sensitivities to different atomic physics approximations, such as
fully and semi-relativistic approaches.
\end{abstract}




\section{Introduction}
\label{sec:intro}

The merger of two neutron stars has been proposed both as the
source of short-duration gamma-ray bursts \citep{1992ApJ...395L..83N} and
the site of
$r$-process production \citep{1982ApL....22..143S}.
Simulations confirmed that the neutron-rich, dynamical ejecta from neutron
star mergers (NSMs) robustly produced
$r$-process elements from the second through third $r$-process
peaks \citep{1999A&A...341..499R,rosswog14,just15}.
With the gravitational wave \citep{abbott17h} and subsequent follow-up 
observations in the electromagnetic spectrum \citep{abbott17a} of a nearby
merger event, astronomers were able, for the first time, to validate  
these theories and, to some extent, determine the yields from these NSMs.

\begin{figure}
\centering
\includegraphics[clip=true,angle=0,width=0.8\columnwidth]
{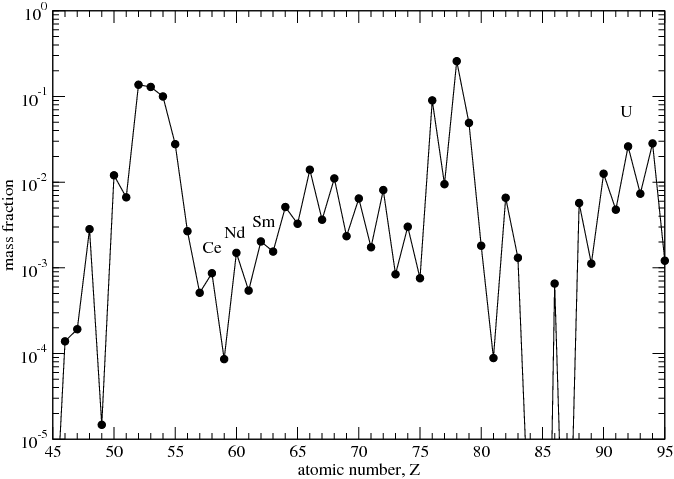}
\caption{Predicted elemental abundances for the ejecta produced
in a $1.4+1.4\,{\rm M_\odot}$ neutron star merger~\citep{rosswog14,grossman14},
approximately one day after merger.
A few labels are provided for easy identification of various lanthanide
and actinide elements considered in this work.
}
\label{fig:mfrac}
\end{figure}

On the surface, the light curves and spectra seem to be a validation
of the ejecta predicted by the merger simulations.  Theorists had
proposed a two-component model for the ejecta consisting of a
neutron-rich dynamical ejecta (producing heavy $r$-process elements) and
a higher electron fraction wind ejecta (light $r$-process and iron peak
elements) \citep{metzger12f,metzger17y}. A significant number of lanthanide
and actinide elements are predicted to be present in the ejecta
(see, for example, Figure~\ref{fig:mfrac}).
The high opacities associated with these elements produced in the heavy
$r$-process support an argument for late-time, infra-red emission, whereas
the wind ejecta would produce early-time UV/optical emission.  With
observations of both an early UV/optical and later infra-red emission,
astronomers could place constraints on the amount of ejecta in each
component.  Many groups attempted such studies, producing a wide range 
of results \citep[for a review, see][]{cote18} and,
although all agree that roughly
$0.01$\,M$_\odot$ was ejected in the merger, the exact amount and the exact 
composition of the ejecta still remains a subject of debate.

The differences in the yield predictions arise from different
assumptions in the structure of the ejecta, the composition of the
ejecta and the opacities.  In many simulations, the ejecta was modeled in
spherical symmetry.  By varying the velocity, mass and composition of
the ejecta, these models produce a range of emission spectra that can
be combined to form multi-component models \citep{pian17,cowperthwaite17}.
Multi-dimensional models can include more detailed ejecta profiles.  
For example, a set of models included two-component ejecta properties 
with the dynamical, heavy $r$-process rich ejecta driven along the 
orbital plane and a more spherical wind
ejecta \citep{tanvir17,troja17,kasen17,wollaeger18,tanaka18}.
These structures 
can produce different emission properties for the same production rate 
of heavy $r$-process elements.
Additionally, even for the same model morphology, there remain differences in
the light-curve predictions from the various modeling groups.
For example, a major difference is a redward shift in the peak emission
produced by different groups.
We discuss this shift in Section~\ref{sub:lc}
and show that this behavior is not due to the different ways
in which the opacities are implemented in the radiation-transport
simulations.

The range of opacities used also varies dramatically.  Many models
rely on constant opacity values, varying from 0.2 to 30~cm$^2$/g.
Only a few models used detailed, frequency-dependent opacities based on atomic
physics calculations and these calculations are still limited to a few
lanthanide elements used as surrogates for the heavy $r$-process
elements \citep{barnes16,kasen17,wollaeger18,tanaka18}.
Calculating accurate opacities for lanthanides and actinides pushes
the frontier in atomic physics research due to the complexity
in modeling the interaction of many bound electrons, some of which
occupy a partially filled $f$ shell.
Generating a complete set of opacities requires making approximations
in the atomic physics calculations that can significantly alter NSM
spectral simulations.

In addition, the implementation of these opacities in NSM modeling
also varies from
group to group.  Due to the existence of a strong velocity gradient in the
ejecta, Doppler effects must be included in the opacities. This
requirement typically includes both a correction to the optical depth of each
line \citep{sobolev60} and effects on line broadening, leading to
a variety of expansion-opacity approaches \citep{castor04}.
These methods require line lists that can be difficult to include
in their entirety
in light-curve calculations due to the large number of lines associated with 
lanthanide and actinide elements.
In this work, we propose using precomputed, binned line contributions,
which allows opacities to be generated in a
compact tabular form for convenient use in transport calculations.
This line-binned approach is guaranteed to preserve the integral of
the opacity over frequency and supersedes our earlier,
preliminary attempt to achieve this goal using line-smeared opacities.

This paper presents a first set of lanthanide opacities, as well as
a representative actinide (uranium) opacity, to be used in 
light-curve calculations for NSMs.  In Section~\ref{sec:atomic}
we review the methods and some of the uncertainties in the atomic
physics calculations for these heavy opacities, discussing relevant
approximations, such as semi- versus fully relativistic models
and configuration interaction.
In Section~\ref{sec:opac} we present a range of line-binned opacities for NSM
calculations, illustrating similarities and differences in the
lanthanide and uranium opacities.
Motivation and justification for our line-binned opacity
approach is presented in Section~\ref{sec:bin_just}.
We summarize with a brief
discussion of the implications of our results for observations.

\section{Atomic Physics Considerations}
\label{sec:atomic}

\subsection{Computational Framework}
\label{subsec:compframe}

In this work we use the Los Alamos suite of atomic physics and plasma modeling
codes~(see \citealt{LANL_suite} and references therein)
to generate the fundamental data and opacities needed
to simulate the characteristics (time to peak, spectra, luminosities,
decay times, etc.) associated with neutron star mergers.
For a given element, a model is composed of the atomic structure
(energies, wavefunctions and oscillator strengths) and photoionization
cross sections. Both the fully and semi-relativistic capabilities of the suite
are used in this work.

The fully relativistic (FR) approach is based on bound- and continuum-electron
wavefunctions that are solutions of the Dirac equation, while the
semi-relativistic (SR) approach uses solutions of the 
Schr\"odinger equation with relativistic corrections.
A fully relativistic (FR) calculation begins with the RATS atomic
structure code \citep{LANL_suite} using the Dirac-Fock-Slater
method of Sampson and co-workers \citep{sampson_physrep}.
A semi-relativistic (SR) calculation begins with the CATS atomic
structure code \citep{cats_man} using the Hartree-Fock method
of Cowan~\citep{cowan}.
These calculations produce detailed, fine-structure data that include a
complete description of configuration interaction for the specified list of
configurations.
Two variant, relativistic calculations are also considered that include
incomplete amounts of configuration interaction
(see Section~\ref{subsec:models_base}).
After the atomic structure calculations are complete,
both the FR and SR methods use the GIPPER ionization code
to obtain the relevant photoionization cross sections
in the distorted-wave approximation. The photoionization data are used
to generate the bound-free contribution to the opacity and are not
expected to be too important for the present application, due to
the range of relevant photon energies, but are included
for completeness. Therefore, they are calculated in the configuration-average
approximation, rather than fine-structure detail, in order to minimize the
computational time.

The atomic level populations are calculated with the ATOMIC code
from the fundamental atomic data. The code can be used in either
local thermodynamic equilibrium (LTE) or non-LTE mode
\citep{atomic1,atomic2,hakel06,colgan_oplib,fontes_cr16}.
The LTE approach was chosen for the present
application, which requires only the atomic structure data in calculating
the populations. At the relevant times, the ejecta densities are low enough
that collective, or plasma, effects are not important and simple
Saha-Boltzmann statistics is sufficient to produce accurate level
populations. The populations are then combined with the oscillator
strengths and photoionization cross sections in ATOMIC to obtain
the monochromatic opacities, which are constructed from the standard
four contributions:
bound-bound (b-b), bound-free (b-f), free-free (f-f) and scattering.
Specific formulae for these contributions are readily available in various
textbooks, such as \citet{huebner}.

Here, we reproduce only the expression for the bound-bound
contribution, as it is useful in understanding the subsequent expression  
for line-binned opacities, as well as the discussion
of such quantities in Section~\ref{sec:opac}:
\begin{equation}
\kappa^{\rm b-b}_\nu = \frac{\pi e^2}{\rho m_e c} \sum_i N_i
\, |f_{i}| \, L_{i,\nu} \,,
\label{opac_bb}
\end{equation}
where $\nu$ is the photon energy,
$\rho$ is the mass density, $N_i$ is the number density of the initial
level in transition $i$,
$f_{i}$ is the oscillator strength describing the photo-excitation
of transition $i$, and $L_{i,\nu}$ is the corresponding line profile function.
The line-binned opacities proposed in this work
are comprised of discrete frequency (or wavelength) bins
that contain a sum over all of the lines centered within a bin.
An expression for this discrete opacity is obtained
from the continuous opacity displayed in Equation~(\ref{opac_bb})
by replacing the line profile with $1/\Delta \nu_j$, i.e.
\begin{equation}
\kappa^{\rm bin}_{\nu,j} = \frac{1}{\Delta \nu_j}\frac{\pi e^2}{\rho m_e c}
\sum_{i \in \Delta \nu_j} N_i
\, |f_{i}| \,,
\label{opac_bb_bin}
\end{equation}
where $\Delta \nu_j$ represents the frequency width of a bin denoted
by integer index $j$. So, the summation encompasses all lines $i$ with
centers that reside in bin $j$.
It is straightforward to show that numerical integration
of Equation~(\ref{opac_bb_bin}) over all bins produces
the same value as that obtained by analytically integrating
Equation~(\ref{opac_bb}) because the line profile function is normalized to
one when integrating over all frequencies. The form
of Equation~(\ref{opac_bb}) is clearly independent of the type of expansion,
which is a distinct advantage over methods that assume a homologous
flow, such as the expansion-opacity approach.
However, the line-binned opacities are expected to be significantly
different than those produced with more traditional approaches.
Physical arguments and numerical examples
in support of using these line-binned, area-preserving opacities to model
kilonovae is provided in Section~\ref{sec:bin_just}.

The line-binned opacities in Equation~(\ref{opac_bb_bin}) can be readily
cast in a tabular form, using a discrete temperature/mass-density grid,
that is commonly employed in radiation-hydrodynamics
simulations. In the latter approach, discrete photon frequency groups are chosen
to model the flow of radiation, and the groups are typically much less
resolved compared to the frequency bins. If a
particular group is denoted by integer index $g$, and the group boundaries
exactly align with specific bin boundaries, then (unweighted)
group opacities can be computed in a straightforward manner
from Equation~(\ref{opac_bb_bin}) via the formula
\begin{equation}
\kappa^{\rm group}_{\nu,g} = \frac{1}{\Delta \nu_g}
\sum_{j \in \Delta \nu_g} \Delta \nu_j \, \kappa^{\rm bin}_{\nu,j} \,,
\label{opac_bb_group}
\end{equation}
where $\Delta \nu_g$ is the frequency width of group $g$.
If a bin overlaps with more than one group,
its opacity contribution is distributed across those groups
in proportion to the area in each group.

When constructing opacity tables, a practical consideration involves
the choice of an oscillator strength cutoff value, which we denote by $f_c$.
Rather than evalutating the summation in Equation~(\ref{opac_bb_bin})
over all available lines, we consider only lines with an oscillator strength
above some prescribed value. For the conditions of interest for modeling
kilonovae, we found that a value of $f_c = 10^{-6}$ is typically sufficient
to produce converged results. Unless otherwise noted, this cutoff value was used
when constructing all of the opacity data discussed in this work.

\subsection{Baseline Atomic Models}
\label{subsec:models_base}

As mentioned in Section~\ref{sec:intro}, atomic models were created for
all 14 lanthanides and one representative actinide (uranium).
In order to obtain converged opacities for the range of
temperatures and densities in our simulations, only the first four
ion stages of each element were considered, similar to the choice
made by \citet{kasen13}. A list of configurations chosen for each element
is provided in Table~\ref{tab:configs} of Appendix~\ref{app:config_list}.
Since Nd was used as a representative element
in the recent study by \citet{kasen13}, we chose an identical
list of configurations for that element in order to make meaningful
comparisons. As expected, the number of Nd levels is identical to those
appearing in Table~1 of that earlier work.\footnote{Based on this analysis,
the $4f^4 6s^1 6p^1$ configuration appears to have been left out of
Table~1 of \citet{kasen13}.}
The number of lines is slightly higher in the present listing, possibly due
to the retention of small oscillator strengths that do not affect
the modeling in a significant way.
We did some tests to include higher lying configurations, but found that
the displayed list is sufficient to produce converged opacities due to
the relatively low temperature and densities of the ejecta.
Therefore, the configuration lists for the other elements were chosen
in a similar fashion. The configurations for Ce~{\sc ii} and Ce~{\sc iii}
are also identical to those chosen by \citet{kasen13}. The number of
levels and lines differ strongly in this case. We cross-checked
the values between our FR and SR calculations, which agree well,
so those earlier values appear to be in error.

\begin{figure}
\centering
\includegraphics[clip=true,angle=0,width=0.9\columnwidth]
{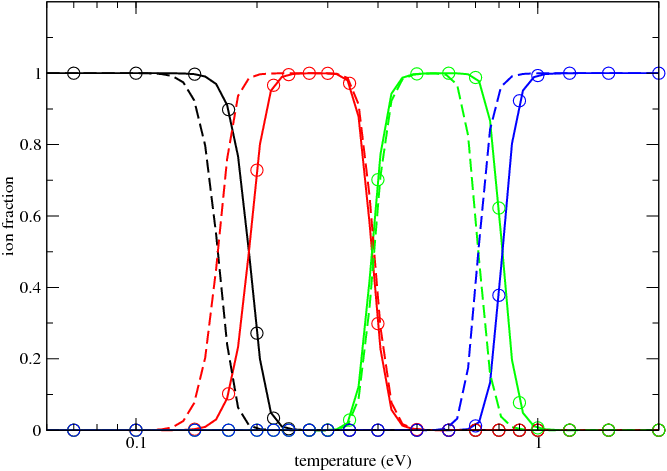}
\caption{Ionization-stage fraction versus temperature for Nd at a typical mass
density of $\rho = 10^{-13}$~g/cm$^3$, calculated with the fully relativistic
(FR) approach. The black curves and circles refer to Nd {\sc i}, the
red ones to Nd {\sc ii}, the green ones to Nd {\sc iii}, and
the blue ones to Nd {\sc iv}. The solid curves use NIST-corrected
ionization energies~\citep{nist}, while the dashed curves use
uncorrected values. The circles
indicate explicit temperatures at which (NIST-corrected) opacities were
calculated for use in the simulation of spectra and light curves.
}
\label{fig:ionfrac_nd}
\end{figure}

As an indication of the quality of our atomic structure calculations,
the ionization energies for the FR and SR models are presented
in Table~\ref{tab:ionpot}, along with the values from the NIST
Atomic Spectra Database (ASD) \citep{nist},
for the following four representative elements:
Ce $(Z=58)$, Nd $(Z=60)$, Sm $(Z=62)$ and U $(Z=92)$.
\begin{table}
\centering
\caption{\rm Ionization energies for the first three
ion stages of four representative elements considered in this work.
Values are presented for the fully relativistic (FR) and semi-relativistic (SR)
methods described in the text, as well as from the NIST ASD \citep{nist}.}
\vspace*{0.5\baselineskip}
\begin{tabular}{lccc}
\hline
Ion stage   &   \multicolumn{3}{c}{Ionization energy (eV)} \\
\cline{2-4}
            &  FR  &  SR  & NIST \\
\hline
Ce {\sc i}  & 4.91 & 5.24 & 5.54 \\
Ce {\sc ii} & 10.5 & 11.2 & 10.9 \\
Ce {\sc iii}& 18.0 & 19.6 & 20.2 \\
\hline
Nd {\sc i}  & 4.58 & 4.97 & 5.53 \\
Nd {\sc ii} & 10.9 & 11.1 & 10.7 \\
Nd {\sc iii}& 19.2 & 20.5 & 22.1 \\
\hline
Sm {\sc i}  & 4.83 & 5.33 & 5.64 \\
Sm {\sc ii} & 10.6 & 10.7 & 11.1 \\
Sm {\sc iii}& 20.3 & 21.6 & 23.4 \\
\hline
U {\sc i}  & 4.00 & 5.48 & 6.19 \\
U {\sc ii} & 12.1  & 11.7 & 11.6 \\
U {\sc iii}& 17.4 & 19.2 & 19.8 \\
\hline
\end{tabular}
\label{tab:ionpot}
\end{table}
The overall agreement is good, with the worst comparisons occurring
for the neutral ion stage (particularly for uranium), which is typically
the most difficult to calculate
due to the presence of more bound electrons and the need
to accurately describe the correlation between them.
As expected, the SR values are more accurate than the FR values
for these near-neutral ions for two reasons.
First, the Hartree-Fock approach uses a better (non-local) description
of the exchange interaction between the bound electrons than the 
Dirac-Fock-Slater method, which uses the Kohn-Sham local-exchange
approximation \citep{kohnsham,sampson_physrep}.
Second, the SR approach
uses semi-empirical scale factors to modify the radial integrals that
appear in the configuration-interaction calculation \citep{cowan}.
In any event, the inaccuracies
in the ionization energies have been removed in both the FR and SR
models of the present study by replacing the calculated values with
those appearing in the NIST database. All level energies within an
ion stage were shifted by the same amount when implementing this procedure.
Of course, inaccuracies in the calculated ionization energies
are reflected in the individual level energies as well, but the line positions
are determined by taking the difference of energies within the same ion stage.
So some beneficial cancellation is expected in this regard when systematic
shifts are present within a given ion stage.
The NIST ionization-energy correction was applied to all of the models
considered in this work. An illustration of the ionization balance
that is obtained with these improved energies is provided
in Figure~\ref{fig:ionfrac_nd} for Nd at a mass density
of $\rho = 10^{-13}$~g/cm$^3$, corresponding to the ejecta density at
$\sim 1$~day after the merger (see Figure~3 in \citealt{rosswog14}).
Due to the relatively low densities associated with the dynamical ejecta,
a single ion stage is dominant over a broad range of temperatures.
This behavior is typical for all of the elements considered in this work
because the ionization energy of each of their first three ion stages
is similar (see, for example, Table~\ref{tab:ionpot}).

\subsection{Variant Models}
\label{subsec:models_var}

In addition to calculating FR and SR models, two less-accurate (but faster
to compute)
FR models were generated in order to test the sensitivity of the kilonova
emission to the quality of the atomic data. Configuration interaction (CI)
is a method to better describe the correlation between the bound electrons
of an atom or ion, and typically results in improved level energies
and oscillator strengths (for a more detailed explanation, see, for
example, \citealt{cowan,LANL_suite}). The use of CI is crucial for
obtaining reasonably accurate atomic structure data
for the near-neutral heavy elements considered here. However, due to
the smearing of lines caused by the large velocity gradients in the
ejecta, it is possible that differing amounts of CI could produce
similar spectra, which, if true, would provide more confidence in
the fidelity of the simulated spectra, at least from an atomic
physics perspective.

In order to test this concept, we generated two additional FR models
for Nd: one that includes CI between
only those basis states that arise from the same relativistic configuration
and one that includes CI between only those basis states that arise from the
same non-relativistic configuration. These models are referred to
here as ``FR-SCR'' and ``FR-SCNR'', respectively
(see \citealt{LANL_suite,fontes_cr16} for additional details). 
The FR-SCR model is less accurate than the FR-SCNR model, which is less accurate
than the FR model described above. All three FR models 
contain the same number of fine-structure levels, but their energies
differ due to the different CI treatments. Additionally,
each model contains a different number of lines, as displayed
in Table~\ref{tab:linecomp}. The variant models have fewer lines
than the baseline FR model, 
and those transitions that are common to the three models will
typically be described by different oscillator strengths.
\begin{table}
\centering
\caption{\rm Number of lines per ion stage of neodymium for the
FR, FR-SCNR and FR-SCR models (see text).}
\vspace*{0.5\baselineskip}
\begin{tabular}{lrrr}
\hline
Ion stage   &   \multicolumn{3}{c}{\# of lines} \\
\cline{2-4}
            &    FR    &  FR-SCNR   & FR-SCR  \\
\hline
Nd {\sc i}  & 25,224,451 &  14,330,369  & 2,804,438  \\
Nd {\sc ii} &  3,958,977 &   3,222,445  &   783,275  \\
Nd {\sc iii}&    233,822 &     137,192  &    51,036  \\
Nd {\sc iv} &      5,784 &       5,393  &     2,051  \\
\hline
\end{tabular}
\label{tab:linecomp}
\end{table}

\section{Sample Opacities and Tables}
\label{sec:opac}

\begin{figure*}
\includegraphics[clip=true,angle=0,width=1.0\columnwidth]
{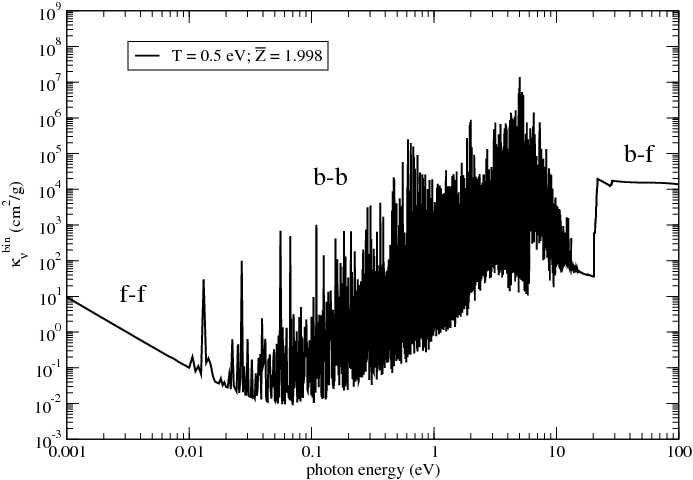}
\includegraphics[clip=true,angle=0,width=1.0\columnwidth]
{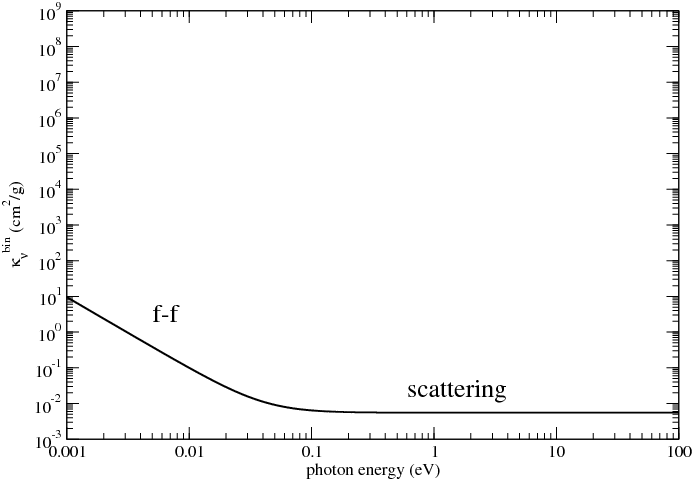}
\caption{
The LTE line-binned opacity for neodymium at $T = 0.5$~eV and
$\rho = 10^{-13}$~g/cm$^3$. The left panel displays the complete opacity,
which includes the bound-bound, bound-free,
free-free and scattering contributions.
The right panel displays only the contributions due to free electrons,
i.e. the free-free and scattering contributions.
The average charge state, $\overline{Z}$, for these conditions is listed
in the legend of the left panel.
}
\label{fig:opac_nd_1}
\end{figure*}

In order to illustrate the basic characteristics of the opacities used
in this study, the LTE monochromatic opacity for Nd is displayed in
Figure~\ref{fig:opac_nd_1}
for typical ejecta conditions of $T = 0.5$~eV and
$\rho = 10^{-13}$~g/cm$^3$.
The left panel displays the complete opacity, with all four
contributions (b-b, b-f, f-f and scattering), while the right panel
shows the contributions that arise only from free electrons (f-f and
scattering) in order to give some indication of the massive
differences that occur when the bound electrons are taken into
account. The b-b contribution was calculated via the line-binned
expression in Equation~(\ref{opac_bb_bin}).
The f-f and scattering contributions were obtained from the
simple, analytic formulas \citep{huebner} associated with Thomson and
Kramers, respectively. The gap between the b-b features and the onset
of the b-f edge occurring at $\sim 20$~eV is due to missing lines that
would be present if more excited configurations had been included in
the model.
Our transport calculations have minimal intensity at these energies.
We note that a mean charge state of $\overline{Z} = 1.998$ is
obtained for these conditions, indicating that the opacity is
dominated by Nd~{\sc iii}.

In this example, the inclusion of the line features dramatically
increases the opacity in the optical range
(1.65--3.26~eV or 0.751--0.380\,$\mu$m) by up to
eight orders of magnitude. The absorption in the near-infrared range below
1.65 eV (0.496--1.65~eV or 10.0--0.751\,$\mu$m) is also greatly increased,
by a few orders of
magnitude in this case, indicating that spectra would more likely be observed
in the mid-infrared range, at least for these specific conditions.

\subsection{Examples of line-binned opacities}
\label{sub:opac_lb}

\begin{figure*}
\includegraphics[clip=true,angle=0,width=1.0\columnwidth]
{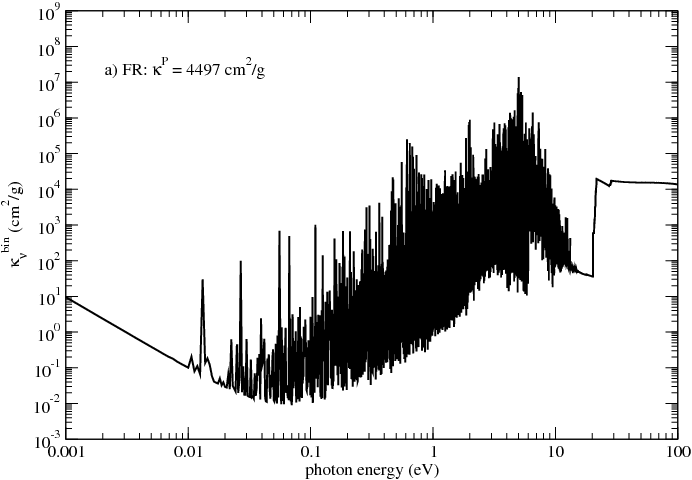}
\hfill
\includegraphics[clip=true,angle=0,width=1.0\columnwidth]
{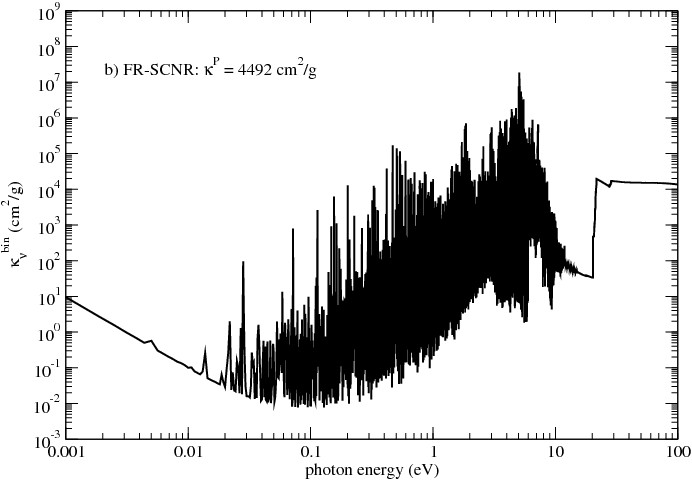}
\includegraphics[clip=true,angle=0,width=1.0\columnwidth]
{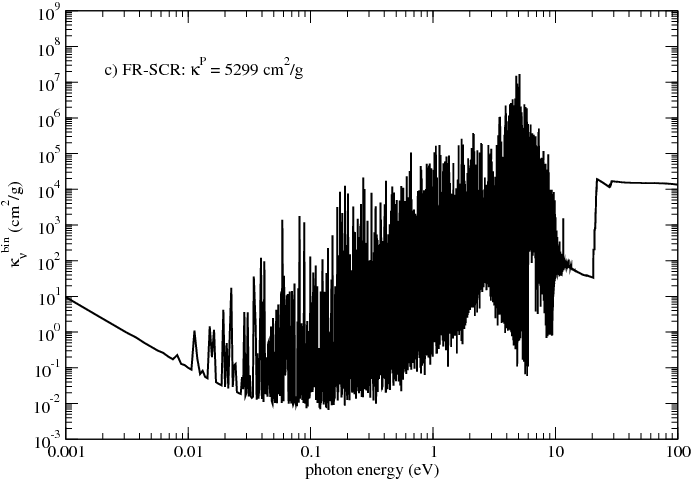}
\hfill
\includegraphics[clip=true,angle=0,width=1.0\columnwidth]
{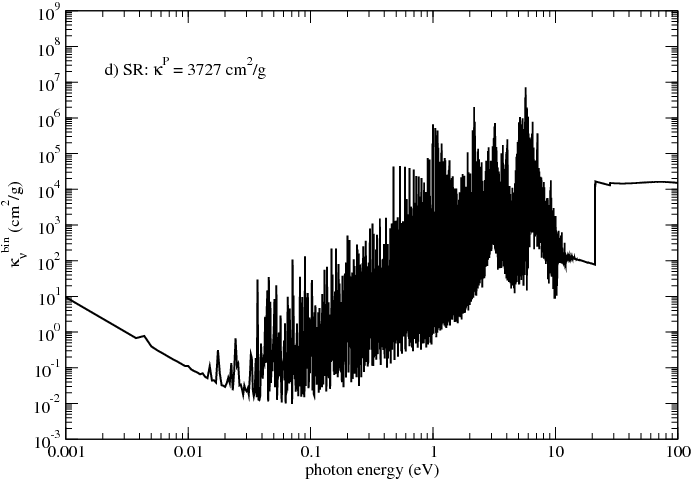}
\caption{
The LTE line-binned opacity for neodymium at $T = 0.5$~eV
and $\rho = 10^{-13}$~g/cm$^3$ using four different models
described in the text:
a) FR, b) FR-SCNR, c) FR-SCR, and d) SR.
The Planck mean opacity, obtained via integration of the line-binned
opacity, is also listed in each panel.
}
\label{fig:opac_nd_3}
\end{figure*}

A more detailed Nd opacity example is provided in Figure~\ref{fig:opac_nd_3},
with the conditions ($T = 0.5$~eV, $\rho = 10^{-13}$~g/cm$^3$) being the
same as those given in Figure~\ref{fig:opac_nd_1}.
Line-binned opacities are presented for the four models described
in Section~\ref{sec:atomic}: FR, FR-SCNR, FR-SCR, SR. The models produce
qualitatively similar results, but there are visible quantitative differences,
as exemplified by the Planck mean opacities displayed in each panel.
The Planck mean opacity is defined in the standard way, i.e.
\begin{equation}
\kappa^{\rm P} \equiv \int_0^\infty B_\nu(T) \kappa_\nu' \,d\nu \Big/
\int_0^\infty B_\nu(T) \,d\nu \,,
\label{planck}
\end{equation}
where $B_\nu(T)$ is the Planck function and $\kappa_\nu'$ indicates
that the scattering contribution is omitted from the
monochromatic opacity. This mean value of the frequency-dependent opacity
permits rough quantitative comparisons between models.
The SR model produces the smallest mean value, given by 3727~cm$^2$/g,
while the least accurate FR-SCR model has a value of 5299~cm$^2$/g,
resulting in a variation of 42\%. The most accurate FR model produces
an intermediate value of 4497~cm$^2$/g, with the FR-SCNR yielding
a similar result. These differences allow us to test the sensitivity
of the kilonova light curves and spectra (see Section~\ref{sec:bin_just})
to changes in the underlying atomic physics models that
are used to construct the opacity.

\begin{figure*}
\includegraphics[clip=true,angle=0,width=0.85\columnwidth]
{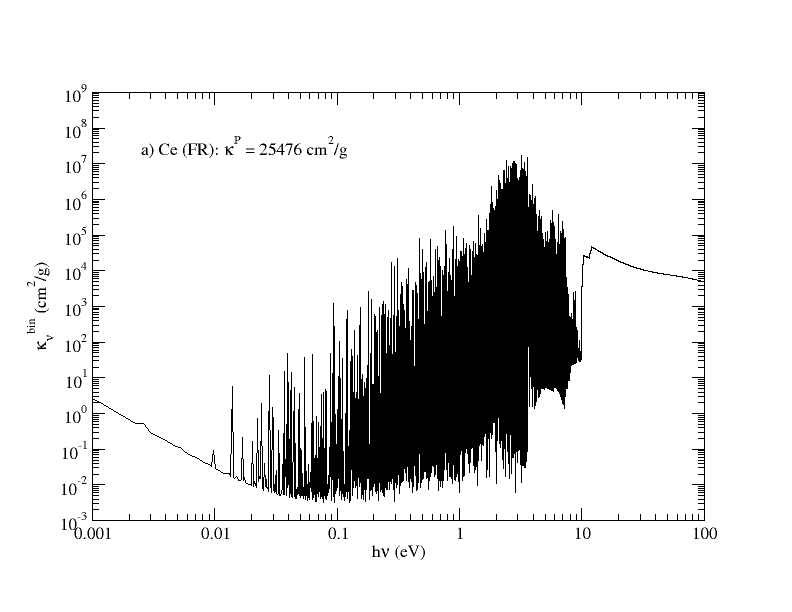}
\includegraphics[clip=true,angle=0,width=0.85\columnwidth]
{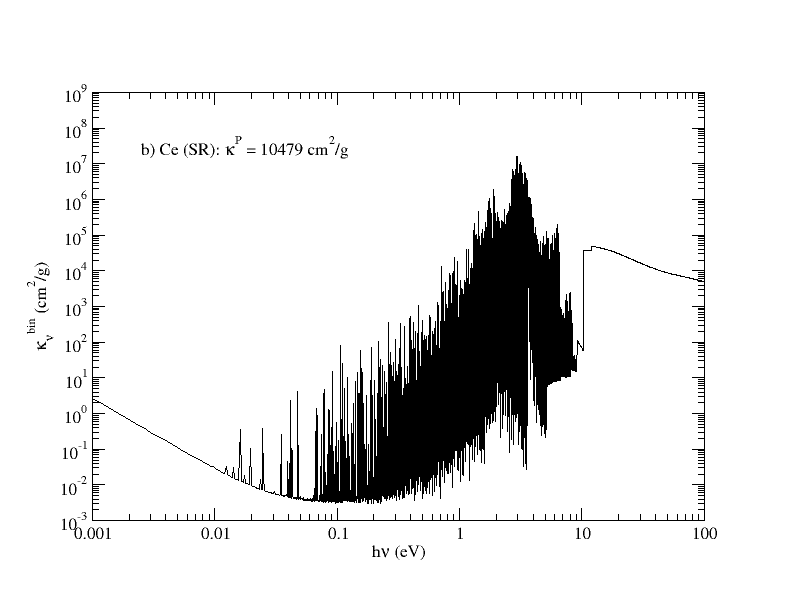}
\hfill
\includegraphics[clip=true,angle=0,width=0.85\columnwidth]
{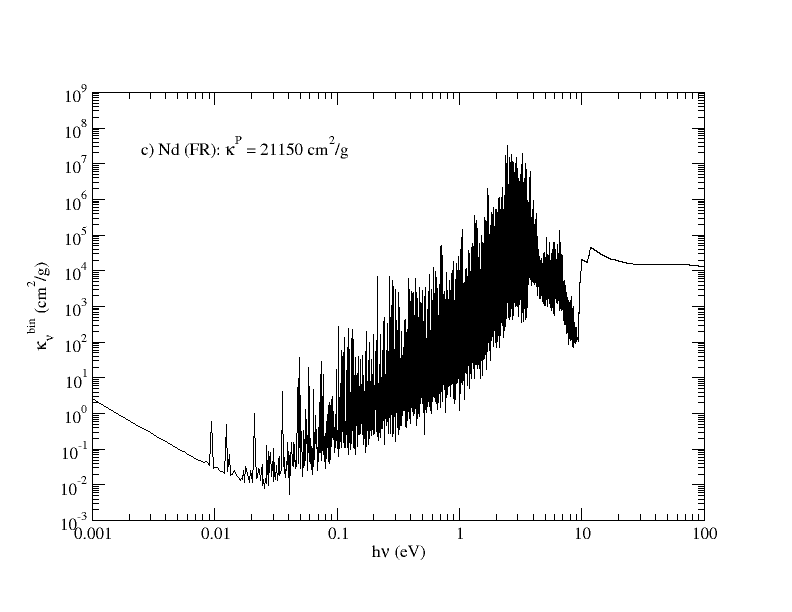}
\includegraphics[clip=true,angle=0,width=0.85\columnwidth]
{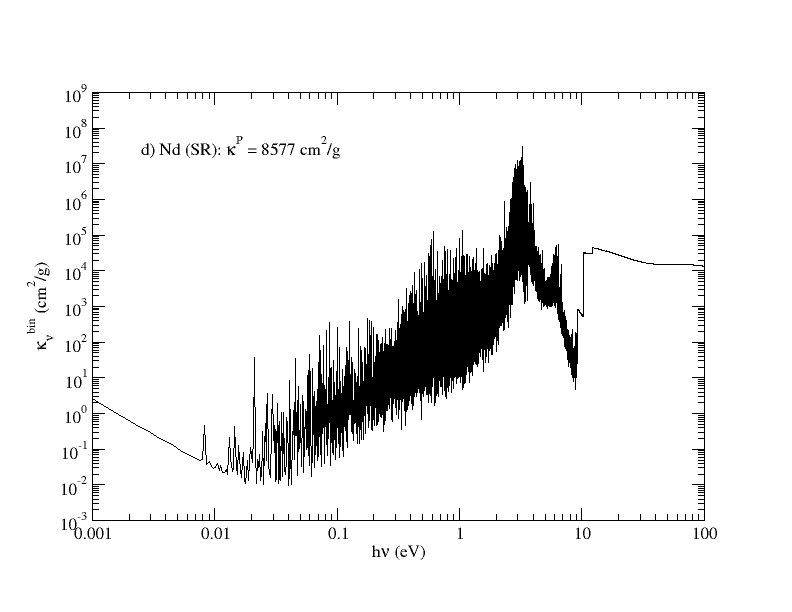}
\hfill
\includegraphics[clip=true,angle=0,width=0.85\columnwidth]
{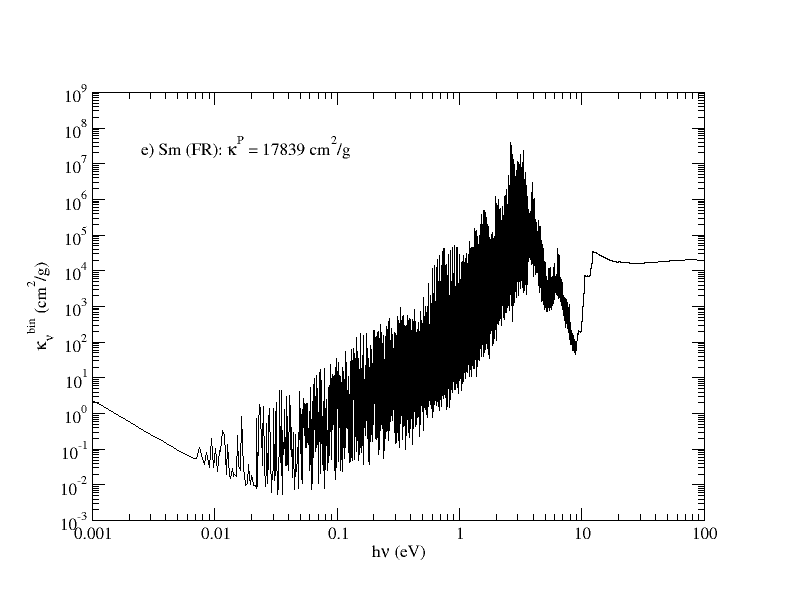}
\includegraphics[clip=true,angle=0,width=0.85\columnwidth]
{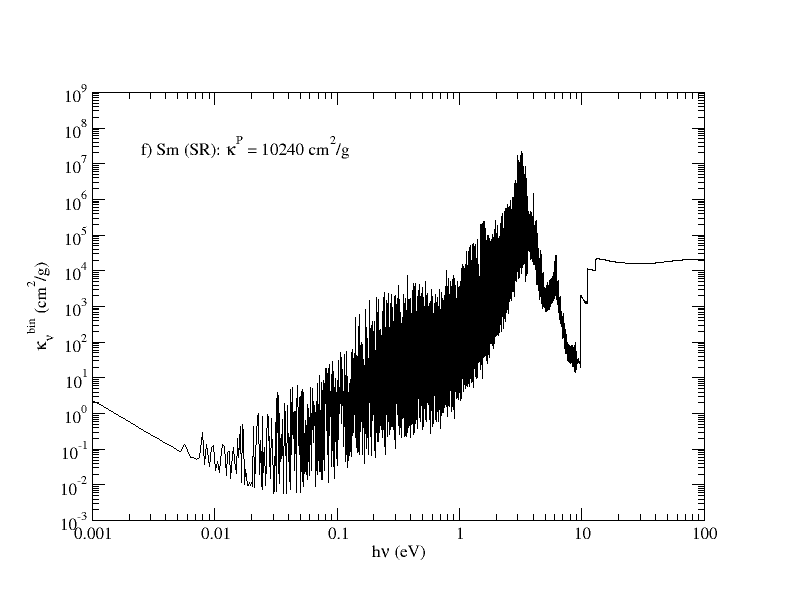}
\hfill
\includegraphics[clip=true,angle=0,width=0.85\columnwidth]
{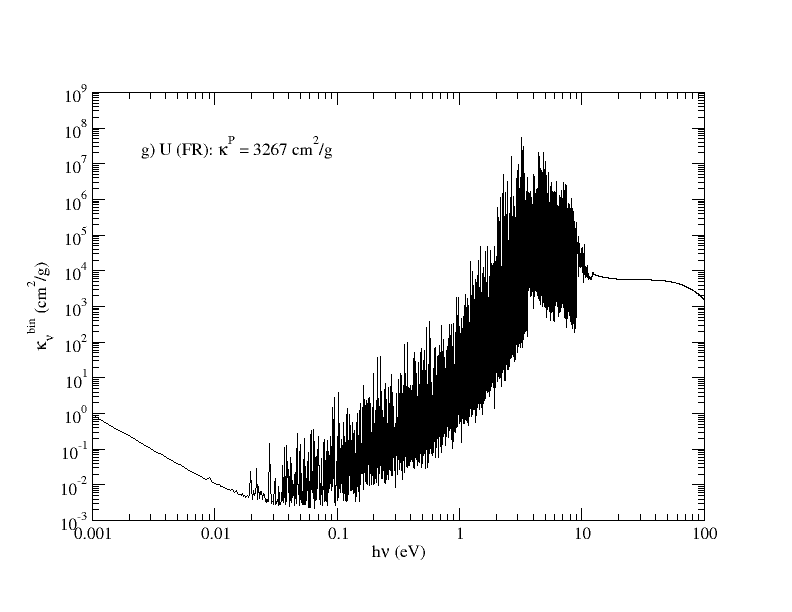}
\includegraphics[clip=true,angle=0,width=0.85\columnwidth]
{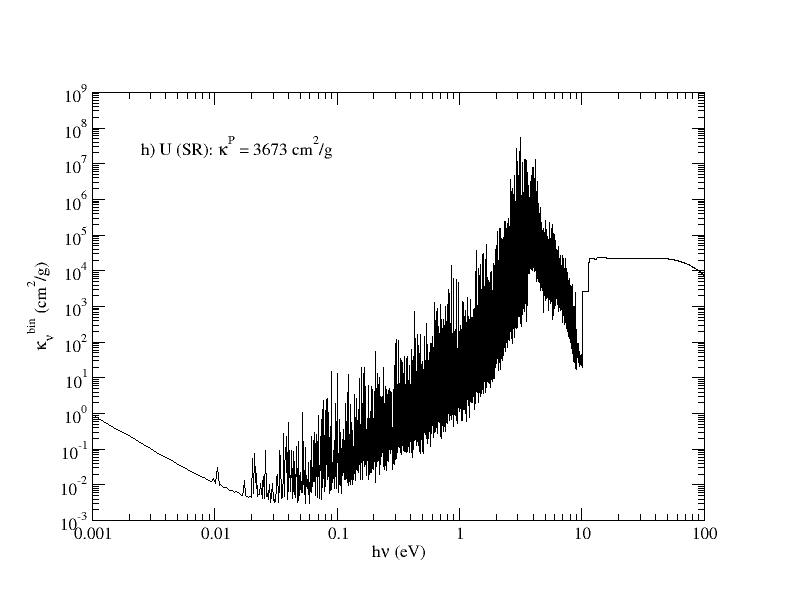}
\caption{
LTE line-binned opacities for four representative elements
(Ce, Nd, Sm and U) at $T = 0.3$~eV and $\rho = 10^{-13}$~g/cm$^3$.
Panels a \& b represent Ce, panels c \& d represent Nd,
panels e \& f represent Sm, and panels g \& h represent U.
The panels in the left column were calculated with the FR approach,
while panels in the right column were calculated with the SR approach.
The Planck mean opacity, obtained via integration of the line-binned
opacity, is also listed in each panel.
}
\label{fig:opac_all}
\end{figure*}

Next, we consider a broader range of elements at a cooler temperature,
presenting line-binned opacities for the four representative elements
(Ce, Nd, Sm and U) at $T = 0.3$~eV, $\rho = 10^{-13}$~g/cm$^3$
in Figure~\ref{fig:opac_all}.
Results are displayed for both FR and SR results in order to compare
these two different atomic physics models for a range of elements.
The opacities for all of these elements display similar qualitative
behaviors: line-dominated absorption that increases with photon energy,
with a peak at $\sim 3$~eV, and a bound-free edge at an energy of $\sim 10$~eV.
These similarities are not surprising because the charge state distribution
for each element is dominated by the second ion stage,
i.e. $\overline{Z} \approx 1$, at these conditions.
Therefore, the bound-bound contribution to the opacity is dominated by lines
associated with the singly ionized stage for each element in this example.
As explained in Section~\ref{subsec:models_base},
and demonstrated in Figure~\ref{fig:ionfrac_nd} and Table~\ref{tab:ionpot},
a single ion stage is dominant over a broad range of temperature at such
low densities for a given element. Furthermore,
the dominant stage is typically the same for all lanthanide (and actinide)
elements due to the similarity in the ionization potentials of their ion
stages.

Despite these comparable trends, an inspection of the frequency-dependent
opacities also reveals the differences inherent in the underlying atomic
energy-level structure for each element. The detailed line structure is
visibly different in each panel of Figure~\ref{fig:opac_all} due
to fundamental atomic physics theory concepts
such as the angular momentum coupling between the various bound electrons,
quantum selection rules for the absorption of photons between different
energy levels, etc. There are also notable differences when comparing
the FR and SR values for a given element. The Planck mean opacity
is, once again, presented in each panel of Figure~\ref{fig:opac_all}
for comparison purposes. The ratio of the FR to SR value of the Planck mean is 
2.43, 2.47, 1.74 and 0.889 for Ce, Nd, Sm and U, respectively.
The variation in this set of ratios gives a basic indication of the uncertainty
in the opacities due to the choice of physics models for this group
of four elements.
Within the FR or SR model, the maximum ratio occurs between Ce and U,
with a value of 7.80 or 2.85, respectively. These two values provide
a rough estimate of how the opacity can vary between different
elements calculated within the same physical framework.
Of course, representing such complex, frequency-dependent absorption features
by a single mean value is a gross approximation. A more realistic comparison
should investigate the use of different opacity models in the simulation
of kilonova emission, which we consider in Section~\ref{sec:bin_just}.

\begin{figure*}
\includegraphics[clip=true,angle=270,width=0.666\columnwidth]
{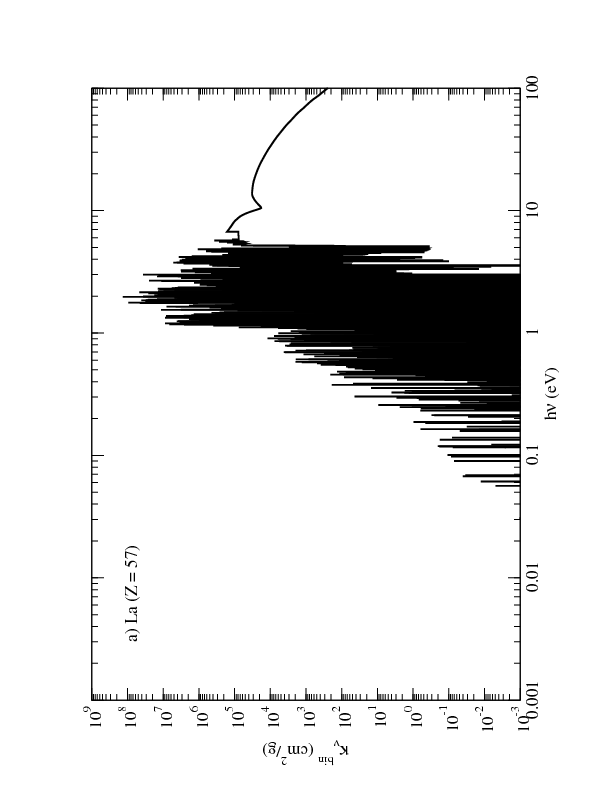}
\includegraphics[clip=true,angle=270,width=0.666\columnwidth]
{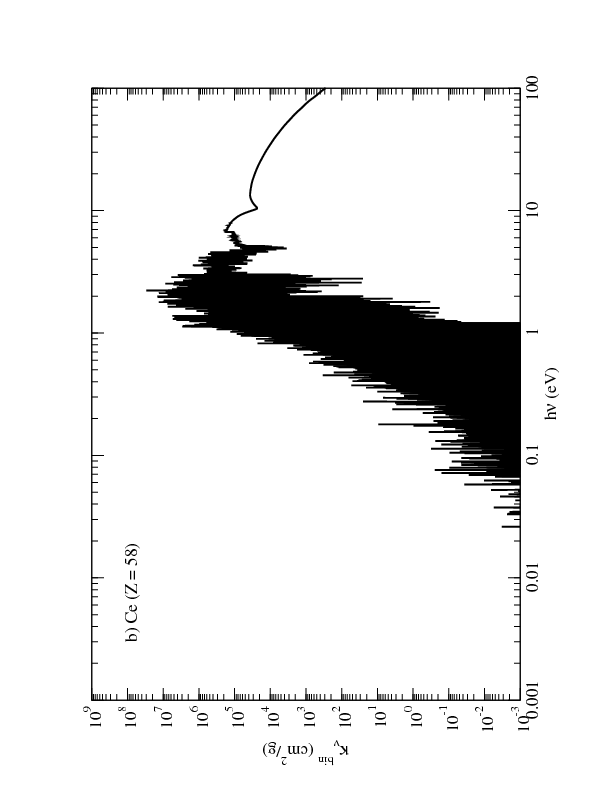}
\includegraphics[clip=true,angle=270,width=0.666\columnwidth]
{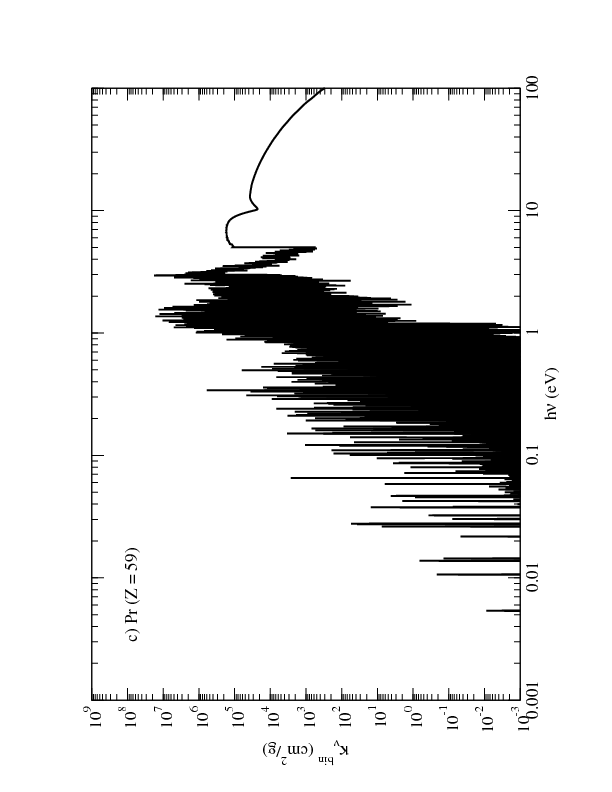}
\includegraphics[clip=true,angle=270,width=0.666\columnwidth]
{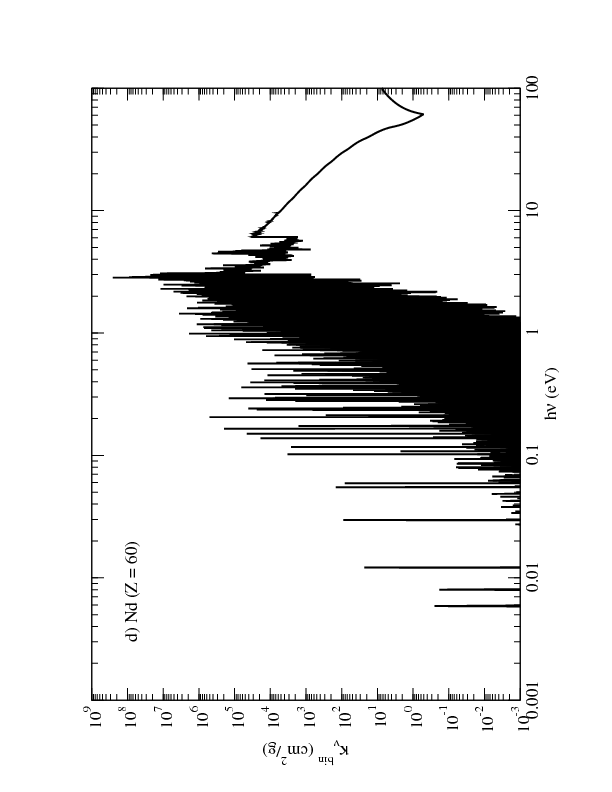}
\includegraphics[clip=true,angle=270,width=0.666\columnwidth]
{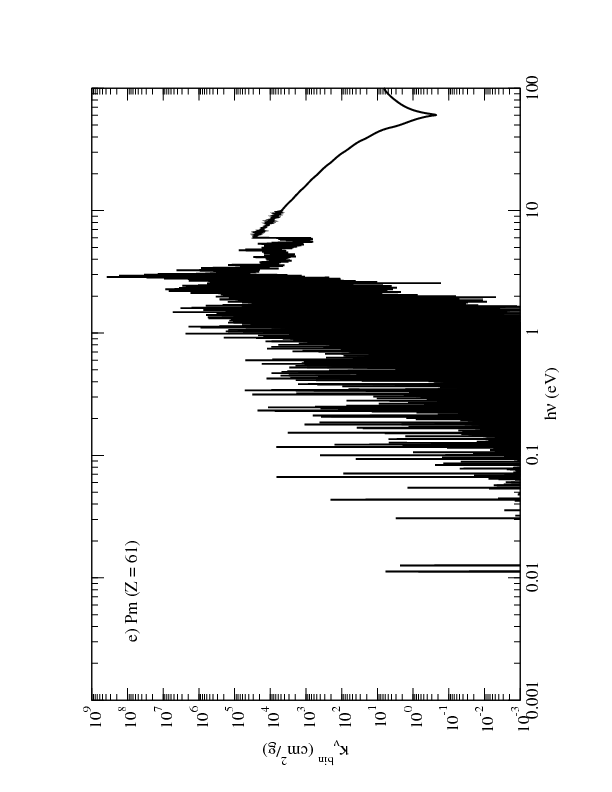}
\includegraphics[clip=true,angle=270,width=0.666\columnwidth]
{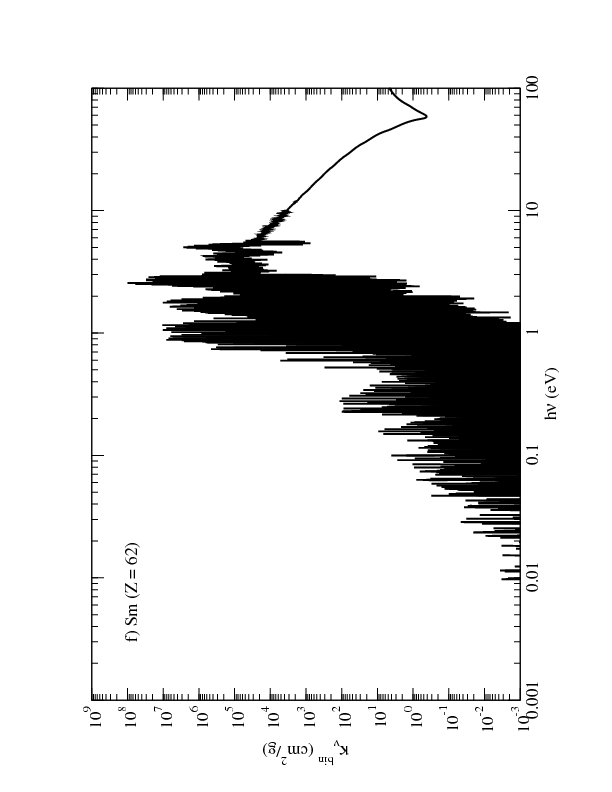}
\includegraphics[clip=true,angle=270,width=0.666\columnwidth]
{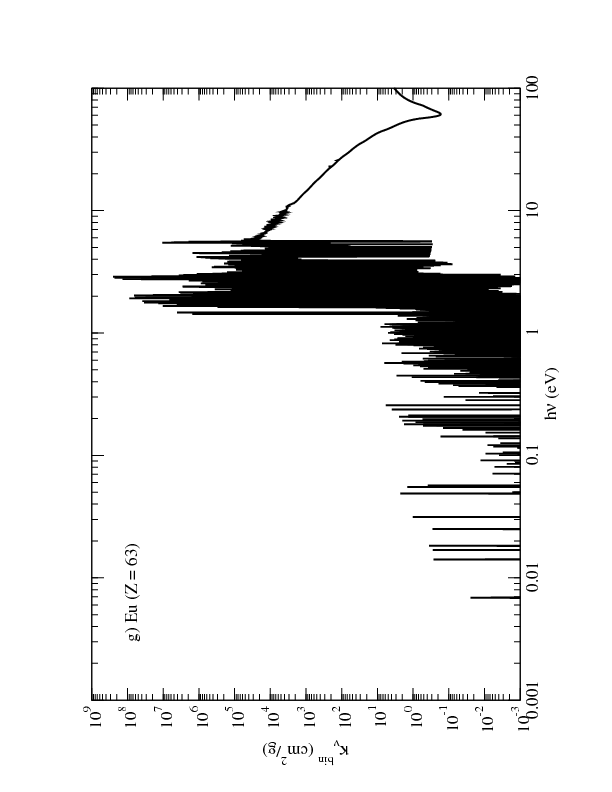}
\includegraphics[clip=true,angle=270,width=0.666\columnwidth]
{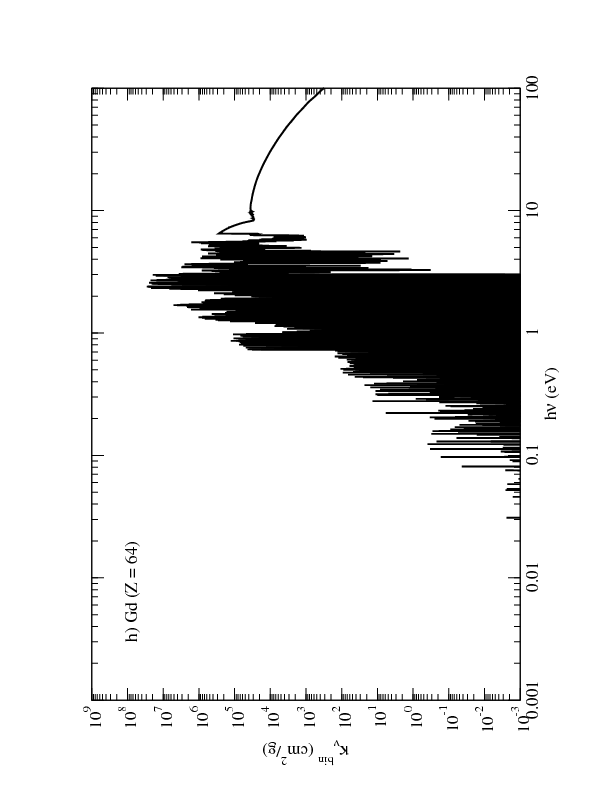}
\includegraphics[clip=true,angle=270,width=0.666\columnwidth]
{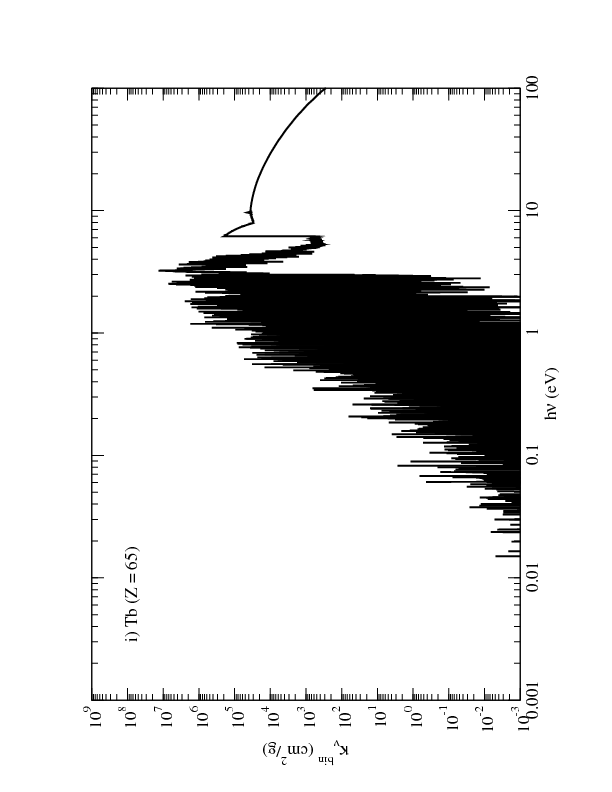}
\includegraphics[clip=true,angle=270,width=0.666\columnwidth]
{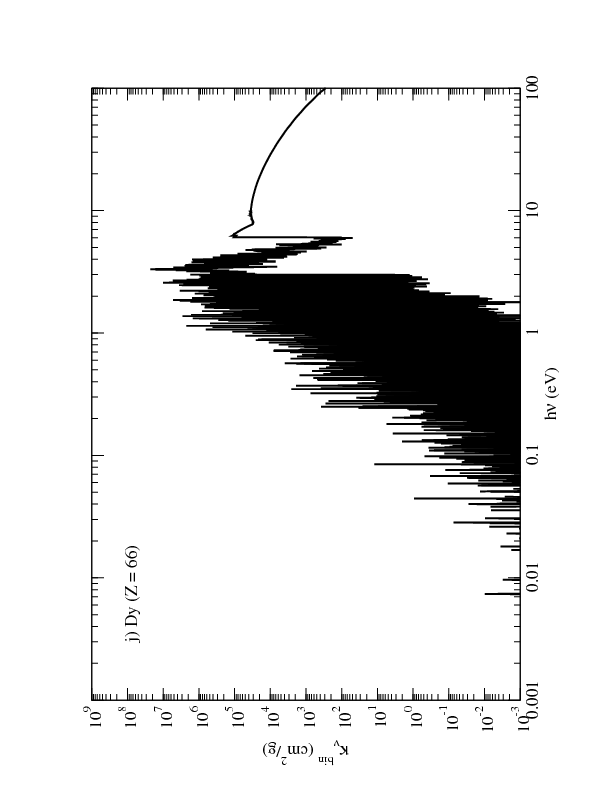}
\includegraphics[clip=true,angle=270,width=0.666\columnwidth]
{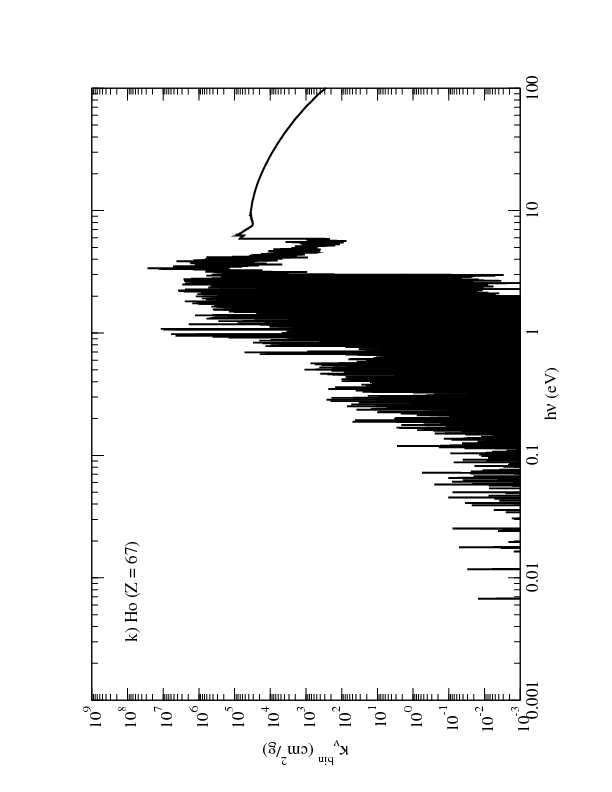}
\includegraphics[clip=true,angle=270,width=0.666\columnwidth]
{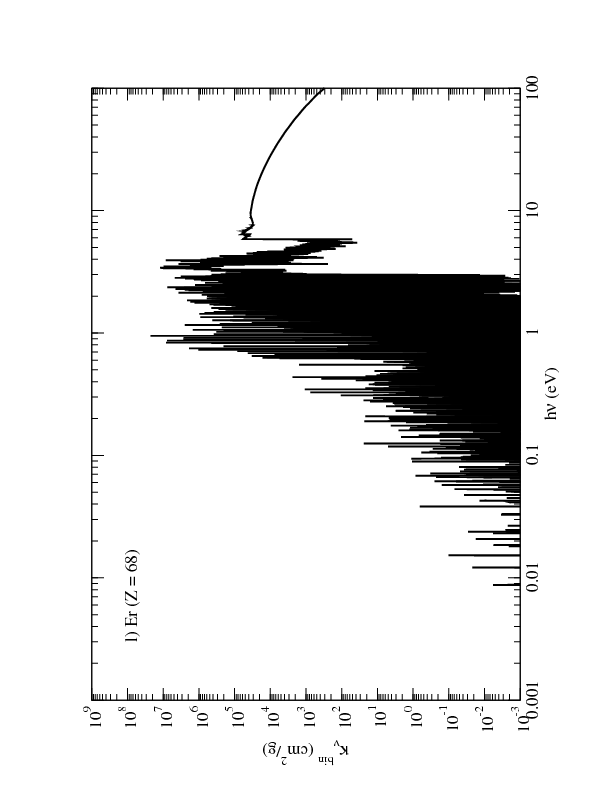}
\includegraphics[clip=true,angle=270,width=0.666\columnwidth]
{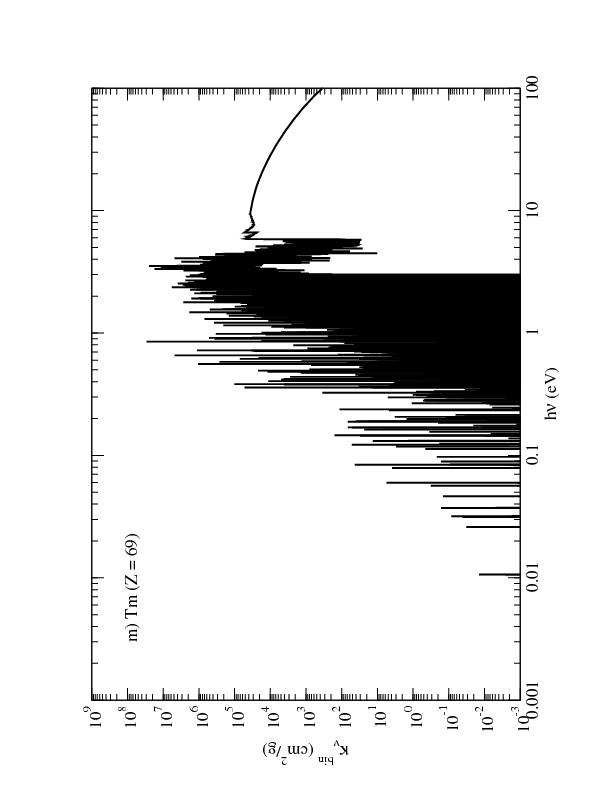}
\includegraphics[clip=true,angle=270,width=0.666\columnwidth]
{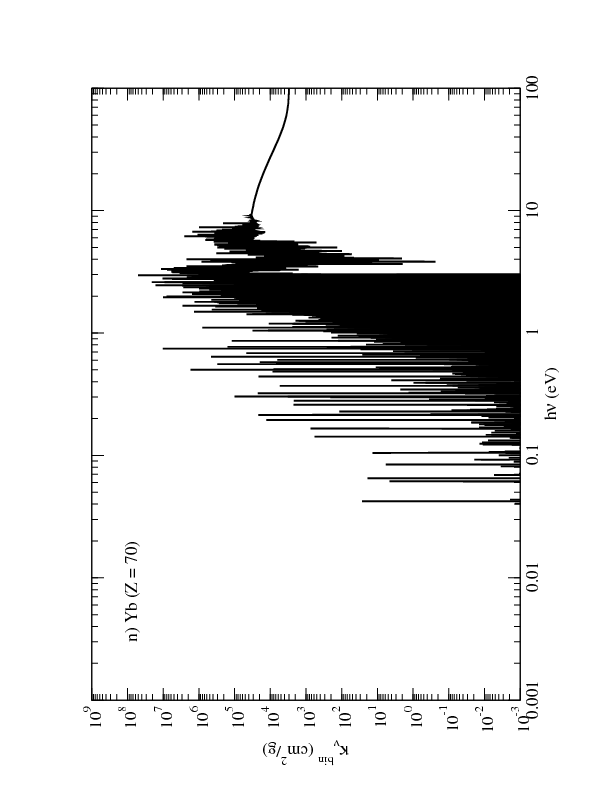}
\includegraphics[clip=true,angle=270,width=0.666\columnwidth]
{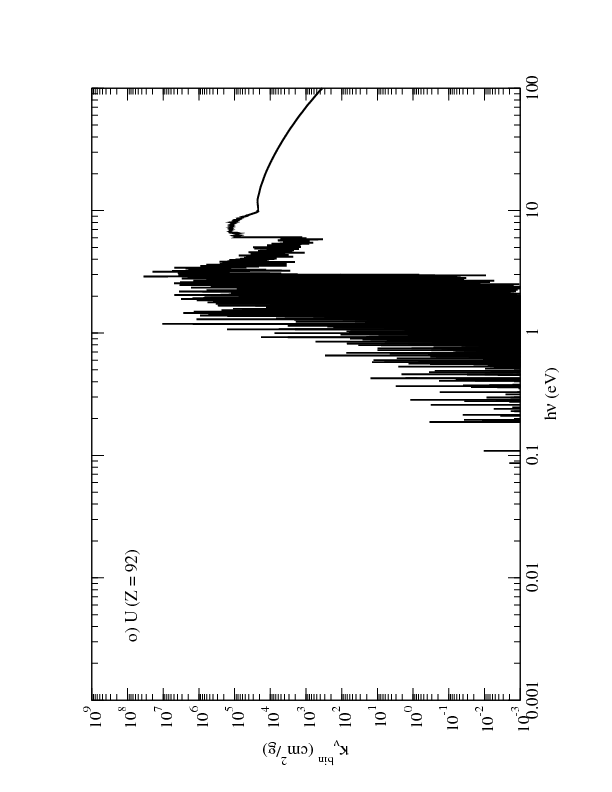}
\caption{The LTE line-binned opacity for all 14 lanthanide elements, as well
as uranium, at $T = 0.1$~eV and $\rho = 10^{-13}$~g/cm$^3$. For these
conditions, the bound-bound contribution to the opacity is dominated by
the first, i.e. neutral, ion stage of each element.
Panels a--n display results for $Z = $57--70, while panel o
displays the uranium ($Z = 92$) result.}
\label{fig:opac_allZ1}
\end{figure*}

\begin{figure*}
\includegraphics[clip=true,angle=270,width=0.666\columnwidth]
{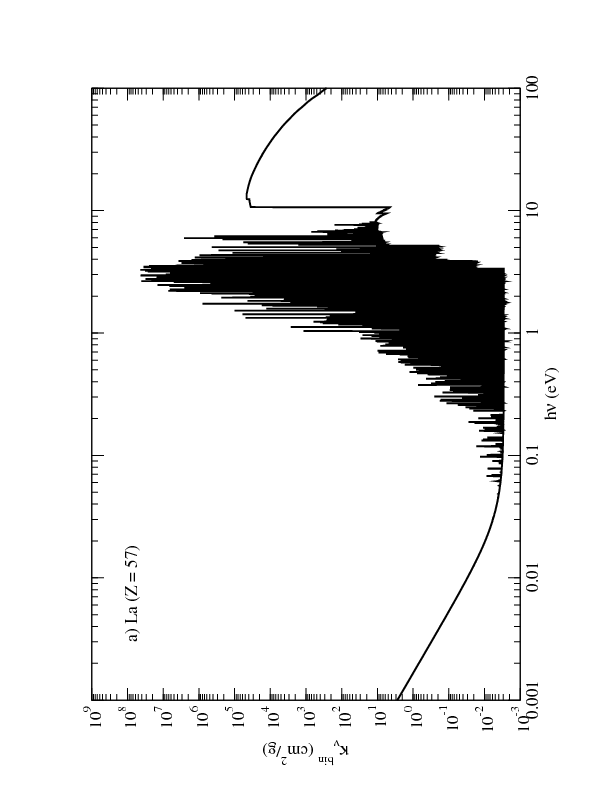}
\includegraphics[clip=true,angle=270,width=0.666\columnwidth]
{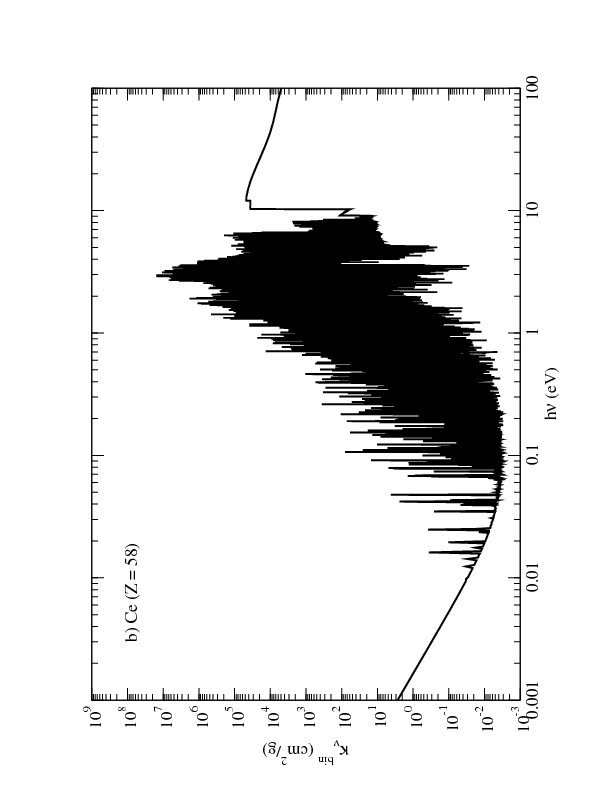}
\includegraphics[clip=true,angle=270,width=0.666\columnwidth]
{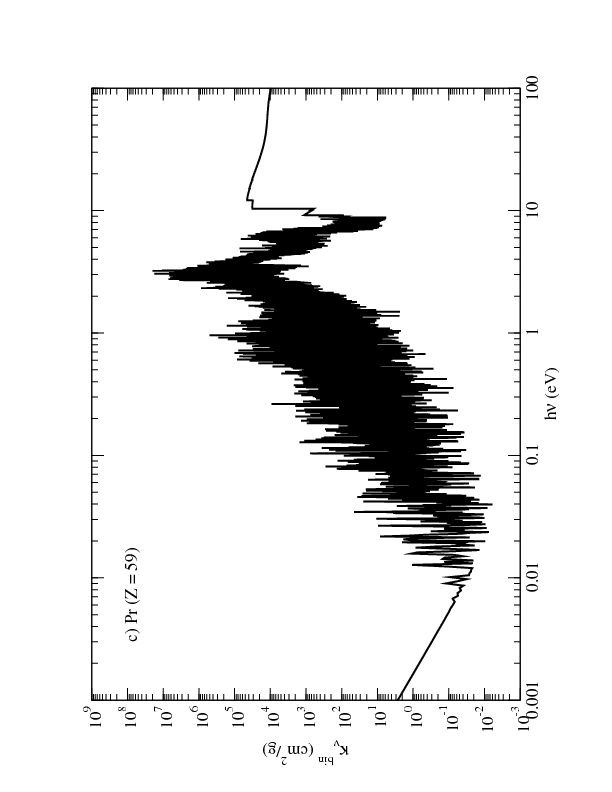}
\includegraphics[clip=true,angle=270,width=0.666\columnwidth]
{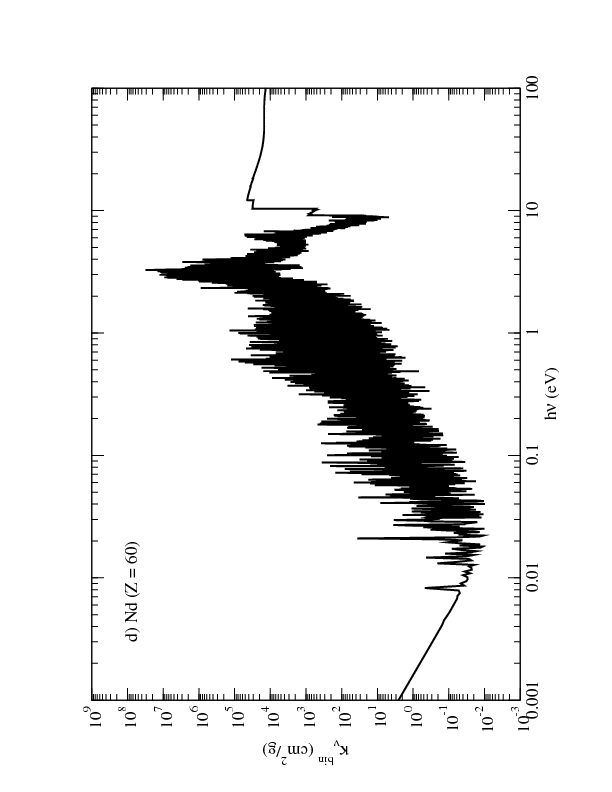}
\includegraphics[clip=true,angle=270,width=0.666\columnwidth]
{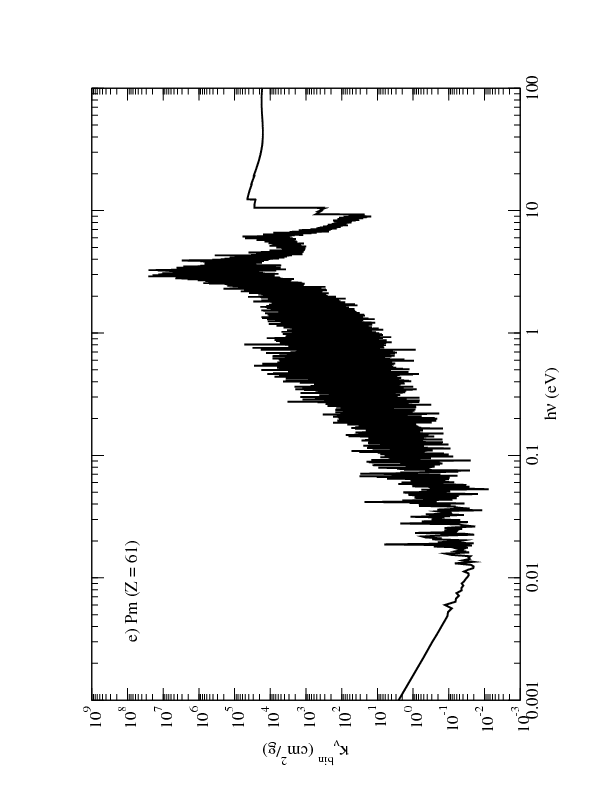}
\includegraphics[clip=true,angle=270,width=0.666\columnwidth]
{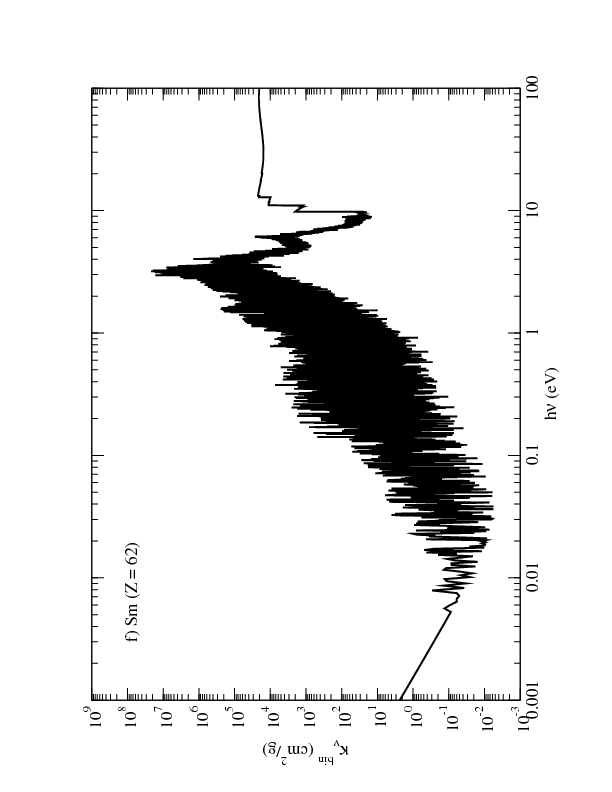}
\includegraphics[clip=true,angle=270,width=0.666\columnwidth]
{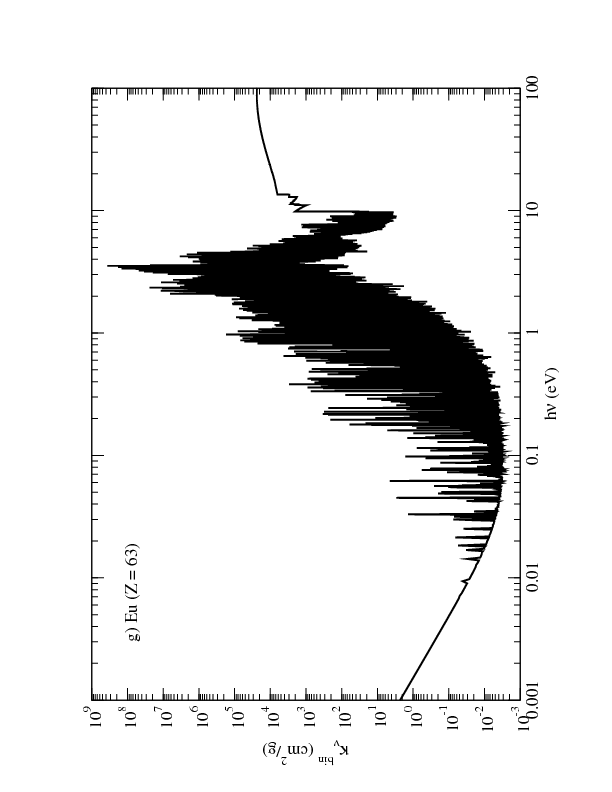}
\includegraphics[clip=true,angle=270,width=0.666\columnwidth]
{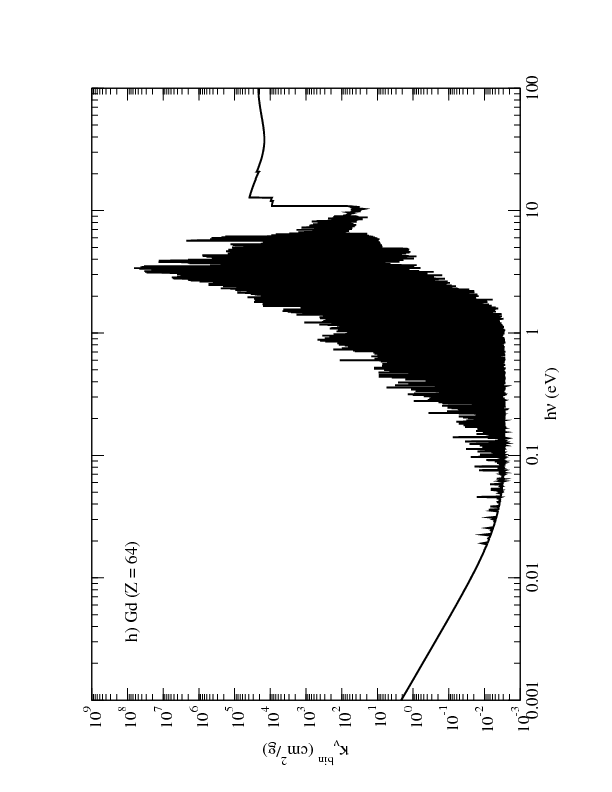}
\includegraphics[clip=true,angle=270,width=0.666\columnwidth]
{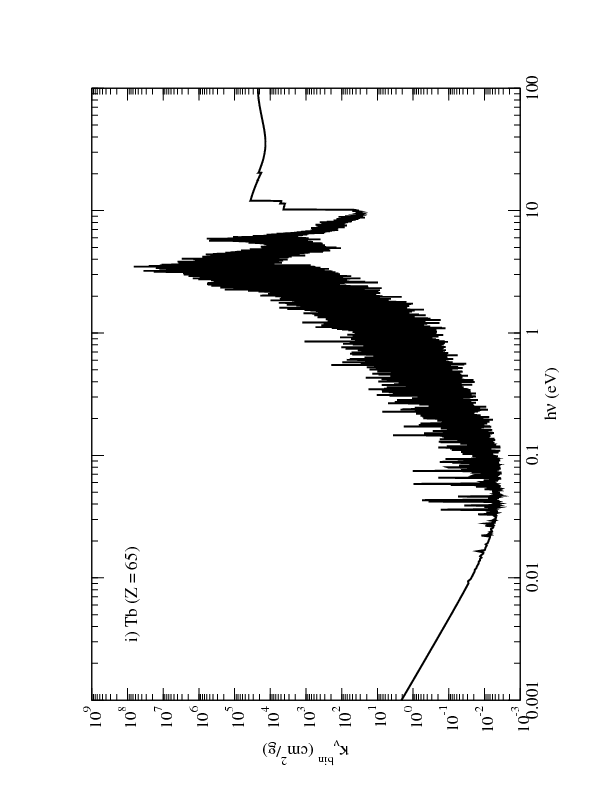}
\includegraphics[clip=true,angle=270,width=0.666\columnwidth]
{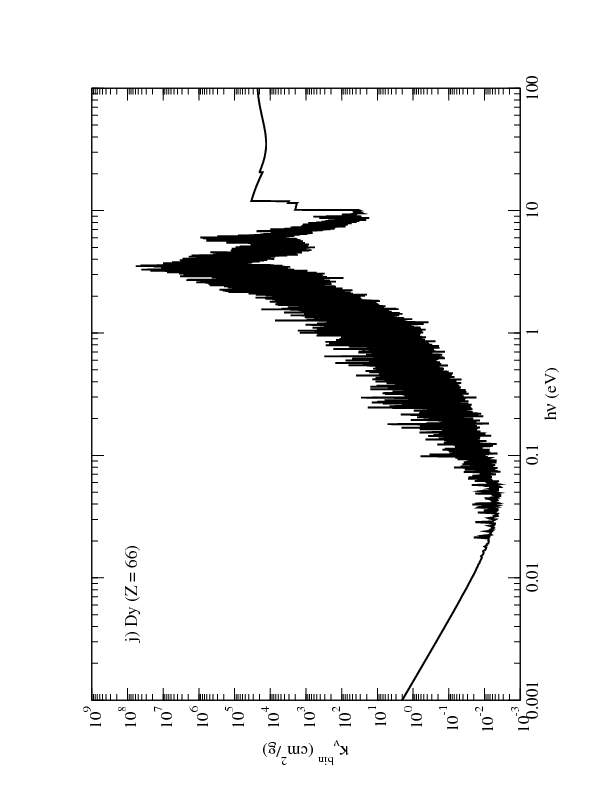}
\includegraphics[clip=true,angle=270,width=0.666\columnwidth]
{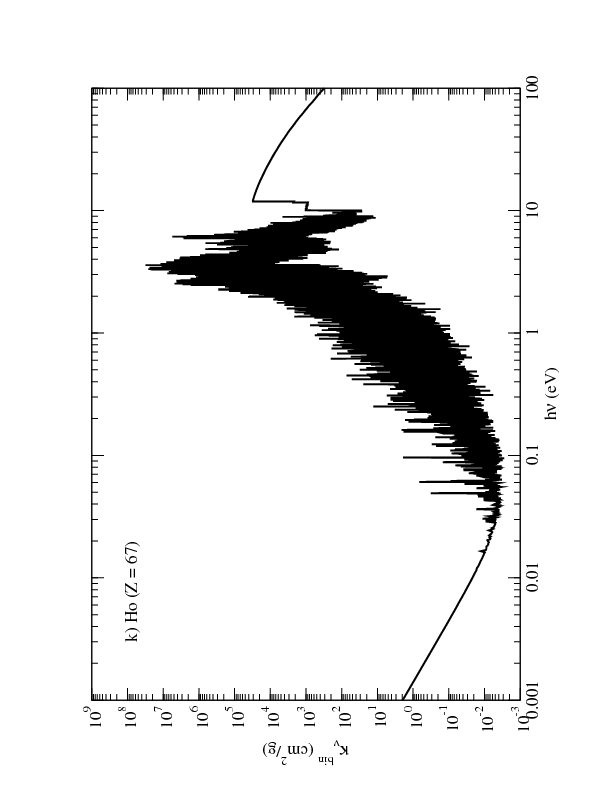}
\includegraphics[clip=true,angle=270,width=0.666\columnwidth]
{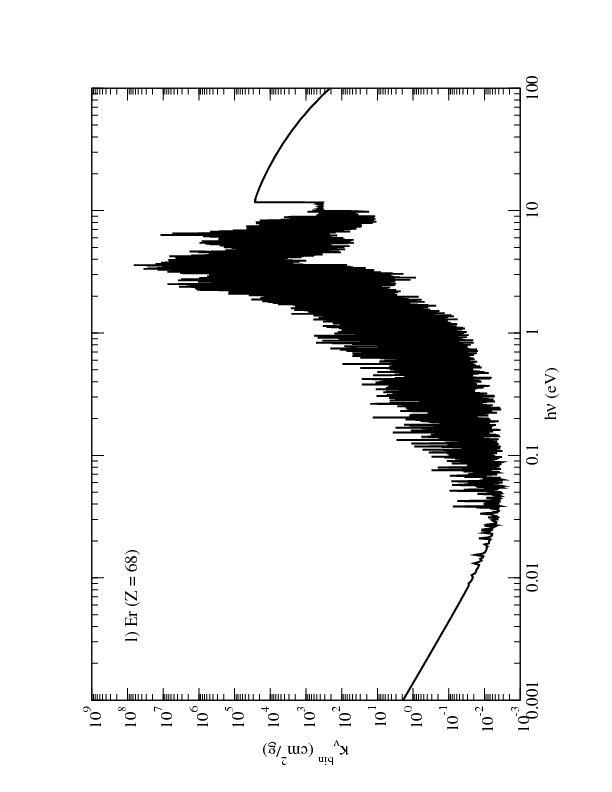}
\includegraphics[clip=true,angle=270,width=0.666\columnwidth]
{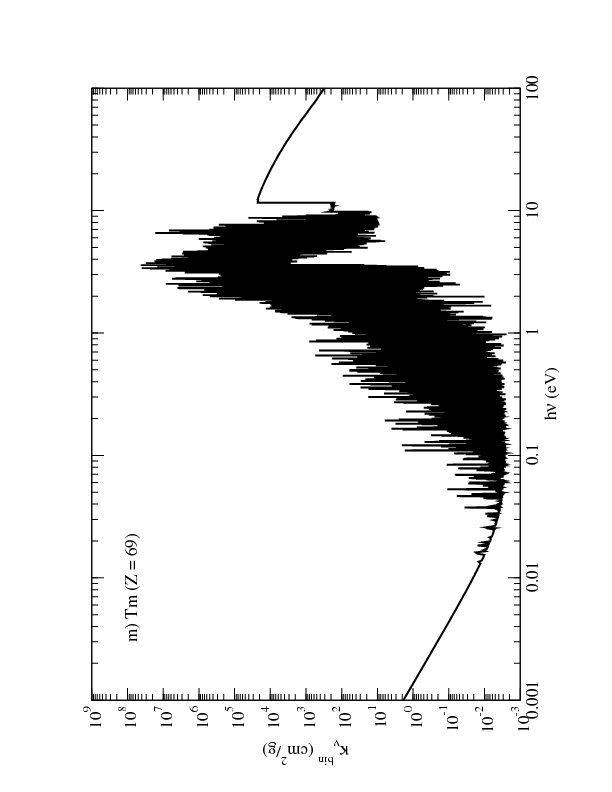}
\includegraphics[clip=true,angle=270,width=0.666\columnwidth]
{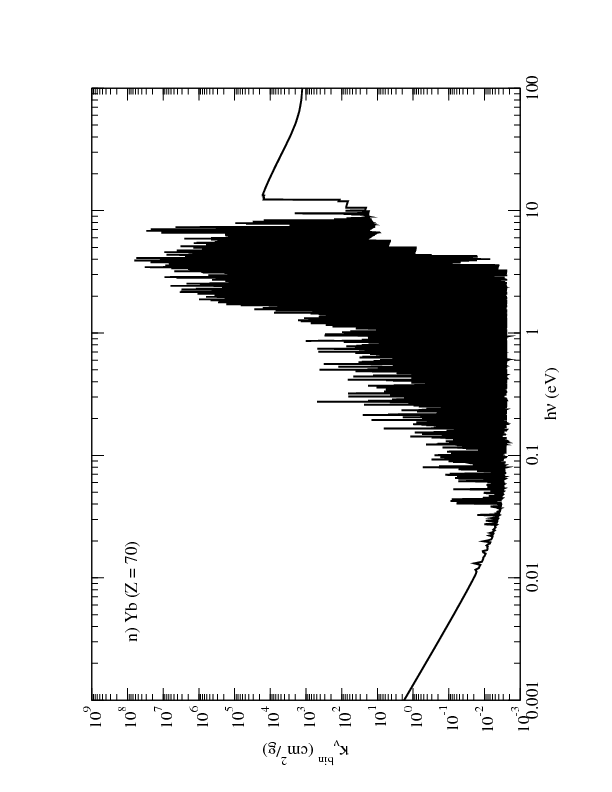}
\includegraphics[clip=true,angle=270,width=0.666\columnwidth]
{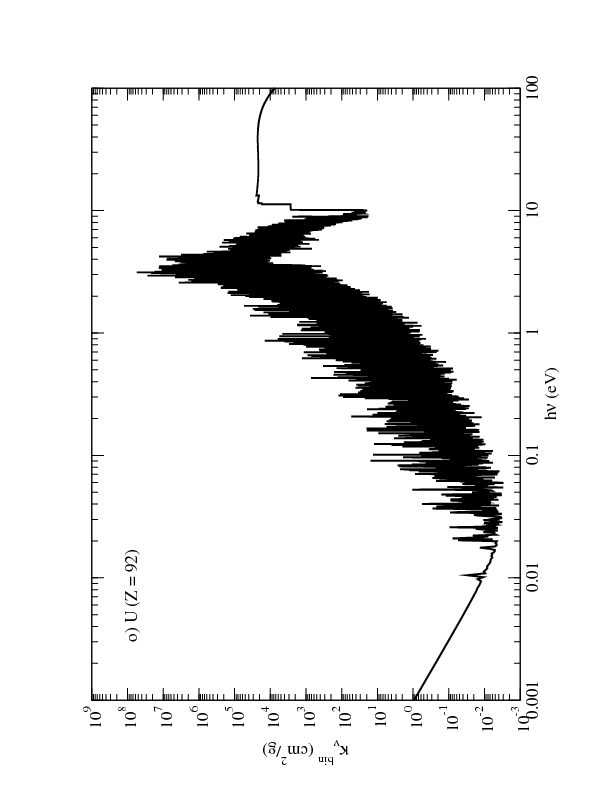}
\caption{The LTE line-binned opacity for all 14 lanthanide elements, as well
as uranium, at $T = 0.3$~eV and $\rho = 10^{-13}$~g/cm$^3$. For these
conditions, the bound-bound contribution to the opacity is dominated by
the second, i.e. singly ionized, ion stage of each element.
Panels a--n display results for $Z = $57--70, while panel o
displays the uranium ($Z = 92$) result.}
\label{fig:opac_allZ2}
\end{figure*}

As a final illustration, we present in Figures~\ref{fig:opac_allZ1}
and \ref{fig:opac_allZ2} the line-binned opacity for all 14 lanthanide
elements, as well as uranium, at two sets of characteristic conditions.
These results were calculated with the SR model.
Once again, a characteristic ejecta density of $\rho = 10^{-13}$~g/cm$^3$
was chosen for both figures.
A temperature of $T = 0.1$~eV was chosen in Figure~\ref{fig:opac_allZ1}
to highlight opacities with a b-b contribution that is
dominated by the first, i.e. neutral, ion stage for all 15 elements.
A higher temperature of $T = 0.3$~eV was chosen in Figure~\ref{fig:opac_allZ2}
to compare opacities with a b-b contribution that is 
dominated by the second, i.e. singly ionized, ion stage of each element.

The qualitative trends in these two figures resemble those discussed above
for Figures~\ref{fig:opac_nd_3} and \ref{fig:opac_all}. For example,
the opacity for all 15 elements in Figures~\ref{fig:opac_allZ1} and
\ref{fig:opac_allZ2} displays line-dominated absorption that
increases with photon energy, peaking at
an energy of about 2--3~eV. In Figure~\ref{fig:opac_allZ1}, the signature
f-f and scattering contributions (see Figure~\ref{fig:opac_nd_1})
are absent because the underlying atomic proceseses of inverse bremsstrahlung
and electron scattering, respectively, require the presence
of free electrons. Since the charge state distribuion is dominated by
the neutral ion stage in this case, $\overline{Z}\approx 0$ and the
free electron density is relatively small. On the other hand,
the f-f and scattering contributions are clearly present
in Figure~\ref{fig:opac_allZ2} for which the temperature, and free electron
density, is higher.

\begin{figure*}
\includegraphics[clip=true,angle=0,width=\columnwidth]
{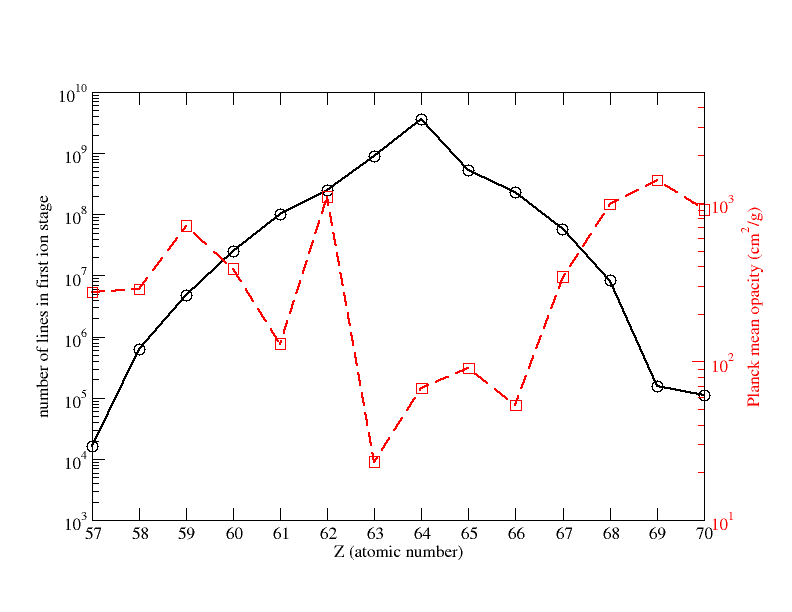}
\hfill
\includegraphics[clip=true,angle=0,width=\columnwidth]
{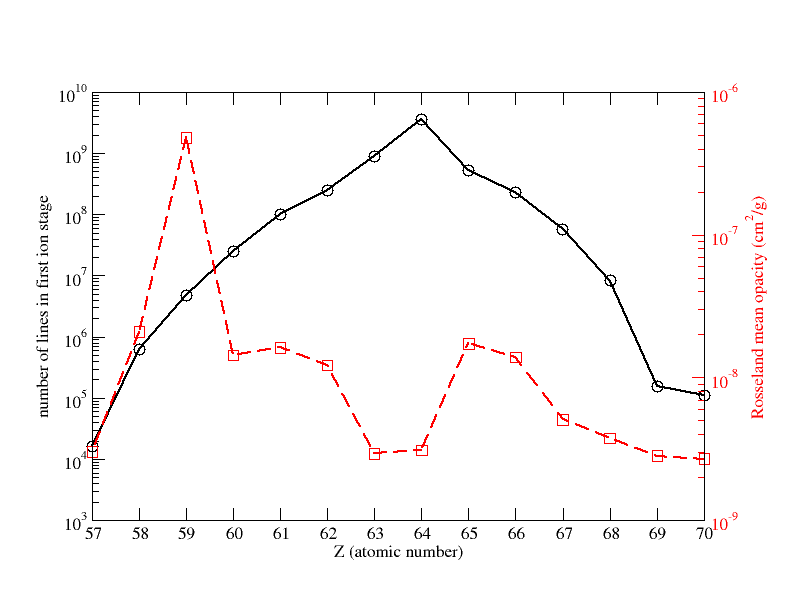}
\caption{The number of lines in the first ion stage versus atomic number, $Z$
(see Table~\ref{tab:configs}).
The mean opacity associated with the line-binned opacities presented
in Figure~\ref{fig:opac_allZ1} is also plotted versus $Z$.
Results are presented for all 14 lanthanide elements ($Z = 57$--70).
In both panels, the number of lines for the first ion stage is represented
by the black solid curve (with circles). This curve is associated with
the left-hand $y$ axis in each panel. The red dashed curves (with squares)
are assocated with the right-hand $y$ axis in each panel, and represent
the Planck mean opacity in the left panel and the Rosseland mean opacity
in the right panel. The mean opacities were calculated at
$T = 0.1$~eV and $\rho = 10^{-13}$~g/cm$^3$, corresponding to the conditions
used in Figure~\ref{fig:opac_allZ1}.
}
\label{fig:opac_allZ1_anal}
\end{figure*}

\begin{figure*}
\includegraphics[clip=true,angle=0,width=\columnwidth]
{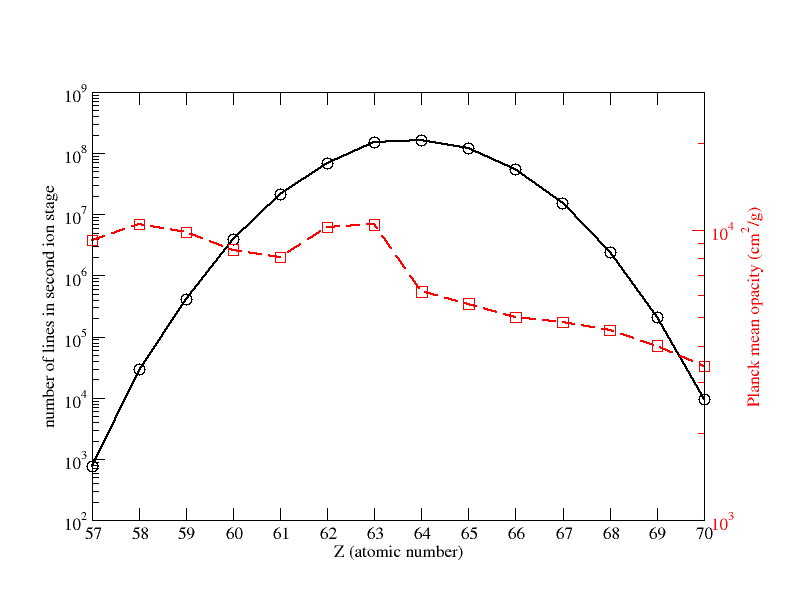}
\hfill
\includegraphics[clip=true,angle=0,width=\columnwidth]
{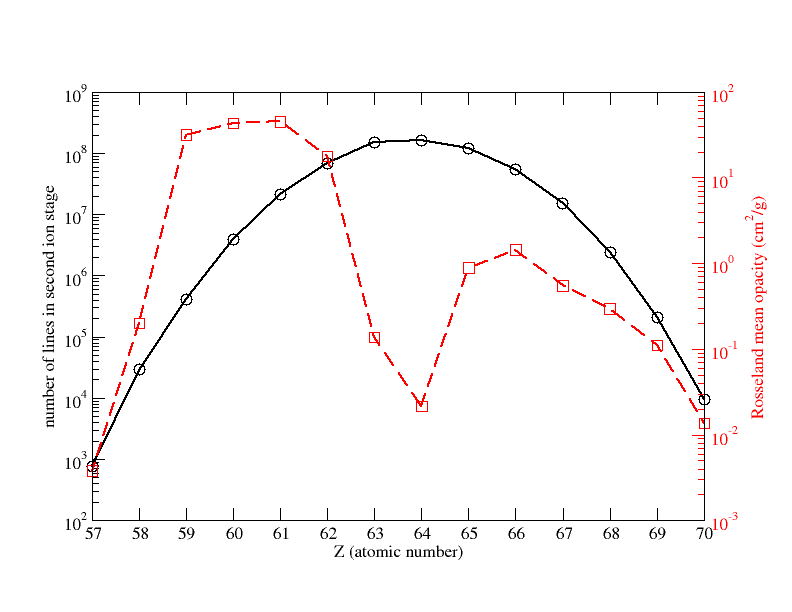}
\caption{The number of lines in the second ion stage versus atomic number, $Z$
(see Table~\ref{tab:configs}).
The mean opacity associated with the line-binned opacities presented
in Figure~\ref{fig:opac_allZ2} is also plotted versus $Z$.
Results are presented for all 14 lanthanide elements ($Z = 57$--70).
In both panels, the number of lines for the second ion stage is represented
by the black solid curve (with circles). This curve is associated with
the left-hand $y$ axis in each panel. The red dashed curves (with squares),
assocated with the right-hand $y$ axis in each panel, represent
the Planck mean opacity in the left panel and the Rosseland mean opacity
in the right panel. The mean opacities were calculated at
$T = 0.3$~eV and $\rho = 10^{-13}$~g/cm$^3$, corresponding to the conditions
used in Figure~\ref{fig:opac_allZ2}.
}
\label{fig:opac_allZ2_anal}
\end{figure*}

In order to provide some qualitative analysis of the line-binned
opacities for the 14 lanthanide elements displayed
in Figures~\ref{fig:opac_allZ1} and \ref{fig:opac_allZ2},
we present Figures~\ref{fig:opac_allZ1_anal} and
\ref{fig:opac_allZ2_anal}, respectively.
In each of the two panels in Figure~\ref{fig:opac_allZ1_anal},
the solid black curve (with circles) represents the number of
lines in the first ion stage, which is the dominant stage for
these conditions, for each element. The number of lines for the various
ion stages can also be found in Table~\ref{tab:configs}. 
Superimposed on this black curve is a red dashed curve (with squares)
that represents the mean opacity obtained from the frequency-dependent
opacities in Figure~\ref{fig:opac_allZ1}. The red dashed curve in the left
panel represents the Planck mean opacity displayed in Equation~(\ref{planck}).
The red dashed curve in the right panel represents the Rosseland mean opacity
defined by the standard harmonically averaged expression
\begin{equation}
[\kappa^{\rm R}]^{-1} \equiv \int_0^\infty B_\nu'(T) [\kappa_\nu]^{-1}
\,d\nu \Big/ \int_0^\infty B_\nu'(T) \,d\nu \,,
\label{rosseland}
\end{equation}
where $B_\nu'(T)$ is the partial derivative of the Planck function
with respect to temperature.

As expected from basic atomic physics considerations,
the number of lines in the first ion stage of the 14 lanthanide elements
peaks at Gd $(Z = 64)$, near the center of the range. The energy level
structure of Gd is the most complicated due to the presence of
the half-filled $4f^7$ subshell, as well as a $5d^1$ subshell,
in the ground-state configuration. This combination of open subshells
results in a maximum in the number of fine-structure levels that are allowed
by quantum physics, i.e. according to the rules of angular momentum coupling
of the bound electrons and the Pauli exclusion principle. This large number
of levels corresponds to a maximum in the number of lines (or transitions)
displayed in Figure~\ref{fig:opac_allZ1_anal}. However, this peak in the
number of lines corresponds to relatively low values in the Planck and
Rosseland mean opacities. This is a counterintuitive result since the
existence of more lines is expected to increase the chances
of photon absorption, which corresponds to higher opacities.
In addition, note that the adjacent element Eu $(Z = 63)$, which
also contains the $4f^7$ subshell in its ground configuration,
similarly corresponds to relatively low values of the two mean opacities.
This behavior can be understood from the fact that
the half-filled $4f^7$ subshell is semi-stable with respect to energy,
i.e. it takes more energy to excite an electron from this type
of configuration than it does from the adjacent ground configurations
that contain the $4f^6$ or $4f^8$ subshell. Thus, the energy-level structure
associated with an element possessing a $4f^7$ subshell in its ground state
can be somewhat different than the other lanthanides, and a significant
fraction of the allowed radiative transitions can occur at higher energies than
for the other lanthandides. For example, according to the NIST database
\citep{nist}, the first excited state of Eu {\sc i} occurs at an energy
of $\sim 1.6$~eV. This value is a factor 2--60 higher than the first
excited state in the neutral stage of most of the other lanthanides,
indicating that the energy-level structure of Eu {\sc i} is different.
Consequently, the number of lines appearing in the Eu (panel g) and
Gd (panel h) opacities displayed in Figure~\ref{fig:opac_allZ1}
is indeed higher than the number for the other elements, but a significant
fraction of those lines occurs at relatively higher energies, above $\sim 1$~eV
in this case. Since the Planck and Rosseland weighting functions peak at
photon energies of $\sim 2.8\times T$ and $\sim 3.8\times T$, respectively,
those higher energy lines do not contribute as much to the mean opacities.
Similar trends are observed in Figure~\ref{fig:opac_allZ2} for which the
second ion stage is dominant. The maximum number of lines occurs for Gd,
but the Rosseland mean value, represented by the red dashed curve in
the right panel, is at a minimum. The Planck
mean does not display such a definitive minimum value, but there is a
significant drop at Gd when traversing the red dashed curve displayed
in the left panel from lower to higher values of $Z$. 

The above trends suggest that, counter to conventional wisdom,
the elements in the middle of the lanthanide range might not produce the
strongest contributions to the opacity of dynamical ejecta in kilonovae.
We emphasize that the use of Planck and Rosseland mean values in the above
analysis is for illustrative purposes and should be interpreted
with a measure of caution.
A study of the relative importance of the various lanthanide elements to the
ejecta opacity could be investigated with kilonova simulations that employ the
frequency-dependent opacities, which is beyond the scope of this work.

\subsection{Line-smeared opacities}
\label{sub:smeared}

In previous work \citep{fontes15,fontes17}, we presented a preliminary attempt
to generate tabular opacities that also preserved the integral of
the monochromatic opacities over frequency. This effort employed a
line-smeared approach to artificially broaden the lines in such a manner
that they could be sufficiently resolved with a typical photon energy grid
(see Section~\ref{sub:opac_tab})
employed in a tabular framework. These line-smeared opacities were
also used in a detailed study of NSM light curves and
spectra \citep{wollaeger18}. Thus, some commentary about how
the line-smeared and line-binned opacities compare is provided here.

\begin{figure}
\includegraphics[clip=true,angle=0,width=\columnwidth]
{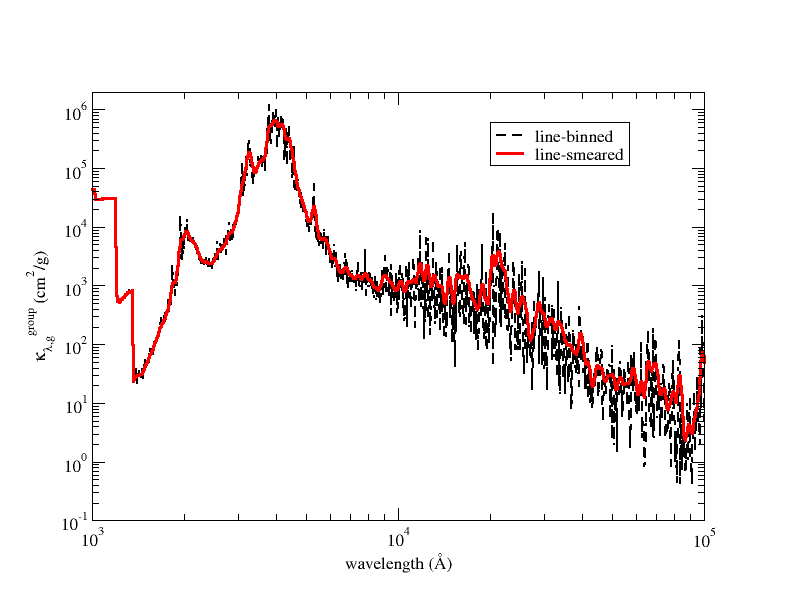}
\caption{
The LTE multigroup opacity of Nd at $T = 0.3$~eV
and $\rho = 10^{-13}$~g/cm$^3$. Results are presented for the
line-binned (dashed black curve) and line-smeared (solid red curve) methods
using 1,024 wavelength groups (see text for details).
}
\label{fig:opac_bin_smear_nd}
\end{figure}

The present line-binned approach
accomplishes the desired goal of preserving the area under the
opacity curve without the concern of losing b-b opacity due to a lack
of photon energy resolution. The use of a discrete sum in the line-binned
approach, see Equation~(\ref{opac_bb_bin}),
ensures that the area will be conserved and, furthermore, allows
fine-structure detail to be more easily captured in a tabular representation
of the opacities.
As a specific example, we present a comparison of the LTE Nd multigroup opacity,
$\kappa^{\rm group}_{\lambda,{\rm g}}$,
generated with the line-binned and line-smeared methods
in Figure~\ref{fig:opac_bin_smear_nd} for the
characteristic conditions of $T = 0.3$~eV and $\rho = 10^{-13}$~g/cm$^3$.
The multigroup opacities were generated using the wavelength version
of the frequency-group opacity displayed in Equation~(\ref{opac_bb_group}),
with a group structure of 1,024 logarithmically
spaced wavelength points (see Section~\ref{sub:lc} for details).
As expected,
the overall agreement between the two curves is good, but the line-binned curve
clearly displays more fine-structure detail at higher wavelengths.
This additional detail in the line-binned curve
is a result of summing the unbroadened lines within a wavelength bin,
rather than smearing the lines across a number of bins, in conjunction with
the fact that the resolution of the groups are on par with the
resolution of the bins at higher wavelengths.
Despite these differences, the line-binned and line-smeared opacities yield
similar kilonova spectra (see Section~\ref{subsub:comp_full} and, specifically,
Figure~\ref{fig:p0_1}).

\subsection{Expansion opacities}
\label{sub:expansion}

In order to provide meaningful comparisons with other works, we also consider
the more traditional approach of expansion opacities when modeling light curves
and spectra in Section~\ref{sec:bin_just}.
Therefore, a brief overview of this approach is provided here.
As mentioned previously,
the expansion-opacity method \citep{sobolev60,castor74,karp77} employed
by \citet{kasen13,barnes13} to simulate kilonovae light curves
applies to the bound-bound contribution to the
opacity and involves a discrete sum over all lines. The approach relies on the
assumption of a homologous expansion and is characterized by an expansion
time, $t_{\rm exp}$. The relevant wavelength range is divided into bins denoted
by index $j$, $\Delta\lambda_j$, and all lines within a bin are summed
to obtain the opacity for that range. The expression for the opacity
associated with bin (or group) $j$ is given by
\begin{equation}
\kappa^{\rm b-b}_{\rm exp}(\Delta\lambda_j) = \frac{1}{\rho c t_{\rm exp}}
\sum_{i \in \Delta\lambda_j} \frac{\lambda_i}{\Delta\lambda_j}
(1 - e^{-\tau_i}) \,,
\label{opac_bb_exp}
\end{equation}
where $t_{\rm exp}$ is the time since mass ejection, the summation index $i$
extends over
all bound-bound transitions that reside in bin $j$,
$\lambda_i$ is the rest wavelength associated with transition $i$,
and $\tau_i$ is the corresponding Sobolev optical depth, i.e.
\begin{equation} 
\tau_i = \frac{\pi e^2}{m_e c}\, N_i\, |f_i|\, t_{\rm exp}\, \lambda_i \,,
\label{tau_sob}
\end{equation}
which is the Doppler-corrected line optical depth.
Due to the presence of $t_{\rm exp}$ in the exponential
of Equation~(\ref{opac_bb_exp}), it is not convenient to construct
tables of the expansion opacities for kilonova modeling.

\begin{figure}
\includegraphics[clip=true,angle=0,width=1.0\columnwidth]
{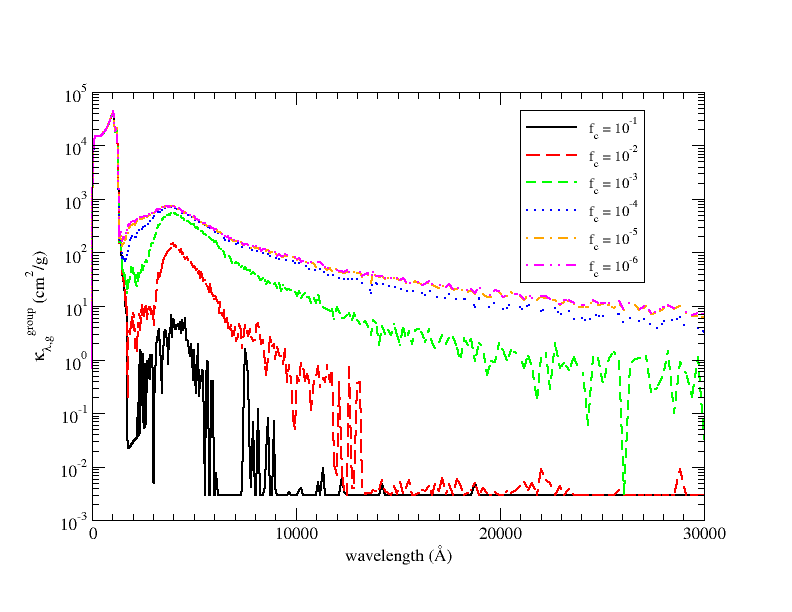}
\caption{
The LTE opacity of Nd at $T = 4,000$~K ($0.345$~eV)
and $\rho = 10^{-13}$~g/cm$^3$ using the expansion-opacity method.
The time since ejection was chosen to be $t_{\rm exp} = 1$~day, along
with a wavelength group structure of $\Delta \lambda = 0.01 \lambda$.
These specifications are the same as those considered in Figure~8
of \citet{kasen13}. The various curves correspond to different values
of the oscillator strength cutoff paramater, $f_c$, as follows: 
solid (black) curve represents $10^{-1}$,
long-dashed (red) curve represents $10^{-2}$,
short-dashed (green) curve represents $10^{-3}$,
dotted (blue) curve represents $10^{-4}$,
dash-dotted (orange) curve represents $10^{-5}$, and
dash-double-dotted (magenta) curve represents $10^{-6}$.
Convergence is obtained for a value of $f_c = 10^{-6}$.
}
\label{fig:opac_exp_nd}
\end{figure}

As a specific example, we present in Figure~\ref{fig:opac_exp_nd} the expansion
opacity of Nd at $T = $ 4,000~K ($0.345$~eV) and $\rho = 10^{-13}$~g/cm$^3$,
generated with the FR model.
In this example, the opacity is plotted versus wavelength instead of energy,
and the particular conditions were chosen in order to facilitate a direct
comparison with Figure~8 of \citet{kasen13}.

There are six curves displayed in Figure~\ref{fig:opac_exp_nd}, corresponding
to different values of the oscillator strength cutoff,
$f_c = 10^{-x}$, with $x =$~1--6.
Convergence is demonstrated for $f_c = 10^{-6}$, which is the same value
that we chose when constructing our tabular, line-binned opacities.
(See Section~\ref{subsec:compframe}.)
The use of such tabular opacities is particularly convenient when exploring
opacity models constructed with different line strength cutoff values.
A sensitivity study of light-curve and spectral modeling to the value
of $f_c$ is presented in Section~\ref{sec:bin_just}.

A comparison of Figure~\ref{fig:opac_exp_nd} with Figure~8 of \citet{kasen13}
allows a relatively direct comparison of the underlying opacity data employed
in each investigation, in contrast to comparisons of spectral quantities
resulting from radiation transport simulations that sample the opacity
over a range of physical conditions. We note that the upper four curves
in Figure~\ref{fig:opac_exp_nd} are qualitatively similar to the curves
displayed in Figure~8 of \citet{kasen13}, with the peak value of the
b-b contribution occurring at $\sim$5,000~\AA\ and monotonically
decreasing at higher wavelengths.
However, the peak value is about three times larger in the present case,
providing a rough measure of the uncertainty in current opacity
calculations as they pertain to kilonova conditions.

This discrepancy is somewhat surprising due to the fact that the same list
of configurations, resulting in the same number of lines (see the
Nd data listed in Table~\ref{tab:configs}), was used in both cases.
The differences are perhaps an indication of how difficult it
is to perform accurate atomic structure calculations for such complicated
atoms and ions. An alternative explanation is that the curves displayed in
Figure~8 of \citet{kasen13} were generated with a less complete set
of lines. This potential explanation is supported by the observation that
the short-dashed (green) curve in Figure~\ref{fig:opac_exp_nd}, which was
generated with a value of $f_c = 10^{-3}$, is in better agreement
with \citet{kasen13} over the entire wavelength range. This improved
agreement includes the large oscillatory behavior at high wavelengths,
although our green curve is still a factor of two higher at the peak value
occurring at $\sim$5,000~\AA.
Another qualitative difference occurs at higher
wavelengths, where there appear to be more points in the curves
of \citet{kasen13}. We were able to obtain similar behavior (not shown) in the
high-wavelength region by employing a linearly spaced wavelength grid,
rather than the logarithmically spaced grid
obtained from the prescription $\Delta \lambda = 0.01 \lambda$.

\begin{figure}
\includegraphics[clip=true,angle=0,width=1.0\columnwidth]
{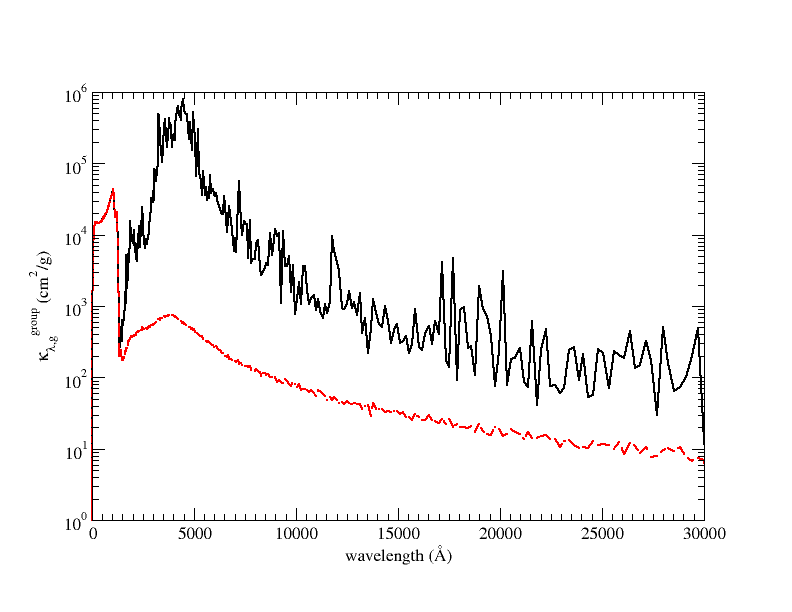}
\caption{
The LTE opacity of Nd at $T = 4,000$~K ($0.345$~eV)
and $\rho = 10^{-13}$~g/cm$^3$ using the line-binned (solid, black curve) and
expansion-opacity (dashed, red curve) methods to obtain the
b-b contribution. For the expansion-opacity calculation,
the time since ejection was chosen to be $t_{\rm exp} = 1$~day.
In both cases, a wavelength group structure of $\Delta \lambda = 0.01 \lambda$
was chosen. These specifications are the same as those considered in Figure~8
of \citet{kasen13} and Figure~7 of \citet{fontes17}.
}
\label{fig:opac_exp_bin_nd}
\end{figure}

As a final opacity comparison, we present in Figure~\ref{fig:opac_exp_bin_nd}
the line-binned and expansion opacities of
Nd at the same conditions used for Figure~\ref{fig:opac_exp_nd}.
There are two curves in the figure,
one representing the bound-bound contribution obtained
from the current line-binned approach, Equation~(\ref{opac_bb_bin}),
and the other from the expansion-opacity method, Equation~(\ref{opac_bb_exp}).
Both curves were generated with the FR model.
(For completeness, we mention that the two
corresponding curves generated with the SR model (not
shown) are quantitatively similar to the two FR curves
displayed in this figure.)
We note that, for these conditions, the line-binned approach
produces an opacity that is one or more orders of magnitude
greater than the expansion method over much of
the wavelength range. As expected, these trends are very similar
to the behavior exhibited in the corresponding comparison of
line-smeared versus expansion opacities in Figure~7 of \citet{fontes17}.
In general, the area-preserving opacities
are expected to be equal to or greater than the expansion opacities
based on a simple examination of the mathematical behavior
of Equations~(\ref{opac_bb_exp}) and (\ref{tau_sob}) versus
Equation~(\ref{opac_bb_bin}). (Compare, also, Equation~(\ref{tau_bb_exp})
to Equation~(\ref{tau_bb_bin}) in the upcoming Section~\ref{sub:tau}.)
As a consequence, the luminosity is expected to be diminished
when line-binned opacities, rather than expansion values, are used
when simulating light curves. However, the deviation
in the resulting light curves depends on a broad range of conditions that are
relevant for a complete simulation, and can not be determined from a simple
opacity comparison carried out at a specific set of conditions.

\subsection{Opacity tables}
\label{sub:opac_tab}

In order to perform radiation-transport calculations in an efficient
manner, opacity tables were generated for the 15 elements discussed above
using prescribed temperature and density grids that span
the range of conditions of interest.
The temperature grid consists of 27 values (in eV):
0.01, 0.07, 0.1, 0.14, 0.17, 0.2, 0.22, 0.24, 0.27, 0.3, 0.34, 0.4, 0.5, 0.6,
0.7, 0.8, 0.9, 1.0, 1.2, 1.5, 2.0, 2.5, 3.0, 3.5, 4.0, 4.5, and 5.0.
Specific temperature values are also indicated by circles in
the ionization balance plot of Figure~\ref{fig:ionfrac_nd}.
The density grid contains 17 values ranging from 10$^{-20}$
to 10$^{-4}$~g/cm$^3$, with one value per decade.
Our photon energy grid is the same 14,900-point grid
that is used in standard Los Alamos tabular opacity efforts,
e.g.~\citealt{colgan_oplib}. The grid is actually a temperature-scaled
$u = h\nu/kT$ grid with a non-uniform spacing that is designed to provide
accurate Rosseland and Planck mean opacities. A description of this grid is
available in Table~1 of \citealt{frey13}.

\section{Motivation and Justification for Line-Binned Opacities}
\label{sec:bin_just}

In this section, we provide motivation for a straightforward approach
to numerical
solutions of the thermal radiation transport equations in kilonovae.
Specifically, we discuss how an approach involving straight discretization of
opacity in the relevant phase space (i.e. space, time, angle
and frequency) may be applicable to the kilonova scenario.  Straight
discretization is a traditional numerical approach that makes good
sense when properties of the phase space are smooth on scales of an
affordable discrete resolution.  That said, bound-bound transitions
dominate the opacity in Type~Ia supernovae (SNe) and kilonovae,
and their frequency
distribution is generally not smooth for affordable energy resolutions.  As a
result, a different type of radiation transport scheme, taken from the
Ia SNe literature, i.e. the expansion-opacity formalism,
has provided the primary path,
to date, for simulating emerging light curves and spectra from kilonova models
\citep{kasen13,barnes13,kasen17,tanaka18}.
While both Ia~SNe and kilonovae
do have line-dominated opacities, as well as homologously expanding ejecta,
we argue that there are key differences between their physical conditions that
permit this alternative, straight-discretization numerical approach,
i.e. Equation~(\ref{opac_bb_bin}),
to be considered in calculating the kilonova emergent light.

The expansion-opacity approach tracks the flow of energy through the
iron- and lanthanide-rich expanding ejecta
of the kilonova.  This approach is aimed at capturing the spatial
diffusion of radiation when the opacity is dominated by a thick forest
of lines (bound-bound interactions).  At deep optical depths within
the kilonova ejecta, a thick forest of lines exists, but it is not
obvious that the energy flow in this region is dominated by the
diffusion of radiation.  In particular, the expected magnitude of the
net outward
radiative flux should be diminished as a result of the uniform spatial
heating in the kilonova ejecta, as compared to the centrally
condensed heating source in a Ia SN.
 
Furthermore, the timescales for energy flow associated with the expansion motion
of the ejecta (which is much higher in a kilonova) can be compared and 
is found to dominate over the timescale for radiation diffusion. 
The typical expansion velocity in a kilonova is
$v_{\rm exp}/c\approx 0.1$--0.2, which translates to a
relatively shallow radiative zone with optical depth $\tau<10$--20.
As such, details of radiative transport at higher optical depths are not
as crucial for computing observables. Indeed, the pre-existing radiation
energy, which scales like $\sim T^4\sim t^{-4}$
(where $t$ represents the time since merger), is quickly decimated
by the expansion and
superseded by instantaneous nuclear heating, which behaves like
$\sim t^{-1.3}$~\citep{korobkin12,grossman14}.

Given these arguments, capturing the accurate diffusion of radiation in the
high optical depth regions may not be of paramount importance to accurately
model kilonova emission.  As alluded to above, this is an important point
since relaxing this
requirement would allow alternative transport methodologies to be valid.
In the remainder of this section, we present a comparison of two approaches
for determining the average opacity used in multigroup transport calculations: 1) the expansion-opacity formalism and 2) a straight discretization,
or line-binned-opacity, formalism.
In the limit of small optical depths, where radiation is expected to carry the
flow of energy, both approaches reduce to the same result.
(See the discussion following Equation~(\ref{tau_bb_bin}).)
In intermediate
optical depth regimes, it is not obvious which approach is more physically
valid, and the line-binned-opacity approach offers a reasonable, alternate
bound on kilonovae opacity.  This alternative approach brings with it
some advantageous numerical properties that we enumerate, following a brief
recap of each approach.
This brief synopsis serves to cast these approaches from an optical depth
perspective, which is often more intuitive than the opacity perspective
presented in Section~\ref{sec:atomic}.

\subsection{Review of optical depth formulae}
\label{sub:tau}

%

Two different expressions for the optical depth
are employed in the radiation transport simulations considered
below. Each of the optical
depths presented here involves a sum over lines
(denoted by summation index $i$) within a wavelength bin,
$\Delta \lambda_j$, which can be related to a spatial zone for a homologous
flow. The optical depth that corresponds to the expansion
opacity in Equation~(\ref{opac_bb_exp}) is given by \citep{pintoeast00}
\begin{equation}
\tau^{\rm b-b}_{\rm exp}(\Delta\lambda_j)
= \sum_{i \in \Delta\lambda_j} (1-e^{-\tau_i}) \,,
\label{tau_bb_exp}
\end{equation}
where, again, $\tau_i$ is the Sobolev optical depth of line $i$ given
by Equation~(\ref{tau_sob}).
This formulation has the property that it limits the
total contribution to the optical depth from a single line to 1, but
uses the full opacity for low optical depths per line.  

The optical depth that corresponds to the line-binned opacity in
Equation~(\ref{opac_bb_bin}) is obtained by requiring the optical depth due
to lines in a wavelength bin to be equal to the sum of the Sobolev optical
depths of all the lines that fall within the wavelength bin.  The result is 
\begin{equation}
\tau^{\rm bin}_{\lambda,j} = 
\sum_{i \in \Delta \lambda_j} \tau_i \,.
\label{tau_bb_bin}
\end{equation}
For small values of $\tau_i$, it is straightforward to use a Taylor
series expansion to show that Equation~(\ref{tau_bb_exp}) reduces
to Equation~(\ref{tau_bb_bin}). For large values of $\tau_i \gg 1$,
it is easy to see that Equation~(\ref{tau_bb_bin}) is always greater
than Equation~(\ref{tau_bb_exp}).

An advantage of this line-binned approach is that the opacities can be
pre-tabulated for any explosion scenario, as it does not include details
of the expansion properties in its definition.  
There are also advantages in computational efficiency 
that come with the ability to pre-tabulate.
The large line lists necessary for a converged opacity are taken into
account only once during the creation of the table, and then all subsequent
simulations amortize that cost. In addition, tabulated opacities are commonly
used by radiation-hydrodynamic codes from fields outside of 
astrophysical light-curve applications (e.g. high energy density experimental
physics at facilities like the Omega Laser at Laboratory for Laser Energetics,
the Z-Machine at Sandia National Laboratory and the National Ignition Facility
at LLNL). Thus, if a tabulated opacity approach can be shown to be valid
for the modeling of kilonova light curves, then other mature
code bases could be more easily brought to bear on kilonova light-curve
challenges.

\subsection{Tests of line-binned opacities for kilonova}
\label{sub:lc}

In this section,
we present light curves and spectra from simulations using line-binned
or expansion opacities for a 1D model with semi-analytic ejecta and
pure Nd, Ce, Sm, or U.
These elements are meant to represent the composition of dynamical ejecta
from the NSM. The problems have an ejecta mass of $1.4\times10^{-2}$ M$_{\odot}$
and mean ejecta speed of 0.125$c$ (maximum speed of 0.25$c$).
We simulate two forms of the problem:
(1) a ``simplified'' version, which is more efficient to simulate
with a direct Sobolev line treatment, is presented
in Section~\ref{subsub:comp_simp} to motivate the utility of the
line-binned opacity treatment and (2) a ``full'' version is presented
in Section~\ref{subsub:comp_full} to explore sensitivities to variations
in the opacity data and photon energy grouping. All problems simulated here
assume LTE, where opacities are calculated using a gas temperature and photon
interaction with a line is purely absorbing and thermally redistributive.
These approximations are the same ones that are made in traditional
expansion-opacity kilonova simulations performed by other authors,
e.g.~\citealt{kasen17,tanaka18}. This option
corresponds to the choice of $\epsilon = 1$ in Equations~(7) and (8),
in conjunction with Equation~(11), of \citet{kasen06_2}.

To model the radiative transfer, we employ the Monte
Carlo code {\tt SuperNu} \citep{wollaeger14},
with some improvements to the accuracy of the discrete
diffusion optimization (Wollaeger et al 2019, in prep).
The span of wavelength simulated is 1,000 to 128,000~\AA\
with 1,024 logarithmically spaced groups, unless otherwise noted.
Consequently, the corresponding group spacings of
$\Delta\lambda_g/\lambda_g = 0.0047$ permit particles
emitted anywhere to traverse multiple groups via redshift
before escaping the ejecta.
For the purpose of this study,
the expansion-opacity formalism was also implemented in {\tt SuperNu}.
This capability allows for self-consistent comparisons to be performed
between line-binned and expansion-opacity simulations, i.e.
the same fundamental atomic physics data are used in both cases,
thereby eliminating any uncertainty
that might occur when comparing with works from other groups.
The radiation-transport approximations associated with the expansion-opacity
implementation are the same as those described at the end of the previous
paragraph, i.e. we make the assumption of complete thermal redistribution
$(\epsilon = 1)$ after a photon is absorbed in a line. Thus, the only difference
between expansion-opacity and line-binned simulations is whether
one uses $1-\exp(-\tau_i)$ or $\tau_i$ in Equation~(8) of \citet{kasen06_2}.

All of the opacity implementations examined here are summarized
in Table~\ref{tab:sims}, which includes definitions of the symbols that
appear in the subsequent figure legends.
Unless otherwise noted in those legends, the chosen element is Nd,
the atomic data are taken from the semi-relativistic model,
and the oscillator strength cutoff value is $f_c=10^{-6}$, i.e. only
oscillator strengths that satisfy $f_i > f_c$ are included in the model.
Variations on these default choices are explained in the table.
Note that we use the term ``line-binned'' to refer to bound-bound opacities that
are generated using both the fully tabulated approach described
in Section~\ref{sub:opac_tab},
and an inline approach, internal to the {\tt SuperNu} code, that
explicitly solves the Saha-Boltzmann equation to obtain atomic level
populations and construct the line opacities.
In the tabular case, {\tt SuperNu} interpolates on temperature and
density to obtain the opacity at required conditions, while no such
interpolation is required for the inline approach.
We differentiate between these two approaches with distinct legend labels:
``Binned-Tab'' for tabulated and ``Binned-Inl'' for inline.
The expansion opacity calculations, labeled ``Expansion-Inl'',
are always performed internally to SuperNu, since the expansion time must
be used to perform the calculation.
For the simulations with element labels, all opacities are fully
tabulated and line-binned. The remaining
bound-free, free-free, and scattering contributions to the opacity are always
calculated from the tables.
As explained in Section~\ref{sub:opac_tab}, the opacity tables employ
a grid with 27 temperature points and 17 density
points, ranging from 0.01 to 5~eV and logarithmically spaced from $10^{-20}$
to $10^{4}$~g/cm$^3$, respectively.

\begin{table*}
  \small
  \caption{Legend symbols for various opacity implementations considered
           in this section, along with the figures in which they appear.
           The Smeared-Tab and Binned-Tab calculations employ pre-tabulated
           opacity data generated on the temperature and density grids
           described in Section~\ref{sub:opac_tab}.
           The Binned-Inl, Expansion-Inl and Sobolev calculations use
           explicitly generated line contributions
           to the opacity within the {\tt SuperNu} code, without interpolation
           on temperature and density.
           SR stands for semi-relativistic atomic models and FR stands for fully
           relativistic atomic models. $f_c$ is the oscillator strength cutoff
           value used in the atomic models.}
  \label{tab:sims}
  \centering
  \resizebox{170mm}{!}{
    \begin{tabular}{|l|p{60mm}|p{45mm}|c|}
      \hline
      \textbf{Symbols} & \textbf{Method} & \textbf{Variations} &
      \textbf{Figures} \\
      \hline
      \hline
      Sobolev &
      \parbox[c]{60mm}{{\tt SuperNu}-internal (inline) opacities using
                       direct treatment of lines with the Sobolev approximation;
                       see Section~\ref{subsub:comp_simp}} &
      (SR, $f_{c}=10^{-3}$) &
      \ref{fig:test} \\
      \hline
      Smeared-Tab &
      \parbox[c]{60mm}{Tabulated opacities with smeared lines;
                       see Section~\ref{sub:smeared}} &
      (SR, $f_c=10^{-6}$) &
      \ref{fig:p0_1} \\
      \hline
      Binned-Tab &
      \parbox[c]{60mm}{Tabulated opacities with binned lines;
                     see Equations~(\ref{opac_bb_exp}), (\ref{opac_bb_group})
                     and (\ref{tau_bb_bin})} &
      (SR, $f_c=10^{-6}$) &
      \ref{fig:p0_1},\ref{fig:p1} \\
      \hline
      Binned-Inl &
      \parbox[c]{60mm}{{\tt SuperNu}-internal (inline) opacities with binned
                     lines; see Equations~(\ref{opac_bb_exp})
                             and (\ref{tau_bb_bin})} &
      (SR, \hbox{$f_{c}\in\{10^{-1},10^{-3},10^{-4},10^{-6}\}$});
      (FR, SR, $f_c=10^{-6}$);
      (FR-SCR, FR-SCNR, $f_c=10^{-6}$) &
      \ref{fig:test},\ref{fig:p2},\ref{fig:p3}; \ref{fig:p4}; \ref{fig:p5}\\
      \hline
      Expansion-Inl&
      \parbox[c]{60mm}{{\tt SuperNu}-internal (inline) opacities with expansion
                       line treatment; see Equation~(\ref{tau_bb_exp})} &
      (SR, \hbox{$f_{c}\in$}\hbox{$\{10^{-1},10^{-3},10^{-4},10^{-6}\}$});
      (FR, SR, $f_c=10^{-6}$);
      (FR-SCR, FR-SCNR, $f_c=10^{-6}$) &
      \ref{fig:test}, \ref{fig:p2},\ref{fig:p3}; \ref{fig:p4}; \ref{fig:p5} \\
      \hline
      Nd, Ce, Sm, U &
      \parbox[c]{60mm}{Element symbols for tabulated, line-binned opacities} &
      SR,FR &
      \ref{fig:p6} \\
      \hline
    \end{tabular}
  }
\end{table*}

\subsubsection{Simplified problem: a comparison of approximate methods
with the fully resolved solution}
\label{subsub:comp_simp}

In order to provide numerical justification for our line-binned
opacities, we consider a simplified version of the
test problem described above, and employ the direct
Sobolev treatment of lines, following \citet{kasen06_2}.
In implementing this more precise line treatment, Equations~(13)
and (15) of \citet{kasen06_2} are used to determine when a photon comes
into resonance with a line and, if so, whether an absorption
occurs, respectively. In addition, we continue to make the assumption
of complete thermal redistribution $(\epsilon=1)$ after a photon is absorbed.
No attempt is made to take into account fluorescence or to use branching
ratios to determine the fate of a photon after absorption occurs.
A description of the implementation of this direct Sobolev method
in the {\tt SuperNu} code is provided in Appendix~\ref{app:dir_sob}.

This particular usage of the direct Sobolev treatment is different
from its application in \citet{kasen06_2}, in which it was employed
to test the validity of the thermal redistribution approximation.
Here, we assume that complete thermal redistribution is already valid, because
that is a primary assumption of nearly everyone who does detailed-opacity
kilonova simulations, and use the direct Sobolev line treatment to provide
an exact solution to that approximate $(\epsilon=1)$ problem.
This unconventional
usage of the direct line treatment allows us to consider a test case that
isolates the particular opacity method of interest, while all other physics
choices remain the same. This characteristic is useful to investigate how
the two approximate opacity methods perform, compared to the exact line
treatment, for the complex range of conditions that are relevent
for kilonova light-curve and spectral modeling.

This simplified version of the problem starts at 4~days post-merger,
uses semi-relativistic oscillator strength data for Nd with
a cutoff value of $f_{c}=10^{-3}$, and includes only bound-bound opacity.
These features make the problem faster to solve on a moderate
number of CPUs ($\sim$30), without discrete diffusion acceleration.
Moreover, solving radiative transfer with only bound-bound contributions
further isolates the potential source of discrepancy
that can arise from the different opacity methods.
We have provided a description of this simplifed problem
in Appendix~\ref{app:test_case} so that other groups can use it
as a test case to produce meaningful comparisons between their
work and the present effort.
While no test problem is perfect, we reiterate that the simplified one being
proposed here has the advantage of exercising the different opacity
implementations over a large range of conditions that occur in an actual
kilonova simulation.

\begin{figure*}
  \centering
  \includegraphics[clip=true,angle=0,width=0.65\columnwidth]
   {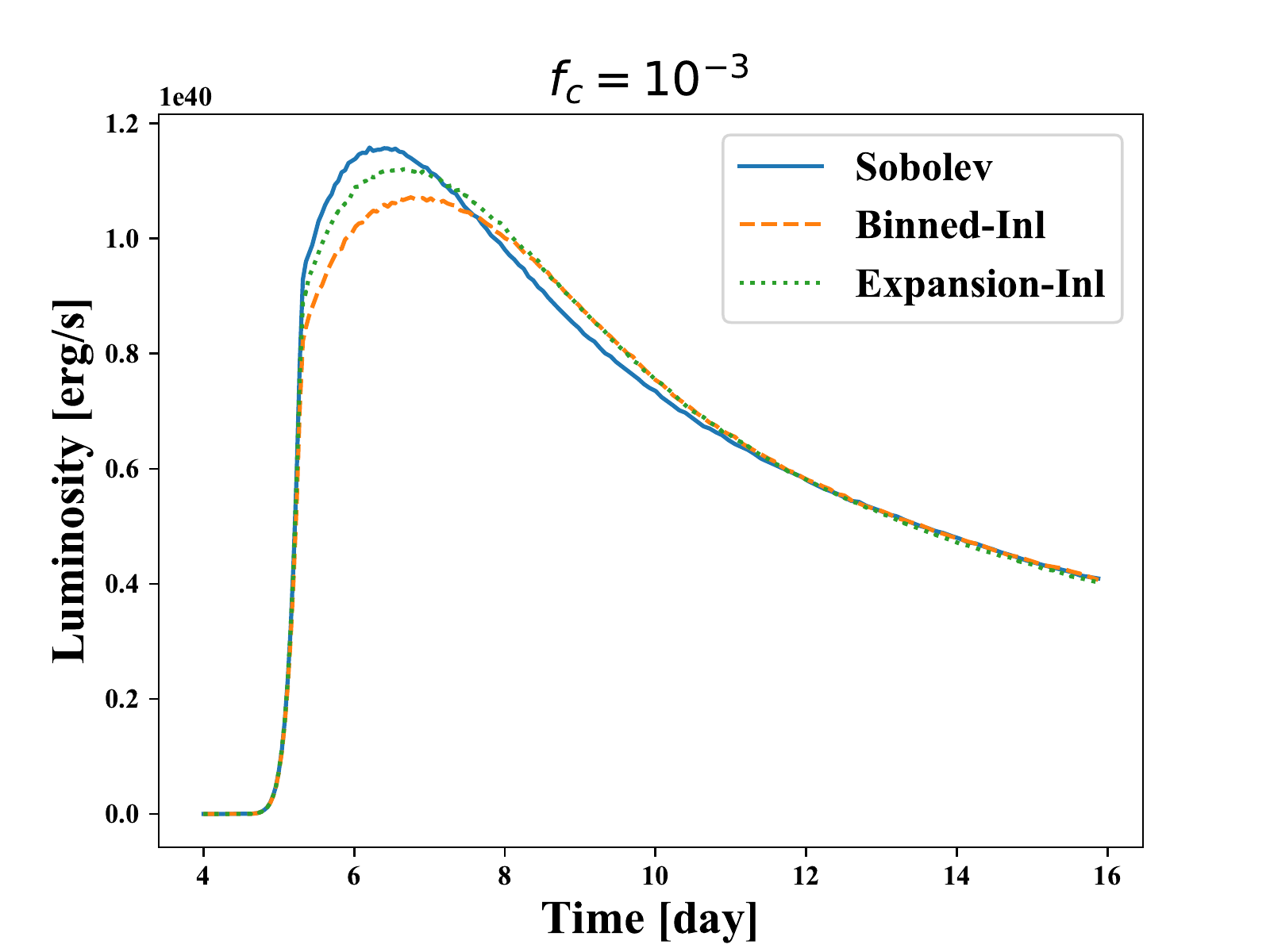}
  \includegraphics[clip=true,angle=0,width=0.65\columnwidth]
   {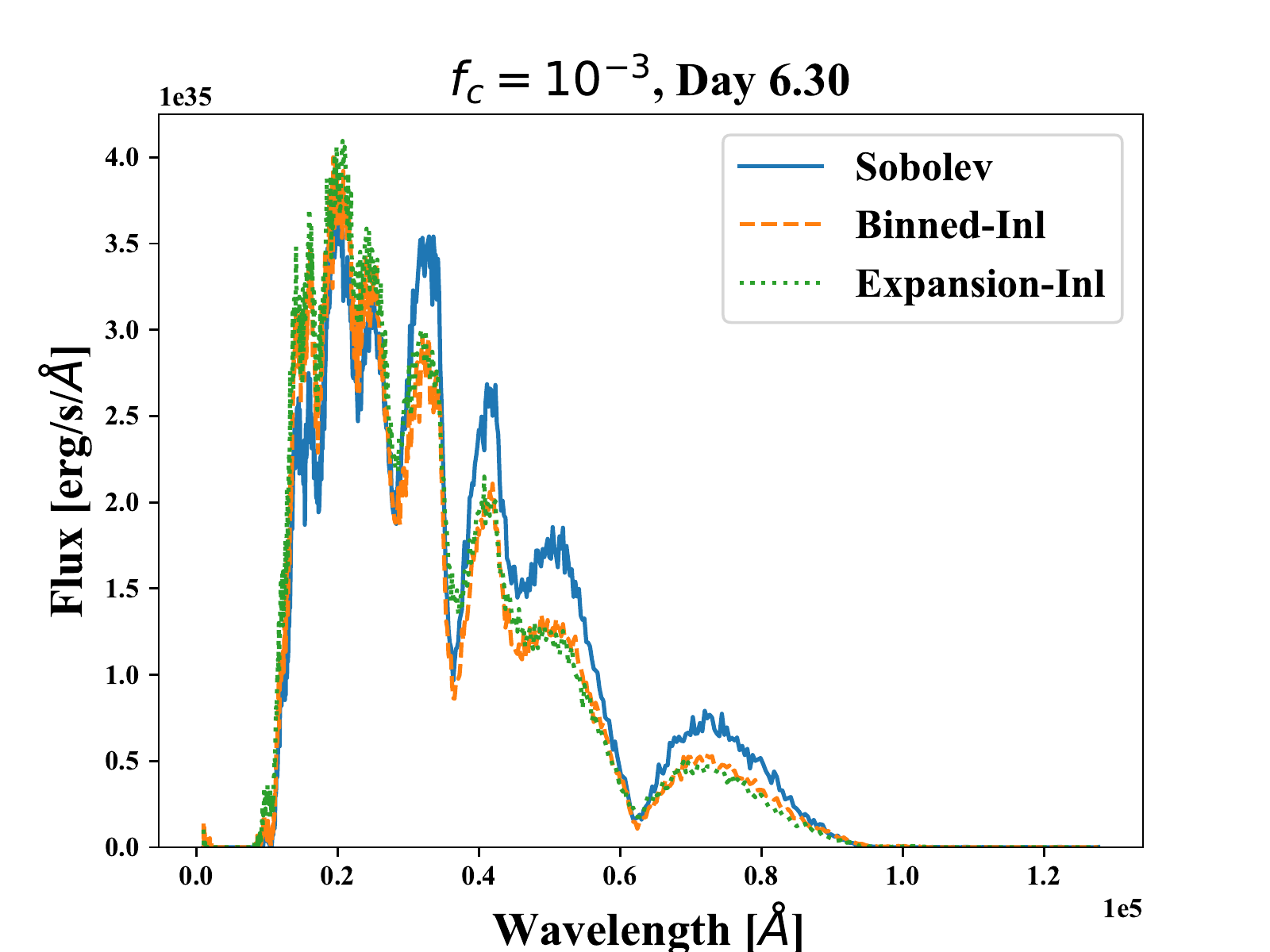}
  \includegraphics[clip=true,angle=0,width=0.65\columnwidth]
   {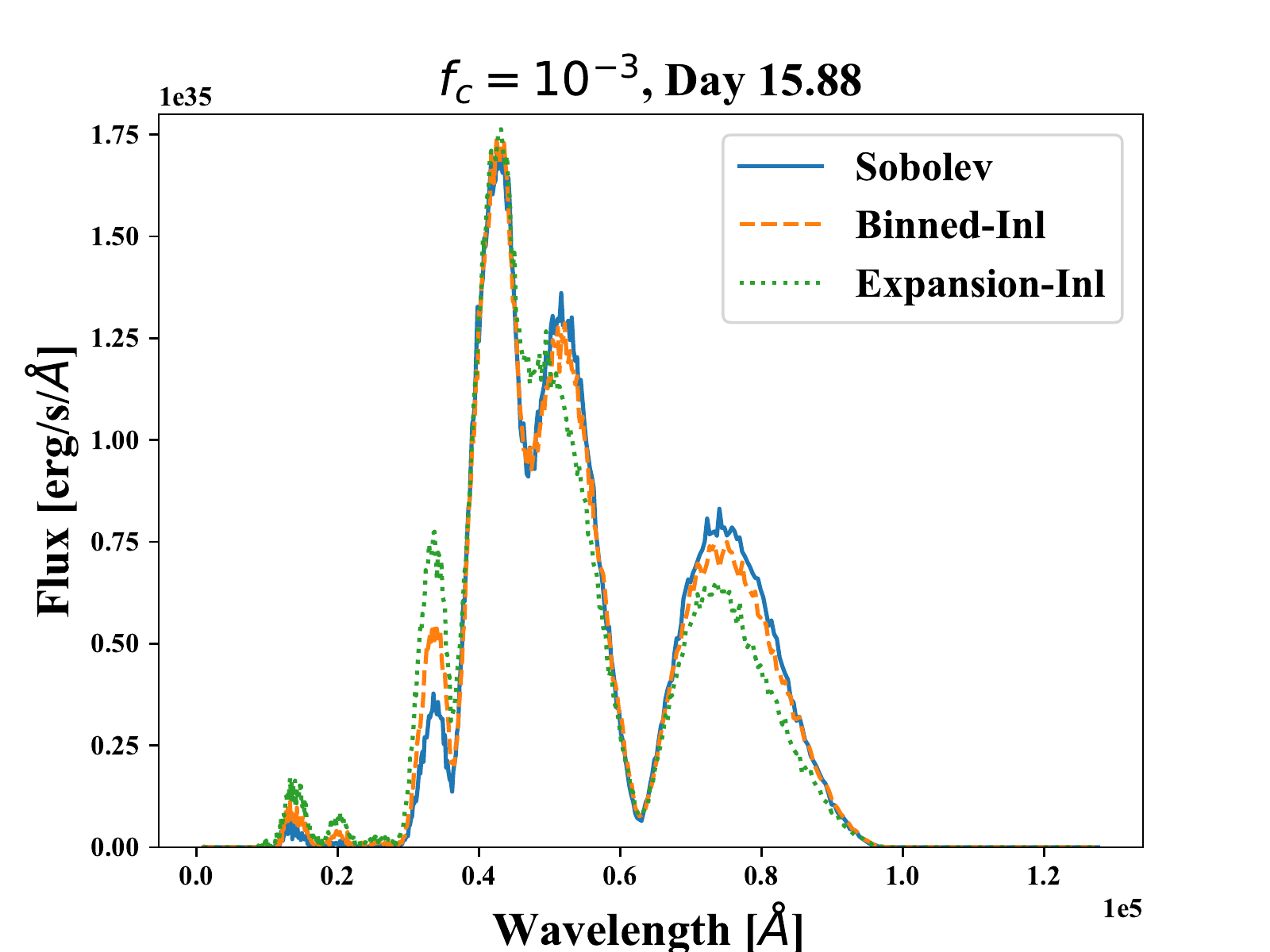}
  \caption{
    Bolometric luminosity (left panel) and radiative flux
    at 6.3~days (middle panel) and 15.88~days (right panel)
    for the simplified test problem described
    in the text. Results are shown for the continuous Monte Carlo Sobolev
    (solid blue), line-binned (dashed orange),
    and expansion-opacity (dotted green)
    methods. For consistency, these methods all employ line contributions
    to the opacity that were generated within the {\tt SuperNu} code, rather
    than via the pre-tabulated approach. Also, an oscillator strength cutoff
    value of $f_c = 10^{-3}$ was used in all cases.
  }
  \label{fig:test}
\end{figure*}

In Figure~\ref{fig:test}, we compare results obtained with the direct Sobolev
treatment to line-binned and expansion-opacity simulations.
In order to avoid confusion, we note that the Sobolev
treatment is not the same as the expansion-opacity approach
and, futhermore, is more accurate than the expansion-opacity approach.
Light curves are displayed in the left panel and spectra in the
middle and right panels
of the figure. In the left panel, we see very good agreement between
all three light curves, with the maximum differences occurring at
the peak. The direct Sobolev
(``Sobolev'') light curve displays the greatest peak value, which occurs
at 6.3~days. The expansion-opacity (``Expansion-Inl'') simulation
displays the next highest peak, and the line-binned (``Binned-Inl'') curve
displays the lowest peak value. The three peak values are within 8\%
of each other. As expected, the expansion-opacity result is
brighter than the line-binned result (see discussion in Section~\ref{sub:tau}),
but they differ by a maximum of only 5\%, which occurs at the peak.
While the direct Sobolev treatment is the most accurate of these three methods,
it is computationally intractable for more realistic simulations.
The expansion and line-binned approaches provide a measure of the uncertainty
arising from different opacity implementations.

As far as the spectral examples are concerned, there are differences
in the heights of several features.
The spectra at the peak of the light curve (middle panel), occurring
at 6.3~days, display an interesting trend: the Sobolev result is the highest
curve above 0.3~\AA, and the lowest curve below that wavelength.
The two more approximate curves agree reasonably well across the entire
wavelength range. On the other hand, at the later time of 15.88~days
(right panel), our line-binned approach agrees better than the
expansion-opacity approach with the more accurate
Sobolev spectrum across much of the wavelength range.
From a theoretical perspective, the lack of a clear trend of agreement when
comparing the two approximate results to the direct Sobolev spectra
suggests that neither one of the approximate methods provides an obviously
better prediction of kilonova spectra,
at least in the context of this simplified problem.
The relatively good agreement between the three calculations provides additional
support for our line-binned approach.
From a practical perspective, these differences in feature heights are
relatively small and are not sufficient to explain discrepancies
between theory and observation. 
We also note that there is no redward shift in the peak emission
of the line-binned curve when comparing the three methods.
(See next section for a detailed discussion about redward shifts.)
However, the simplified problem analyzed here offers only a partial
investigation of the relevant parameter space
and so we consider more physically complete models in the next section.

\subsubsection{Full problem: sensitivity studies}
\label{subsub:comp_full}

For the full problem, the time span is extended, ranging from $10^4$~seconds
to 20~days in the comoving frame ($\sim14$ observer
days), with 400 logarithmically spaced time steps.
All opacity contributions are included as well.
We vary the parameters of this problem to examine how light
curves and spectra are affected by changes to the opacity data and
opacity discretization method.
The variations include:
group averaging methods
(line-smeared versus line-binned versus expansion-opacity),
number of photon energy groups,
tabulated versus inline-generated opacities, oscillator strength cutoff value,
atomic physics model (semi-relativistic versus fully relativitistic variants),
and choice of element.
As mentioned in the previous section, use of the direct Sobolev treatment
is impractical in simulations of the full problem and therefore
is not considered in the present analysis.

\begin{figure*}
  \centering
  {\includegraphics[clip=true,angle=0,width=1.0\columnwidth]{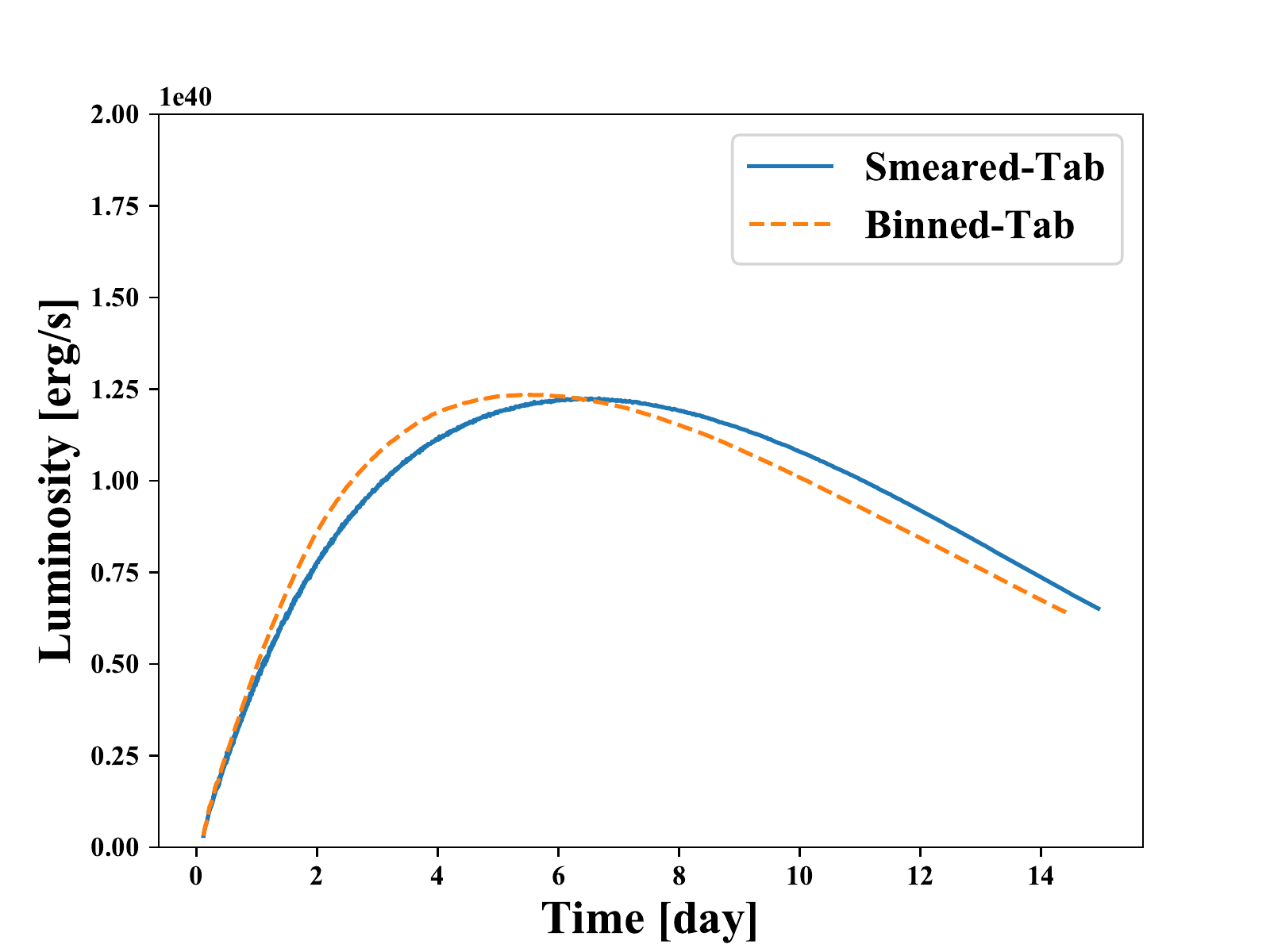}
    \label{fig:p0a_1}}
  {\includegraphics[clip=true,angle=0,width=1.0\columnwidth]{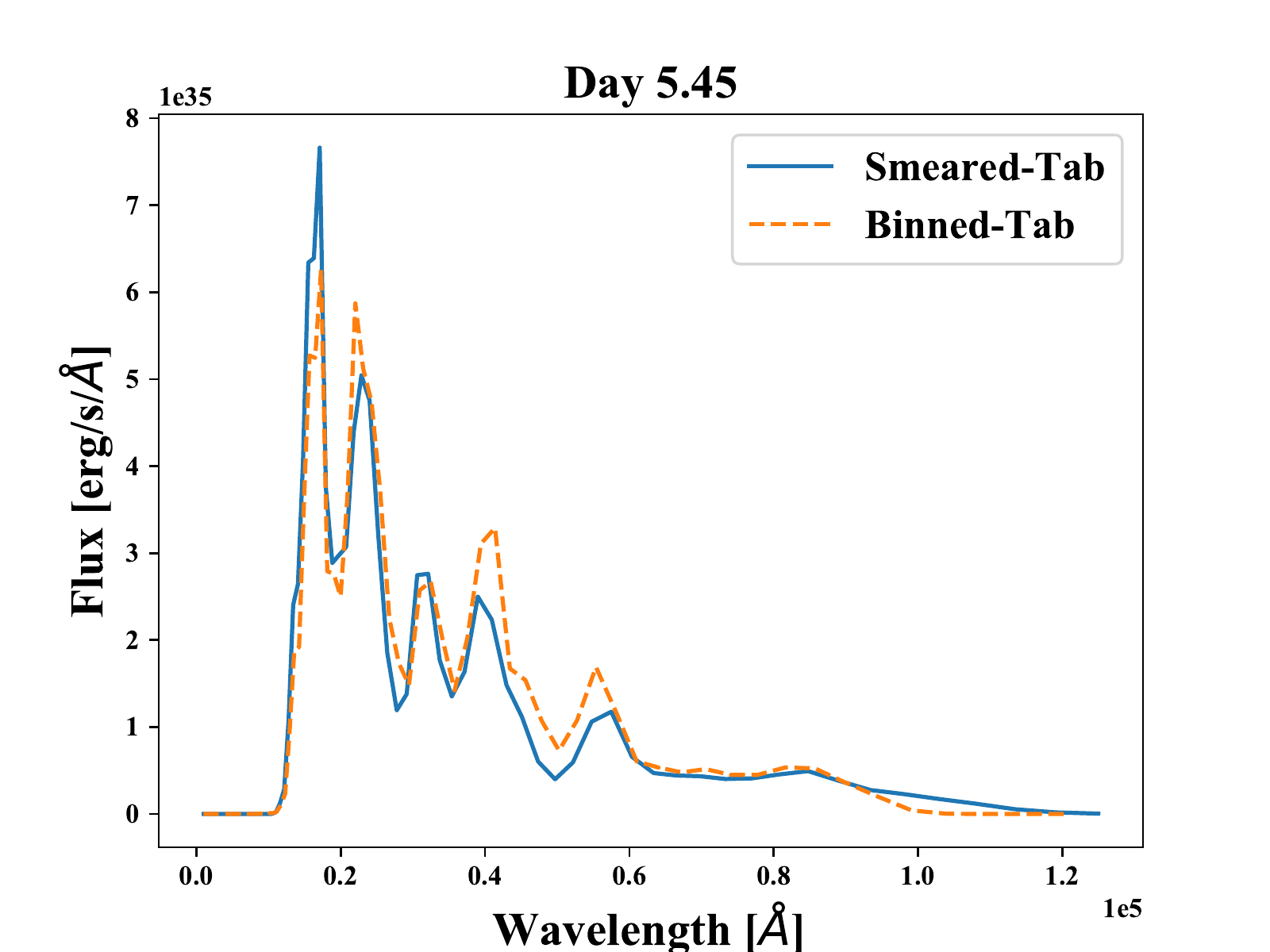}
    \label{fig:p0b_1}}
  \caption{
    For the Nd ejecta test problem, a comparison of
    bolometric luminosities and spectra, at 5.45~days post-merger,
    computed with line-smeared (solid blue curve)
    and line-binned (dashed orange curve) opacities.
    These results were computed with 100 groups, consistent
    with the simulations of \citealt{wollaeger18}.
  }
  \label{fig:p0_1}
\end{figure*}

Figure~\ref{fig:p0_1} displays a comparison of the bolometric luminosity
and spectra, at 5.45~days post-merger, 
calculated with line-binned and line-smeared opacities, both computed
with the tabular approach.
These results were calculated with 100 energy groups, which is the
same resolution used in our detailed study of kilonovae light curves and spectra
\citep{wollaeger18}.
The good agreement between the two calculations indicates that 
the line-smeared approach used in that detailed study produces
similar results to the currently proposed line-binned approach
at this group resolution.

\begin{figure*}
  \centering
  {\includegraphics[clip=true,angle=0,width=1.0\columnwidth]{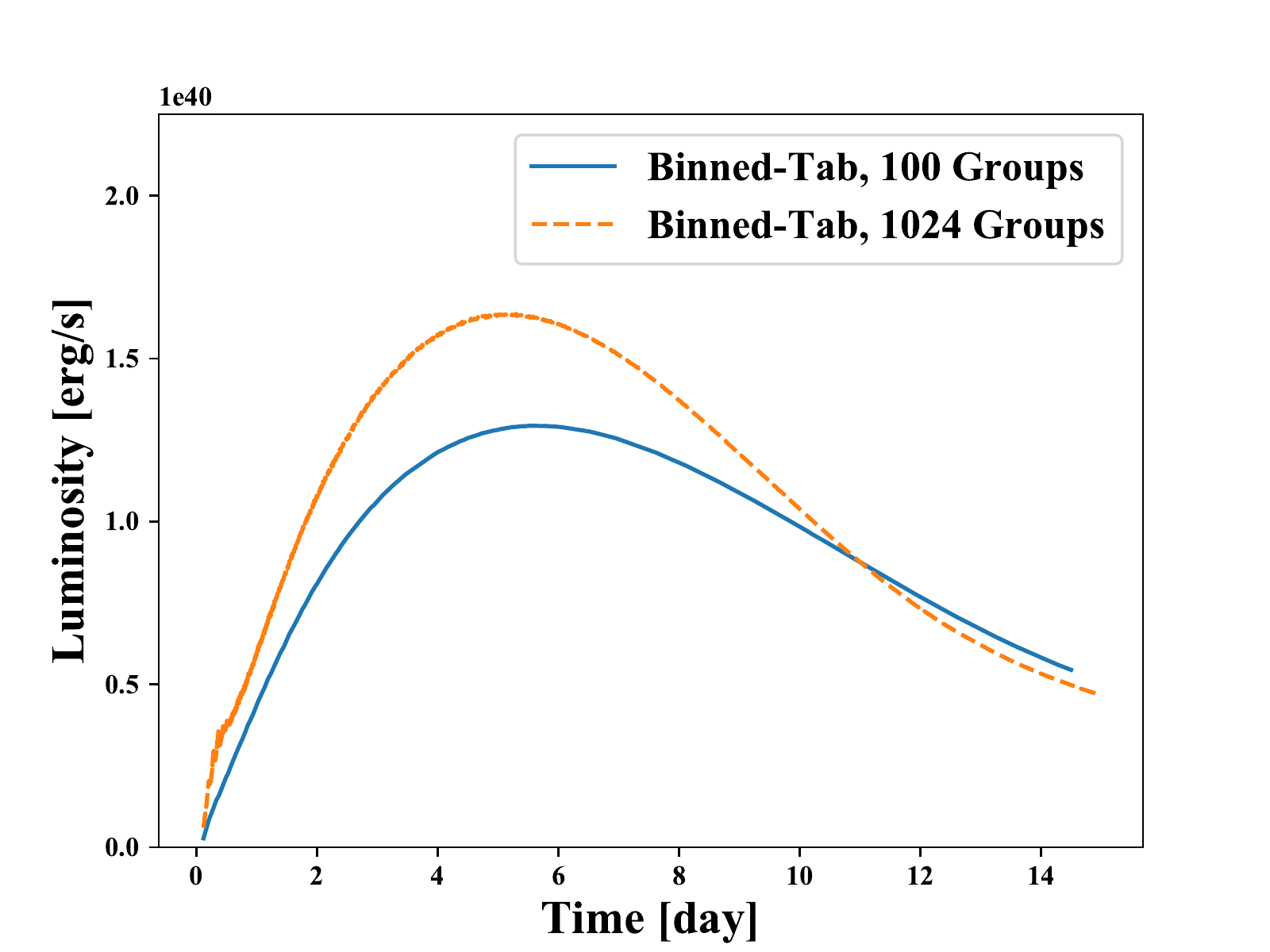}
    \label{fig:p0a_2}}
  {\includegraphics[clip=true,angle=0,width=1.0\columnwidth]{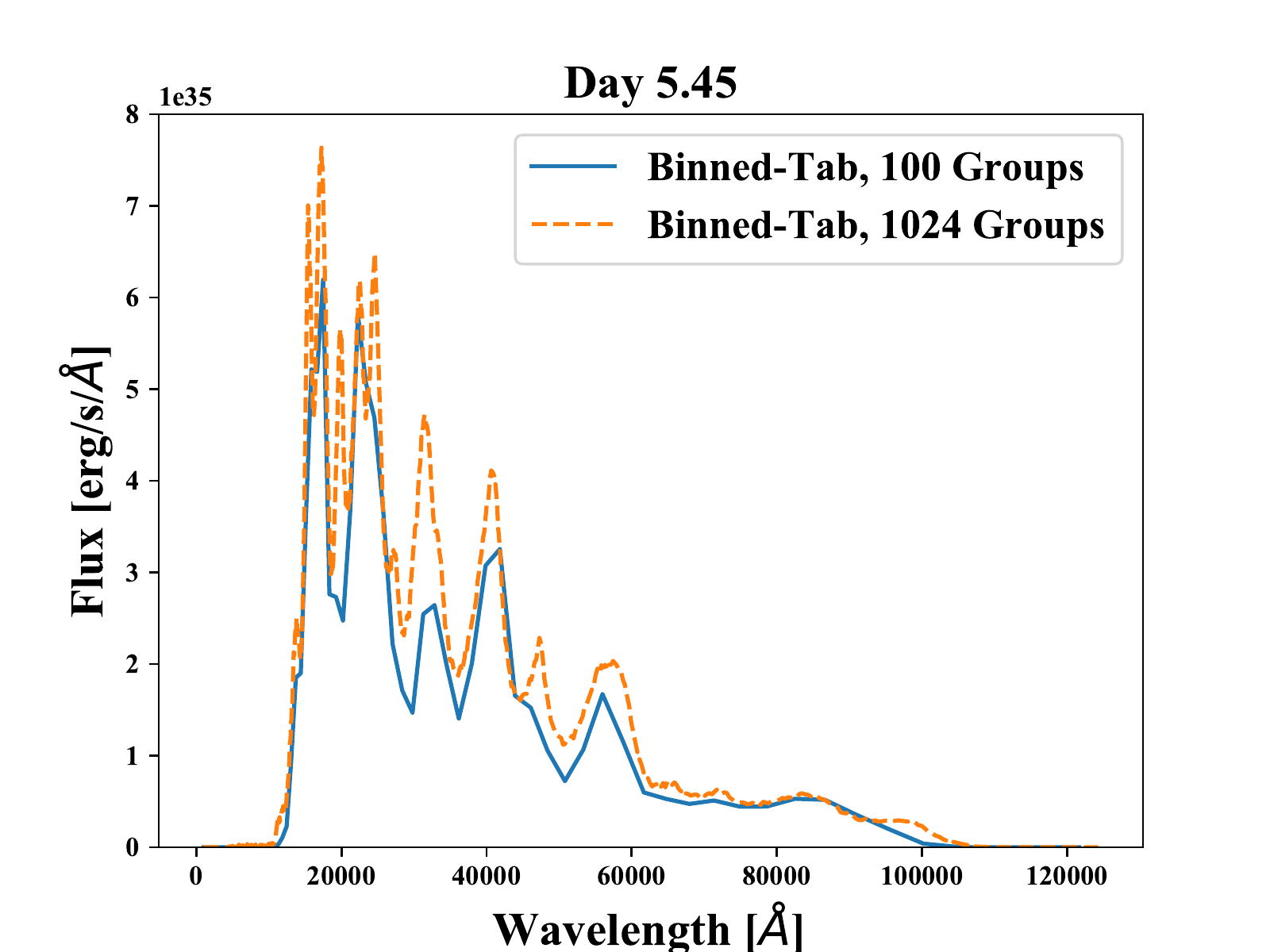}
    \label{fig:p0b_2}}
  \caption{
    For the Nd ejecta test problem, a comparison of
    bolometric luminosities and spectra, at 5.45~days post-merger,
    computed with line-binned opacities. Results are presented
    for 100 (solid blue curve) and 1,024 (dashed orange curve) groups.
  }
  \label{fig:p0_2}
\end{figure*}

In Figure~\ref{fig:p0_2}, we next present a group resolution study
by displaying line-binned results computed with 100 and 1,024 groups.
Once again, bolometric luminosities are presented in the left panel
and spectra, at 5.45~days post-merger, are presented in the right panel.
In this case, we see a difference of about 25\% in the peak luminosity
indicating that the higher number of groups is required to obtain
better convergence. Significant differences are also observed in the spectra
displayed in the right panel of this figure, providing further evidence
that 100 groups is not generally sufficient to produce converged spectra.
This lower group resolution is a deficiency of \citealt{wollaeger18}, but
does not significantly impact the conclusions of that work.  Firstly, the
primary goal of that effort was to establish variations in the kilonova signal
with respect
to changes in morphology, composition, and $r$-process heating. The group
resolution does not affect those trends.  Additionally, the authors showed that
100 groups is virtually converged for the model with mixed dynamical ejecta
composition (see Figure~7 in \citealt{wollaeger18}).
That particular model was used in their section
on detection prospects, and all dependent works.
The 1,024 groups represent the standard resolution
chosen for the test-problem simulations considered in this work
and have been found to produce converged results.
For completeness, we note that additional calculations with 4,096 groups
were carried out (not shown) and the resulting light curves and
spectra agreed well with those produced with 1,024 groups.

\begin{figure*}
  \centering
  {\includegraphics[clip=true,angle=0,width=1.0\columnwidth]{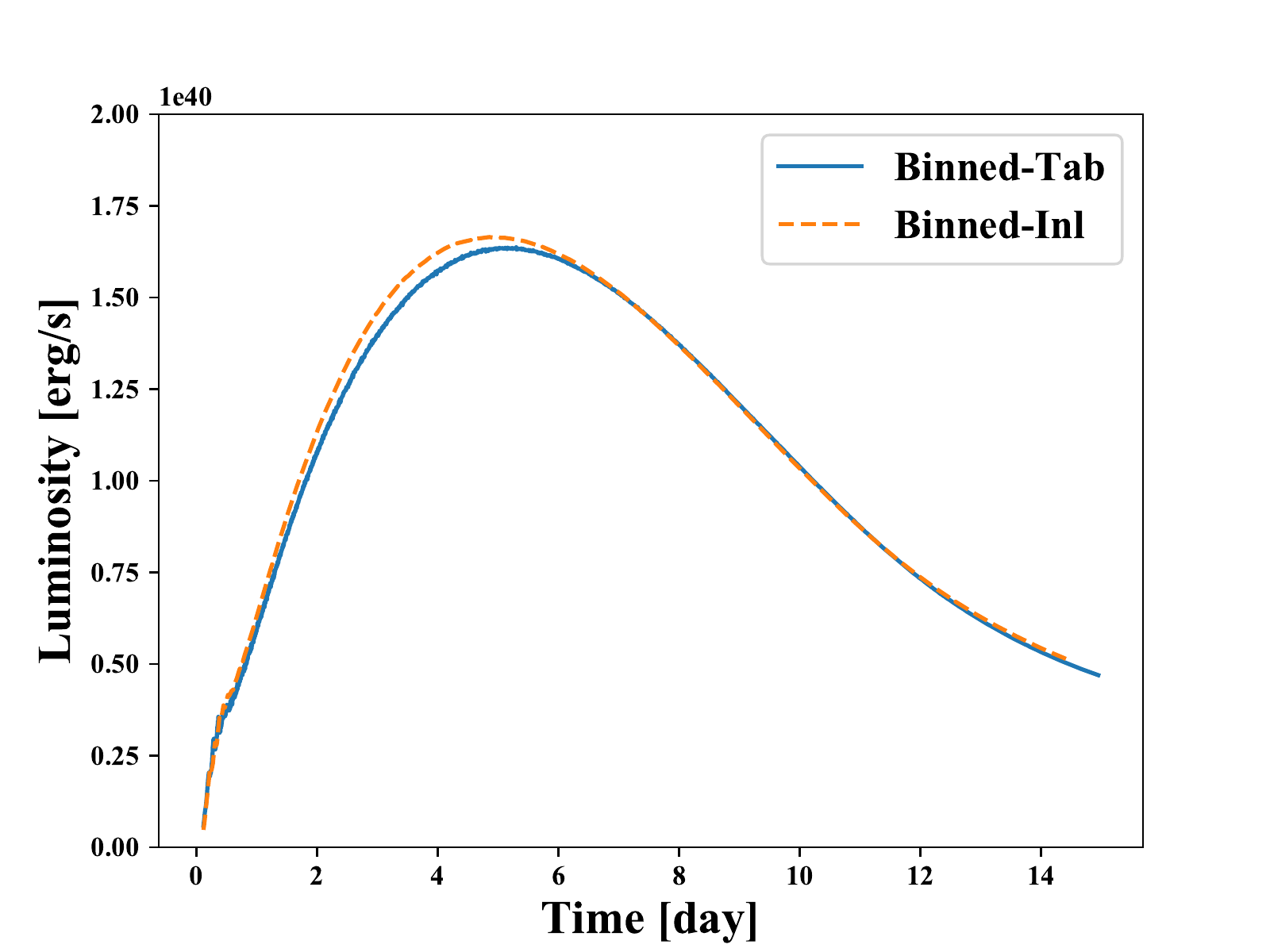}
    \label{fig:p1a}}
  {\includegraphics[clip=true,angle=0,width=1.0\columnwidth]{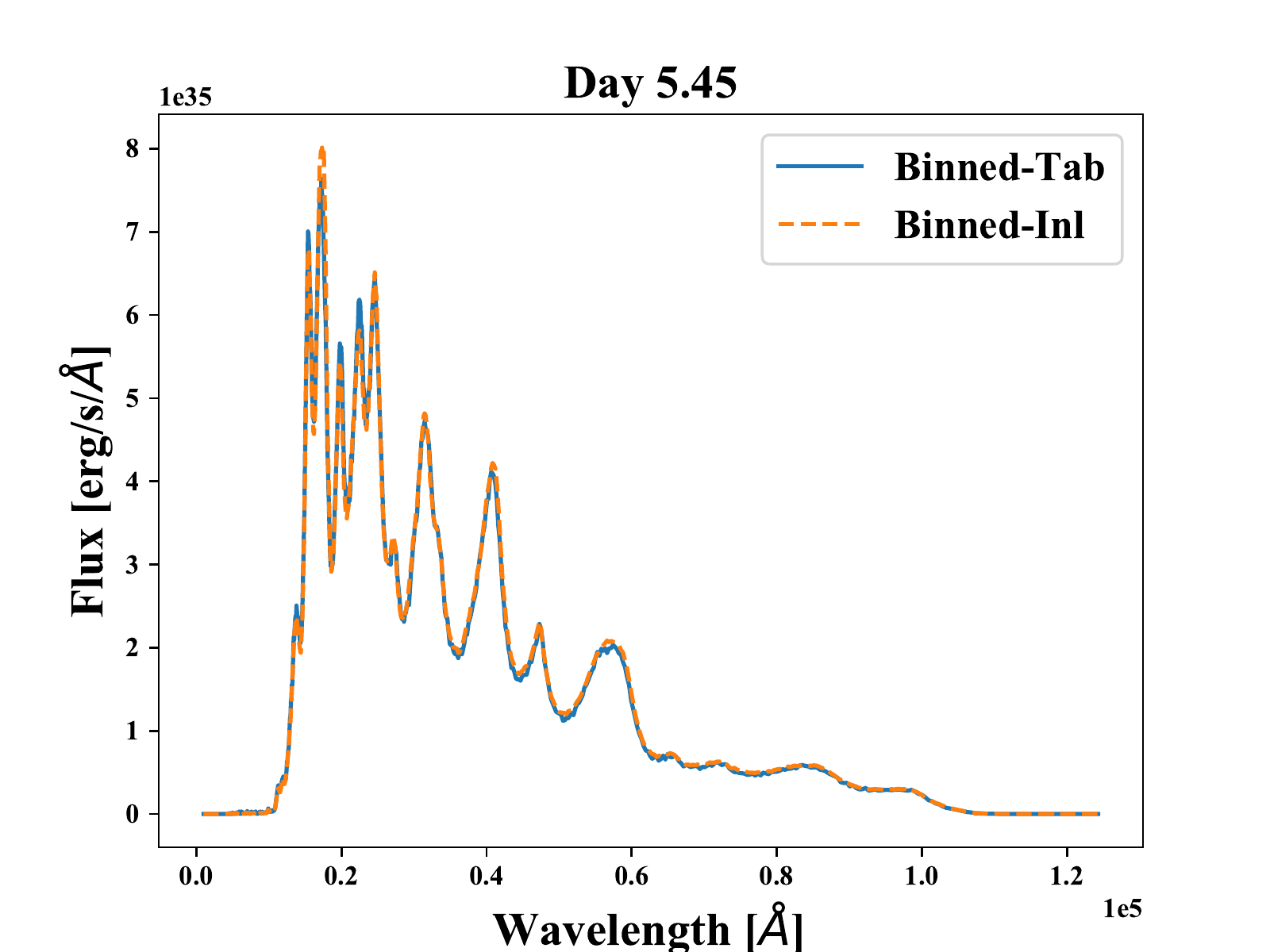}
    \label{fig:p1b}}
  \caption{
    Comparison of tabulated (solid blue curve) and inline (dashed orange curve),
    line-binned opacity implementations.
    Figures~\ref{fig:p1a} and \ref{fig:p1b} display luminosities and
    spectra, respectively.
    The good agreement between these two implementations indicates that
    our method for processing the tabular opacity data does not introduce
    any significant error in the simulations.
  }
  \label{fig:p1}
\end{figure*}

We have also compared the fully tabulated, line-binned approach to the
corresponding inline multigroup implementation, which solves
the Saha-Boltzmann equations using the actual ejecta temperatures and
densities. As mentioned previously,
in the fully tabulated approaches (both smeared and binned), the opacity is
instead obtained within {\tt SuperNu} by interpolation on a precomputed
density-temperature table, which can introduce some inaccuracy.
Figure~\ref{fig:p1} displays the bolometric luminosity and spectra for
tabulated versus inline, line-binned calculations.
The good agreement between these two sets of calculations indicates that
potential errors arising from interpolation of the tabular bound-bound
opacities are insignificant, and, furthermore, that the method for distributing
the tabular, line-binned opacities across group boundaries
(see Section~\ref{subsec:compframe}) is adequate.

\begin{figure*}
  \centering
  {\includegraphics[clip=true,angle=0,width=1.0\columnwidth]{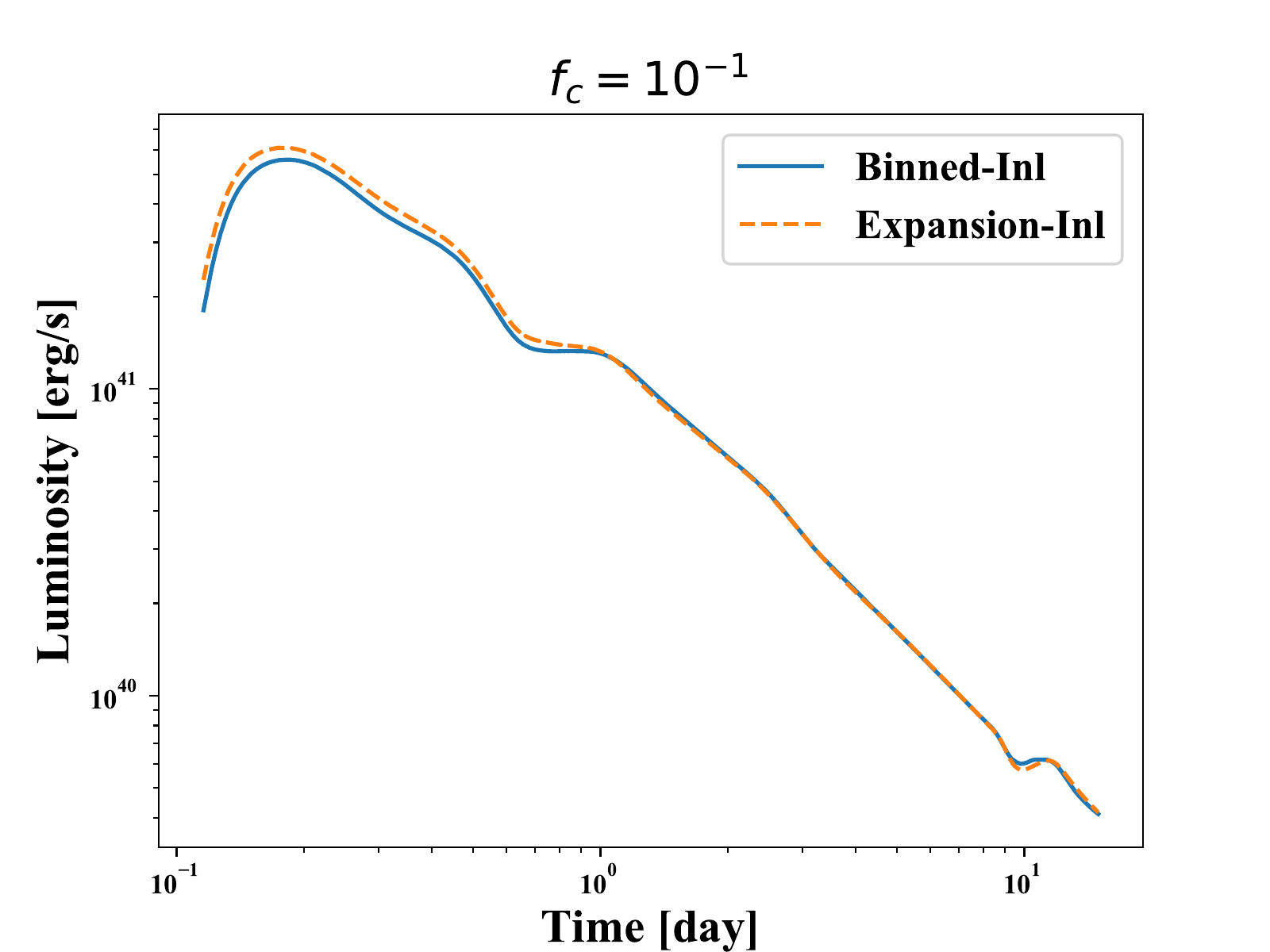}
    \label{fig:p2a}}
  {\includegraphics[clip=true,angle=0,width=1.0\columnwidth]{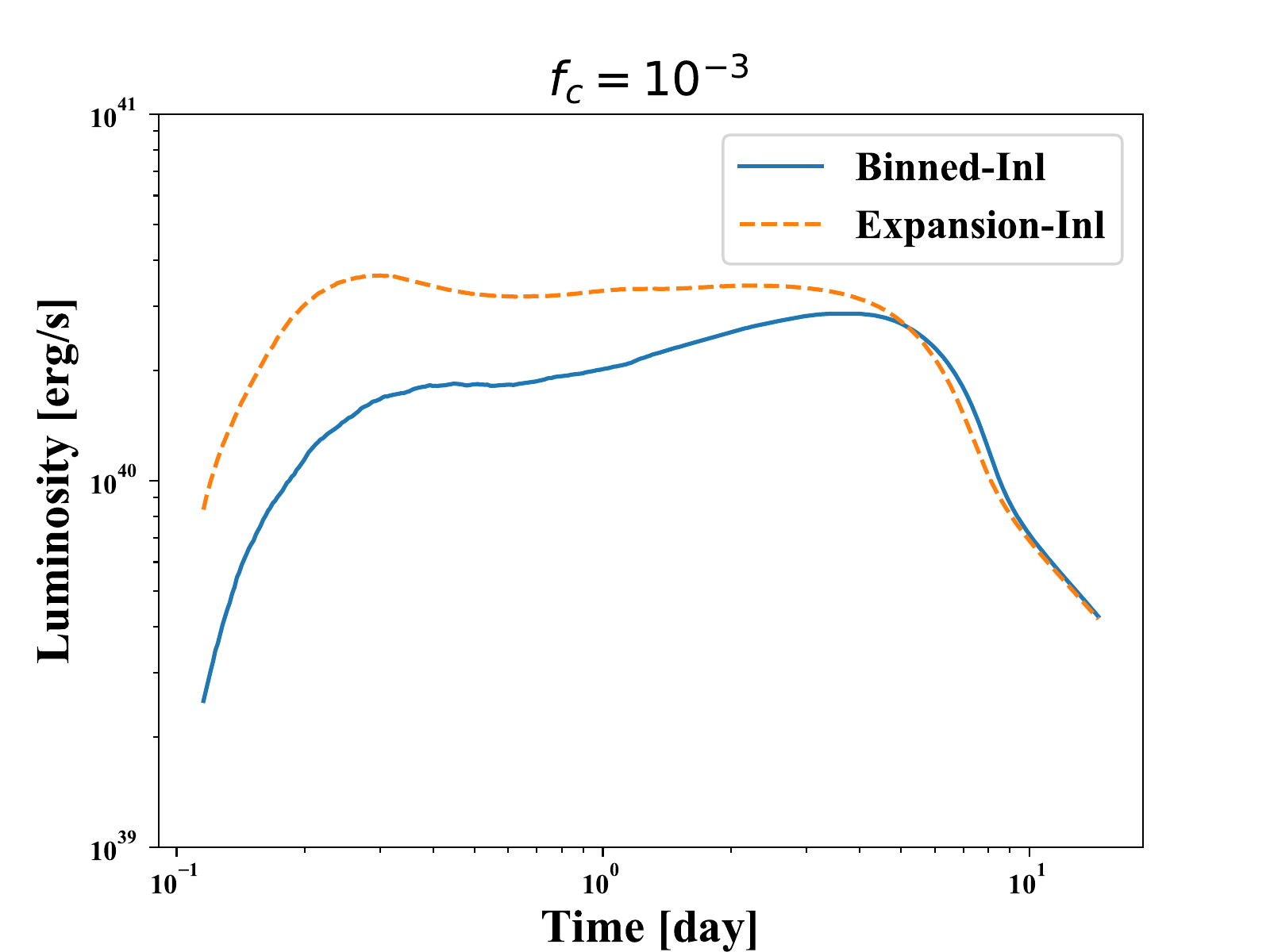}
    \label{fig:p2b}} \\
  {\includegraphics[clip=true,angle=0,width=1.0\columnwidth]{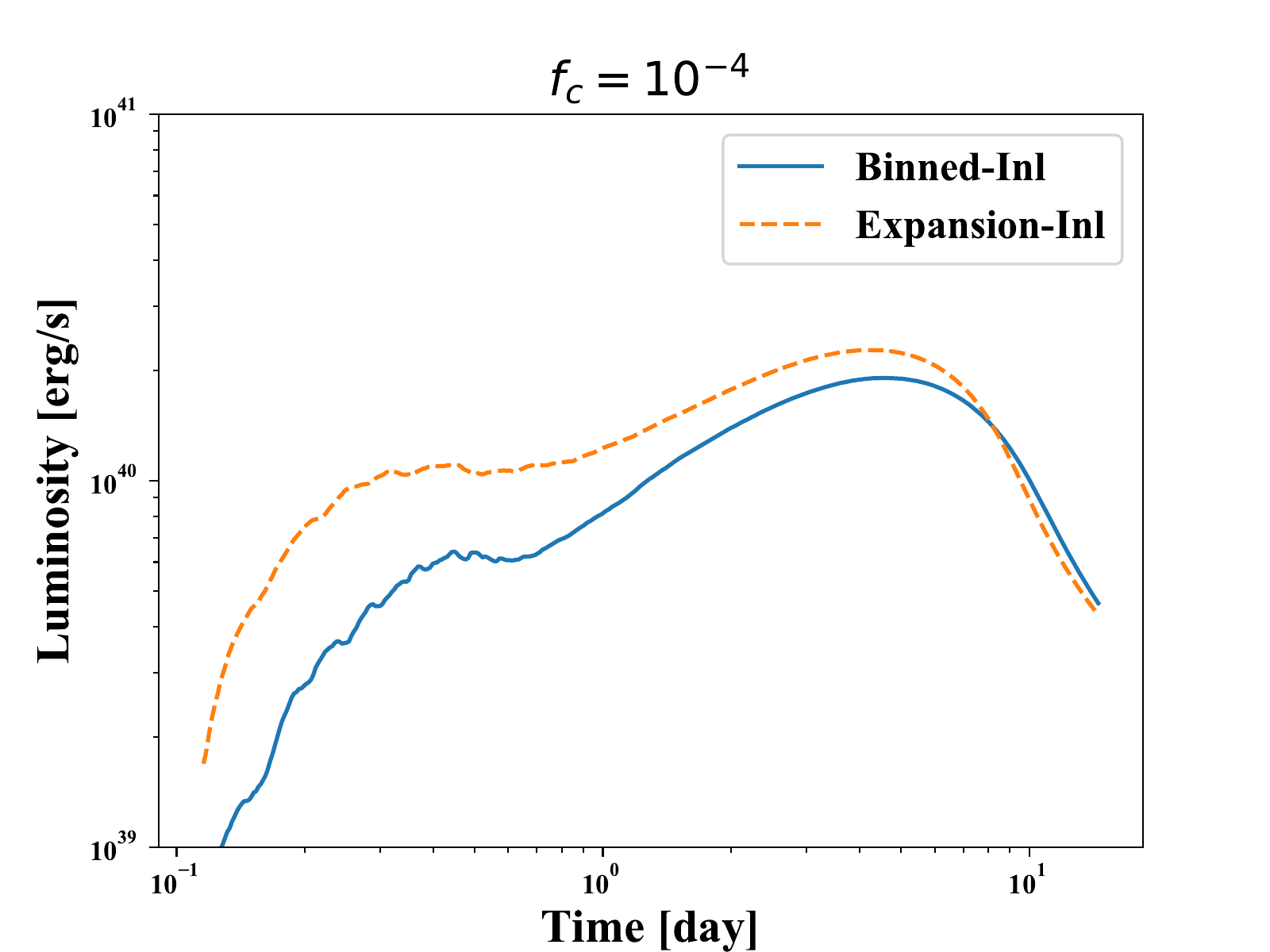}
    \label{fig:p2c}}
  {\includegraphics[clip=true,angle=0,width=1.0\columnwidth]{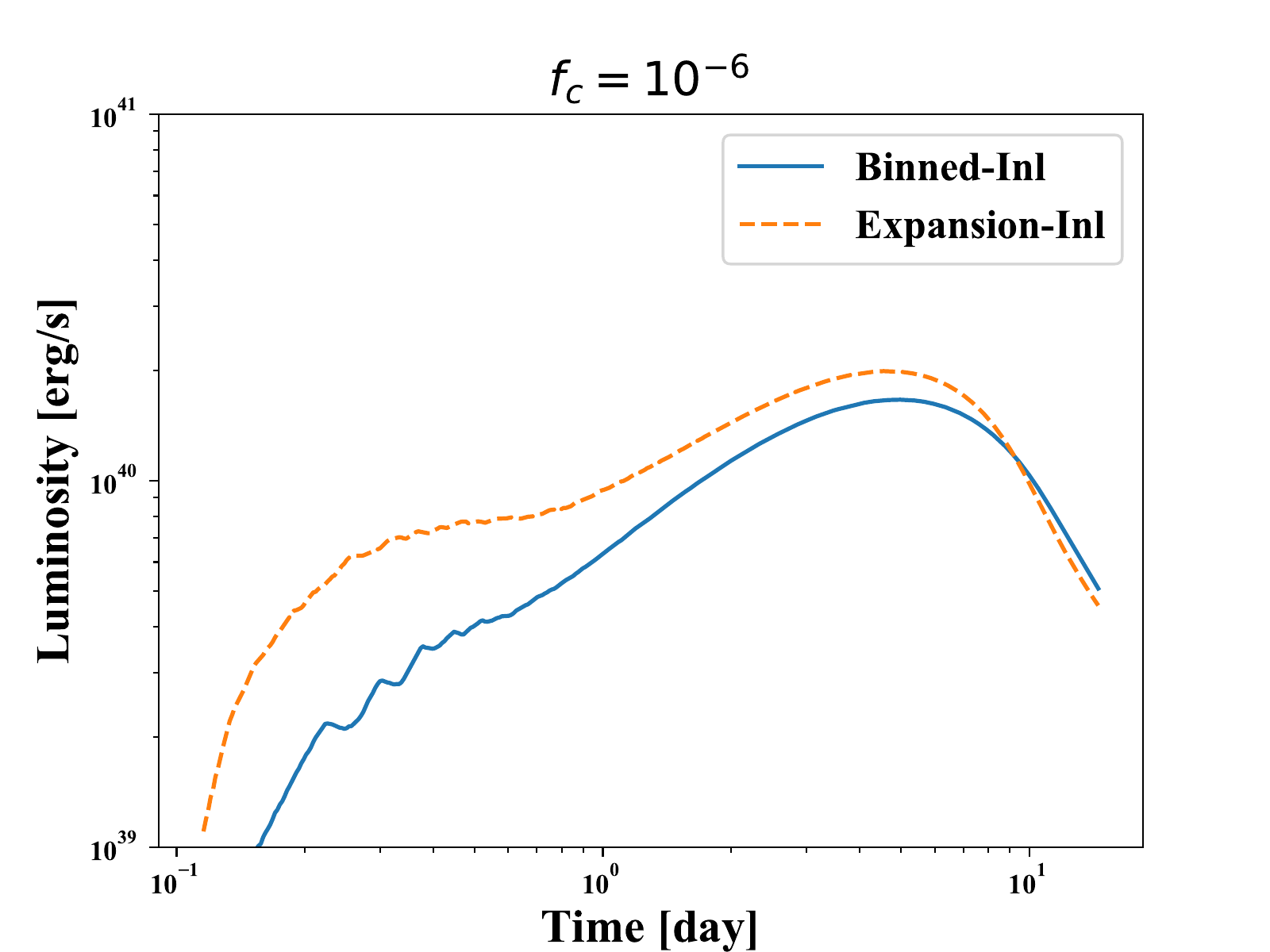}
    \label{fig:p2d}}
  \caption{
    Bolometric luminosity for oscillator strength cutoff values
    of $f_c = 10^{-1}$, $10^{-3}$, $10^{-4}$, and $10^{-6}$ in
    Figures~\ref{fig:p2a},~\ref{fig:p2b},~\ref{fig:p2c} and
    \ref{fig:p2d}, respectively.
    Results are presented for line-binned (solid blue curve)
    and expansion (dashed orange curve) inline opacities.
    There is a relatively strong dependence on $f_c$ down to a value 
    of $10^{-4}$, below which the light curves do not change too much.
  }
  \label{fig:p2}
\end{figure*}

\begin{figure*}
  \centering
  {\includegraphics[clip=true,angle=0,width=1.0\columnwidth]{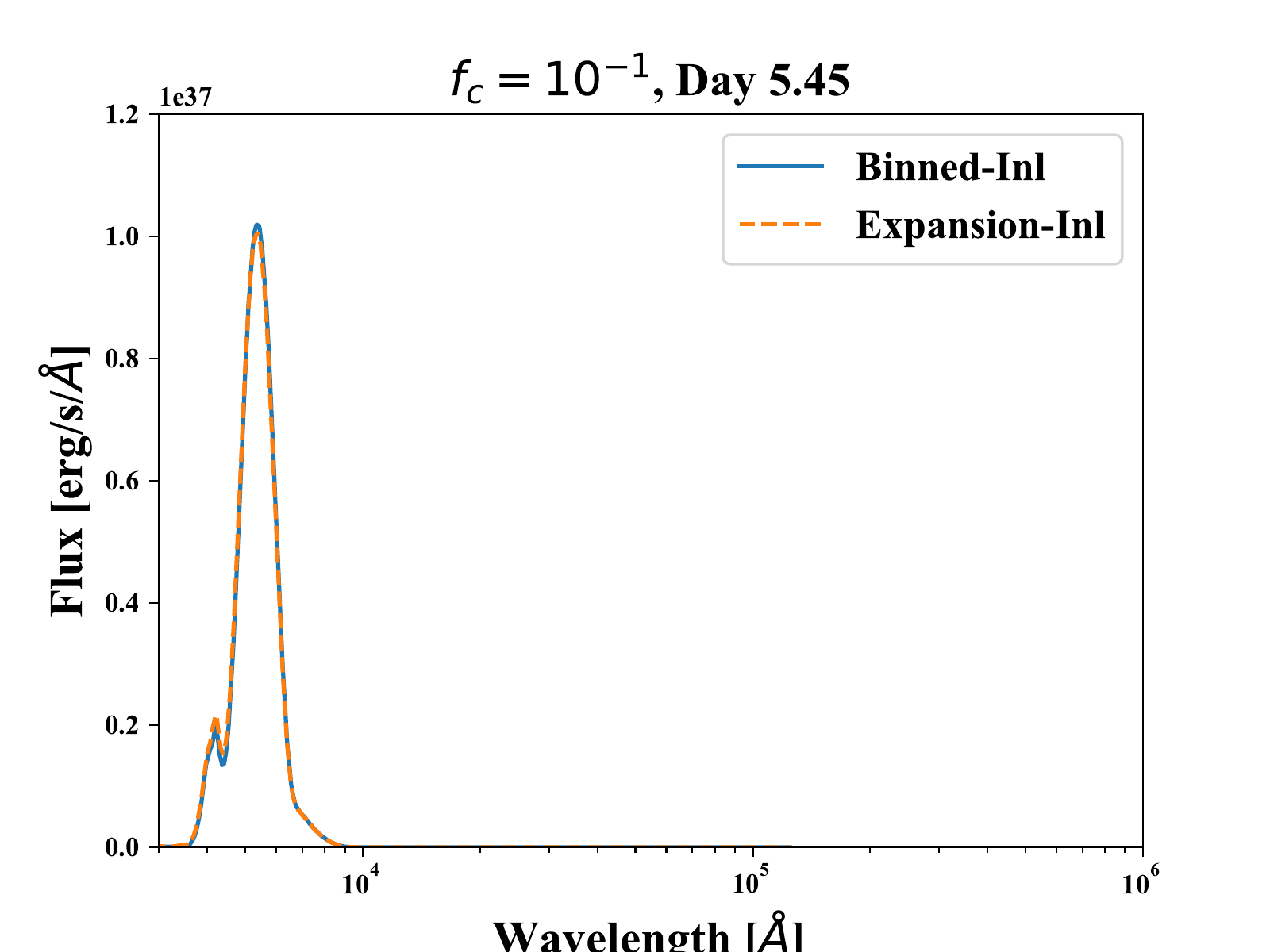}
    \label{fig:p3a}}
  {\includegraphics[clip=true,angle=0,width=1.0\columnwidth]{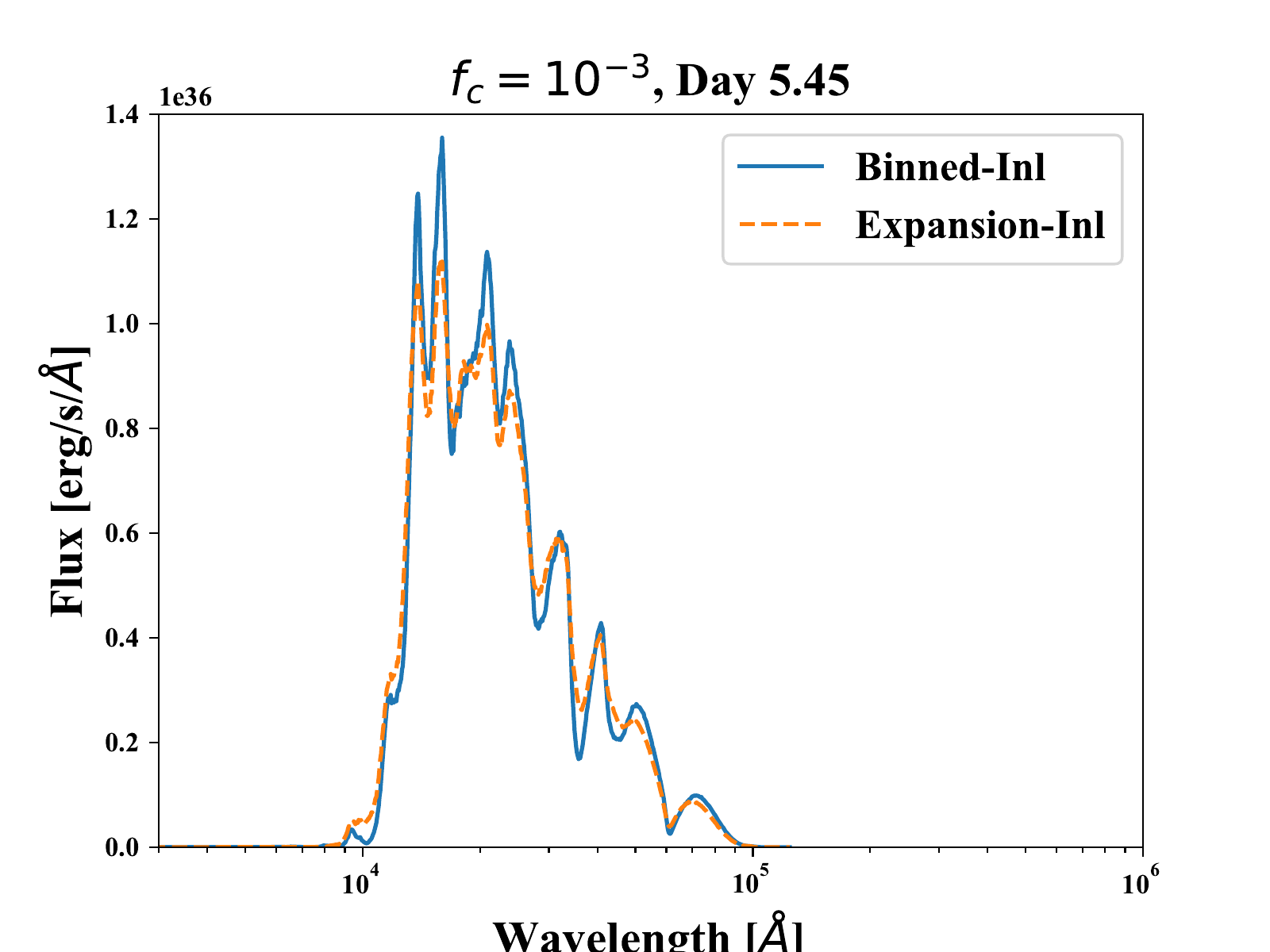}
    \label{fig:p3b}} \\
  {\includegraphics[clip=true,angle=0,width=1.0\columnwidth]{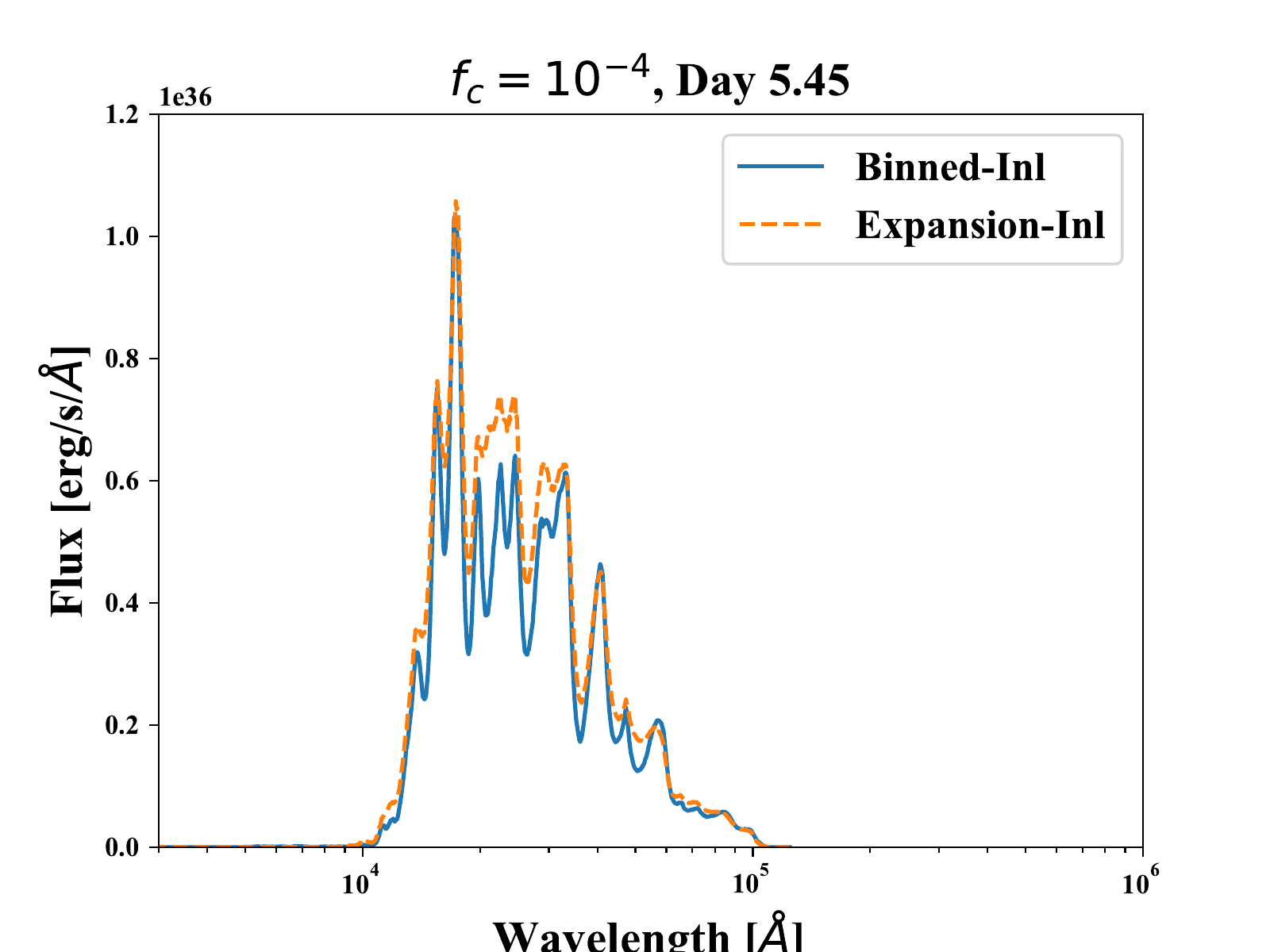}
    \label{fig:p3c}}
  {\includegraphics[clip=true,angle=0,width=1.0\columnwidth]{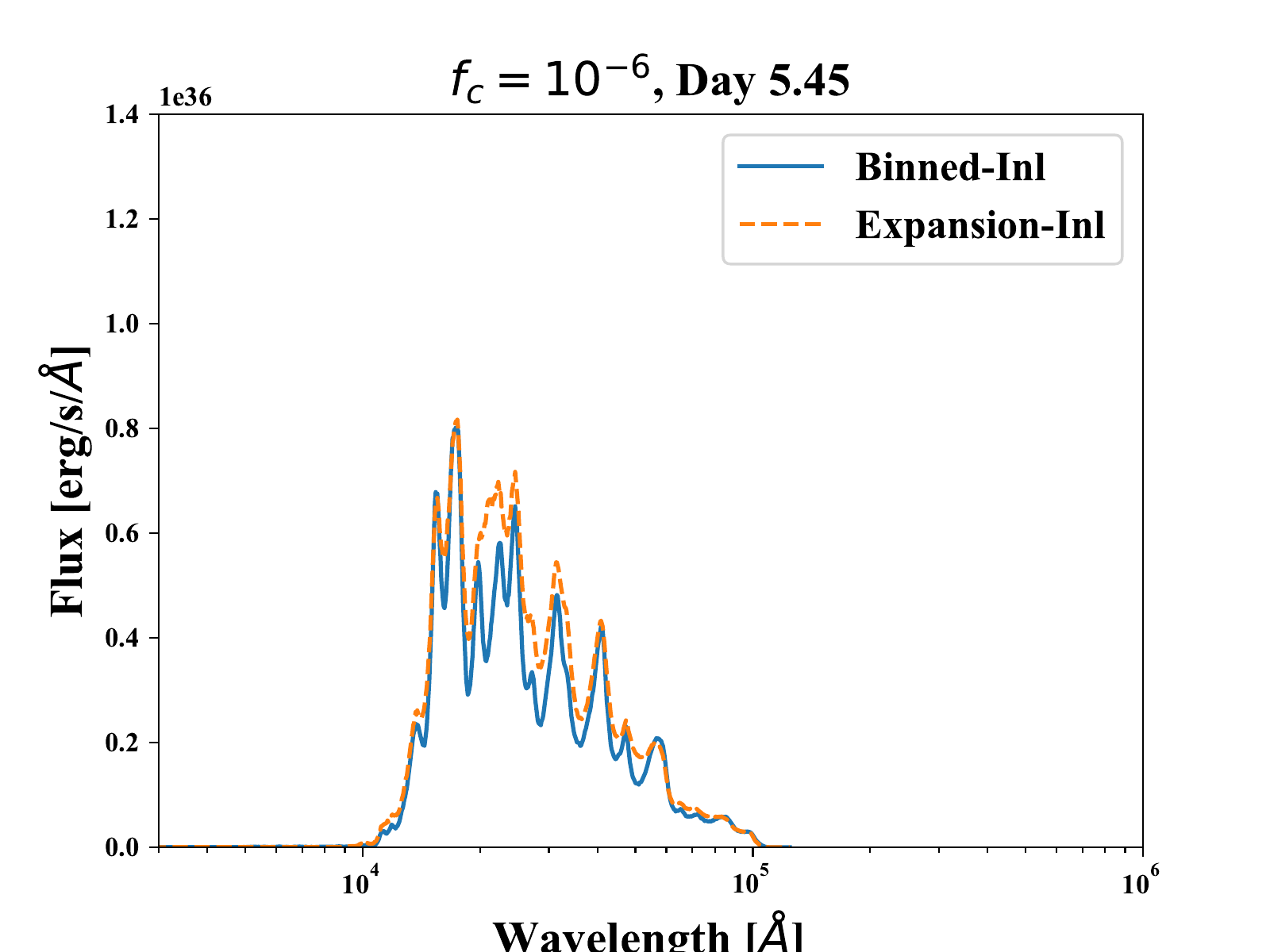}
    \label{fig:p3d}}
  \caption{
    Spectra at day 5.45 for oscillator strength cutoff values
    of $f_c = 10^{-1}$, $10^{-3}$, $10^{-4}$, and $10^{-6}$ in
    Figures~\ref{fig:p3a}, \ref{fig:p3b}, \ref{fig:p3c} and
    \ref{fig:p3d}, respectively.
    Results are presented for line-binned (solid blue curve)
    and expansion (dashed orange curve) inline opacities.
    There is a strong dependence on $f_c$ down to a value
    of $10^{-4}$, with significant differences still occurring in certain line
    features for the most complete model with $f_c = 10^{-6}$.
  }
  \label{fig:p3}
\end{figure*}

We next investigated the sensitivity of the emission to the
oscillator strength cutoff value, $f_c$.
Figures~\ref{fig:p2} and~\ref{fig:p3} display plots of light
curves and spectra, respectively, for different oscillator
strength cutoffs.
Each panel contains two curves that were generated with the inline,
line-binned and expansion-opacity methods.
Results are shown for cutoff values of
$f_{c}\in\{10^{-1},10^{-3},10^{-4},10^{-6}\}$.
These four cutoffs correspond to including a total of
2,814, 1,225,925, 6,447,337, and 22,710,094 lines, respectively.
These results indicate that the opacity changes significantly
when lowering $f_{c}$ from $10^{-1}$ to $10^{-3}$, but the effect of including
additional lines appears to level off at $10^{-4}$, which is consistent
with the behavior exhibited by the opacity curves
in the earlier Figure~\ref{fig:opac_exp_nd}.
We also note that the same convergence behavior is exhibited for both
the line-binned and expansion-opacity methods, although the light
curves and spectra display some quantitative differences,
even for the most inclusive atomic model generated with $f_{c} = 10^{-6}$.
For example, there is about a 30\% difference in the peak luminosity
when comparing expansion-opacity and line-binned light curves for
this full problem, versus only 5\% mentioned in the previous section
for the simplified problem. However, we again note that these disparities
are not sufficient to explain discrepancies between theory and observation.

\begin{figure*}
  \centering
  {\includegraphics[clip=true,angle=0,width=1.0\columnwidth]{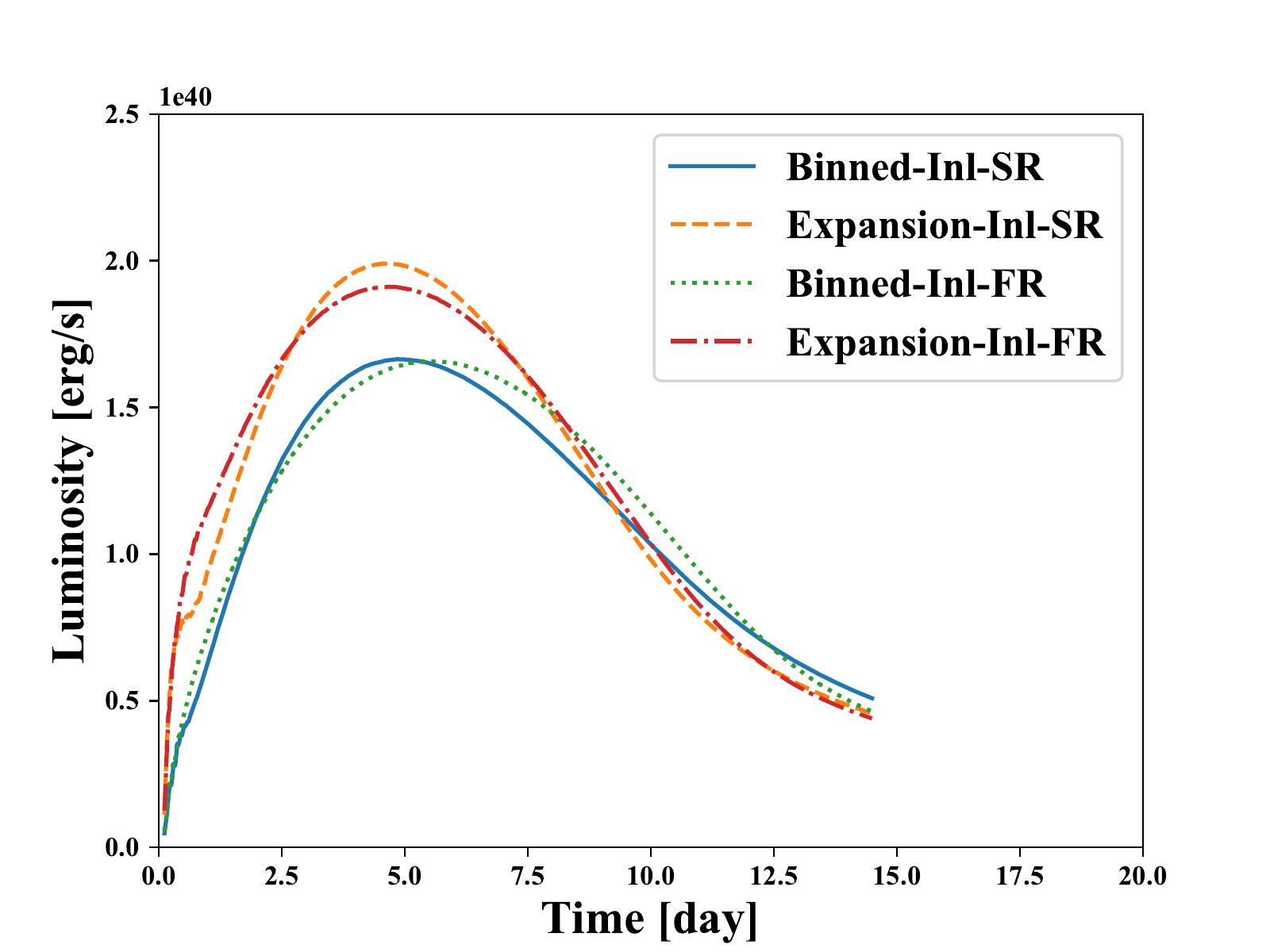}
    \label{fig:p4a}}
  {\includegraphics[clip=true,angle=0,width=1.0\columnwidth]{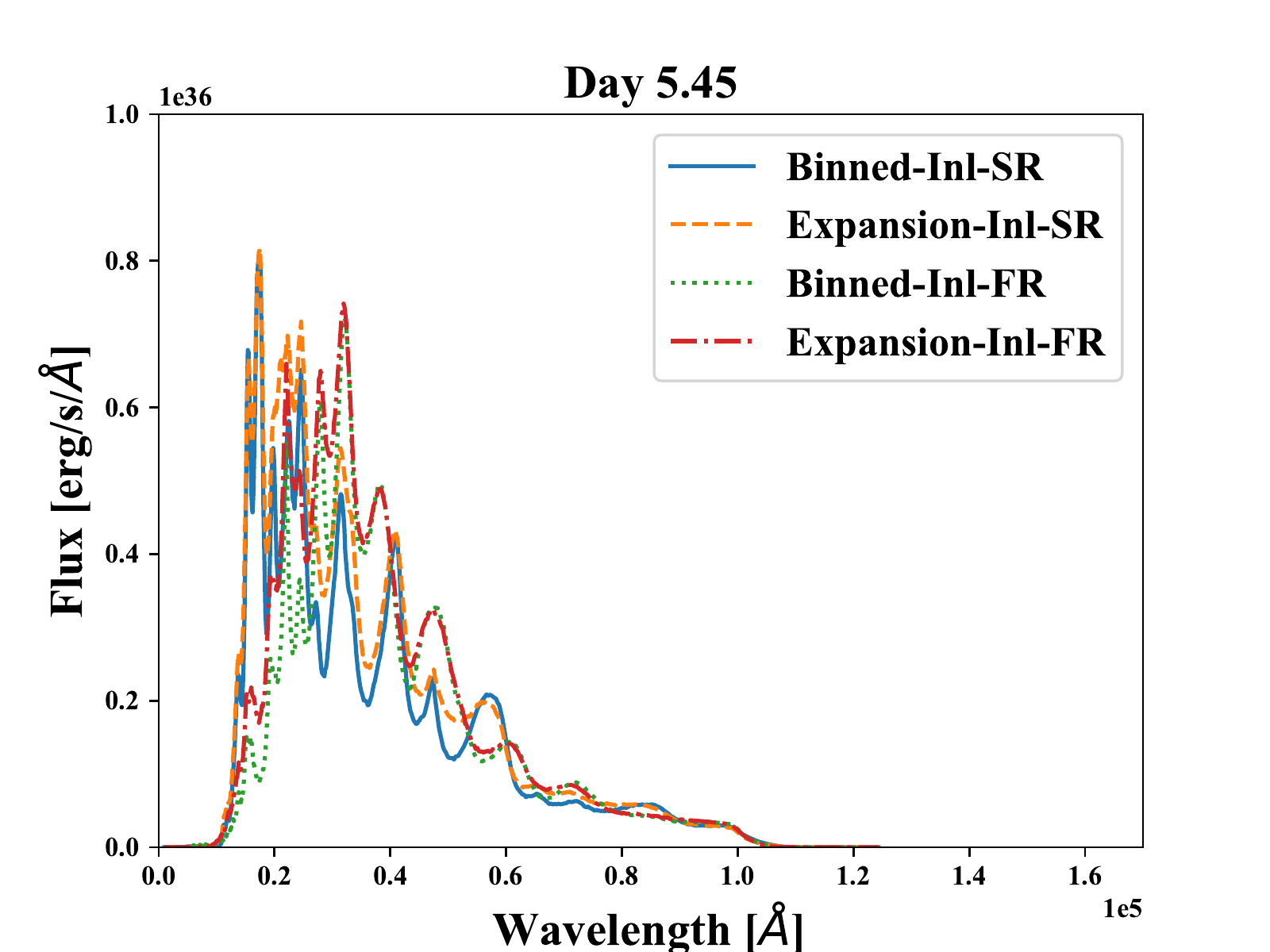}
    \label{fig:p4b}}
  \caption{
    Comparison of line-binned and expansion-opacity
    results using semi-relativistic (SR) and fully relativistic (FR)
    atomic data to calculate opacities.
    Figures~\ref{fig:p4a} and \ref{fig:p4b} display luminosities and
    spectra, respectively.
  }
  \label{fig:p4}
\end{figure*}

\begin{figure*}
  \centering
  {\includegraphics[clip=true,angle=0,width=1.0\columnwidth]{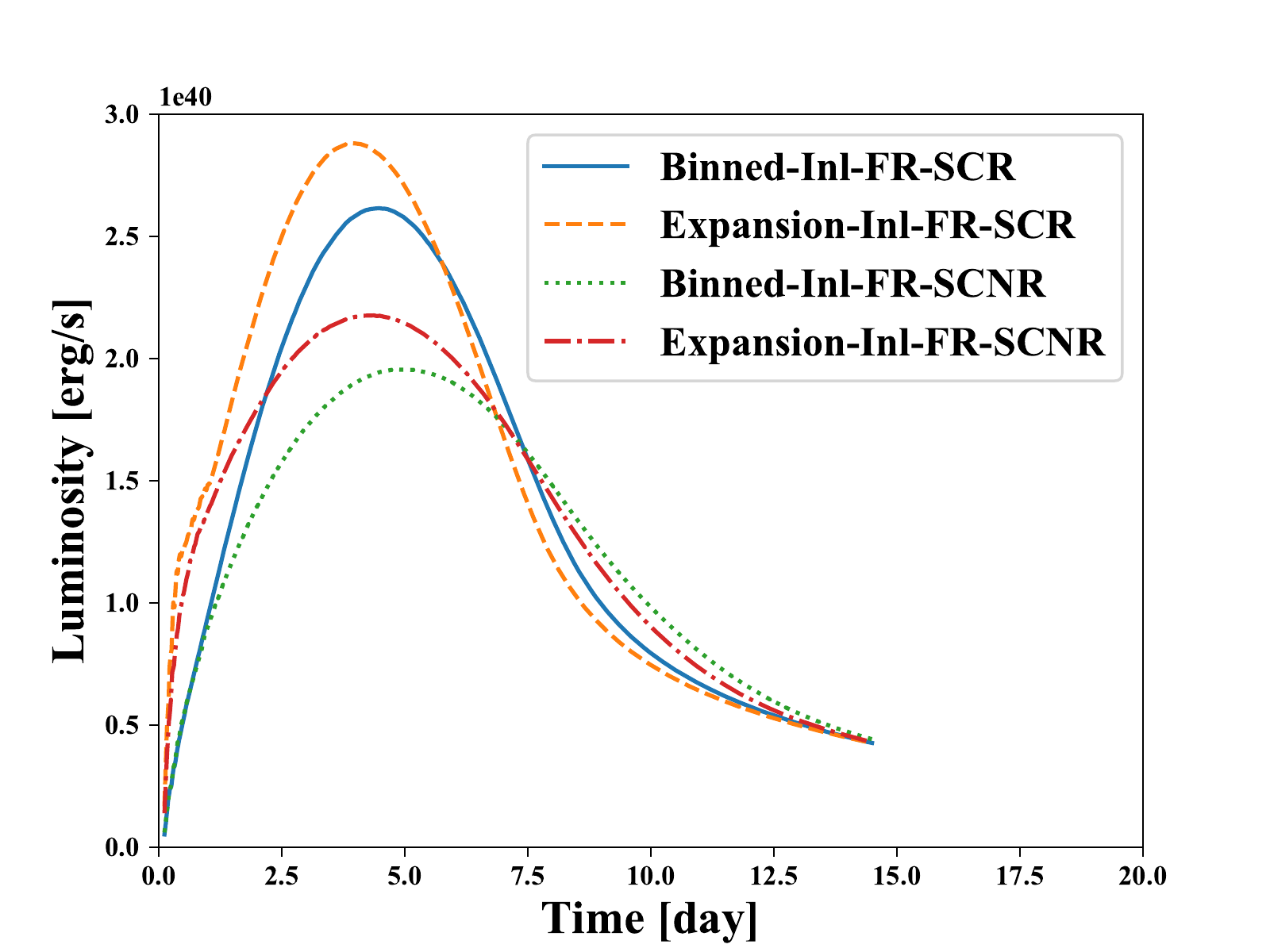}
    \label{fig:p5a}}
  {\includegraphics[clip=true,angle=0,width=1.0\columnwidth]{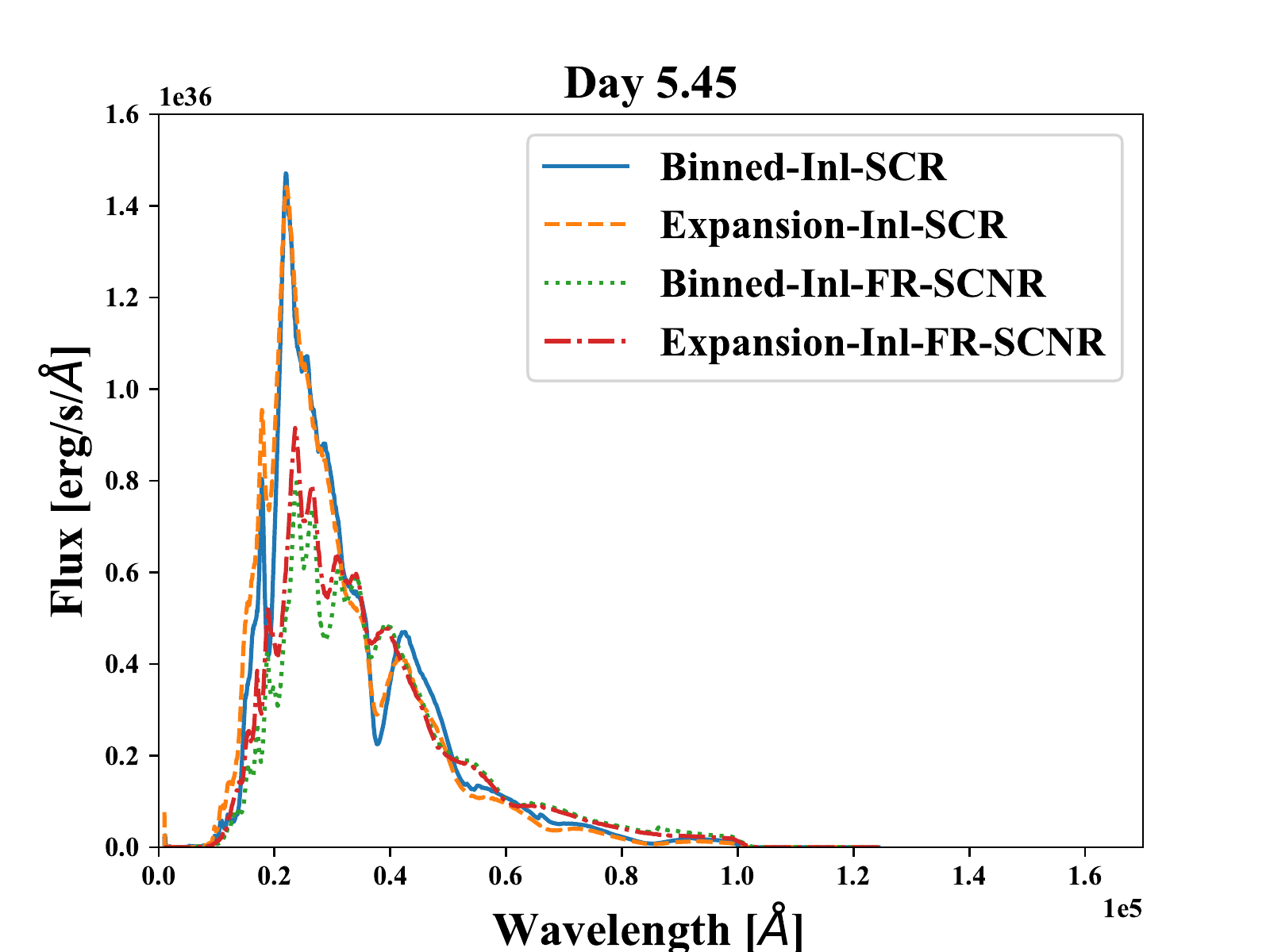}
    \label{fig:p5b}}
  \caption{
    Comparison of line-binned and expansion opacity
    results using the different configuration-interaction variations
    of FR atomic data (SCR and SCNR)
    to calculate opacities.
    Figures~\ref{fig:p5a} and \ref{fig:p5b} display luminosities and
    spectra, respectively.
  }
  \label{fig:p5}
\end{figure*}

In Figure~\ref{fig:p4}, we compare light curves using either
semi-relativistic (SR) or fully relativistic (FR) atomic data,
and either the line-binned or expansion-opacity method.
The bolometric luminosities (left panel) agree reasonably well between opacity
discretization methods, but the spectra agree better
between the two types of atomic data.
Figure~\ref{fig:p5} displays light curves and spectra for the FR-SCR and
FR-SCNR variants of the FR model, which have less fidelity, i.e. less
configuration interaction, than the default FR model.
Unlike the SR versus FR comparison in Figure~\ref{fig:p4},
the bolometric luminosities do not 
group together according to the opacity discretization method,
near peak luminosity.
This behavior indicates a stronger sensitivity to the amount
of fine-structure detail that is included in the atomic physics models,
relative to a sensitivity to the discretization method.
Additionally, the FR-SCNR light curves and spectra are more closely in
agreement with those of the FR and SR models, which is consistent
with theoretical expectations (see Section~\ref{subsec:models_var}
and Table~\ref{tab:linecomp}).

It is interesting to note that the spectral responses to the above suite of
parameter variations do not show significant shifts in the wavelength
band at which peak emission occurs.  This suggests that the
area-preserving opacity technique, and these other numerical resolution
choices employed in our earlier work \citep{wollaeger18}, do not
appear to explain why the spectra presented therein is shifted toward
red wavelengths, as compared to results produced by other authors who
use the expansion-opacity treatment (e.g.~\citealt{kasen17,tanaka18}).
Clearly, the parameter study presented here is not
exhaustive and further methodical, sensitivity studies of other key
properties will be needed (e.g.~ejecta characteristics such as
density, elemental composition and velocity distributions, as well as
details in the atomic data employed by each group).  As discussed in
Section~\ref{subsub:comp_simp}, simplified test problems provide a valuable
path to isolating the impact of various assumptions employed by the full
kilonova modeling codes, and future work should attempt to define such
tests.

\begin{figure*}
  \centering
  {\includegraphics[clip=true,angle=0,width=1.0\columnwidth]{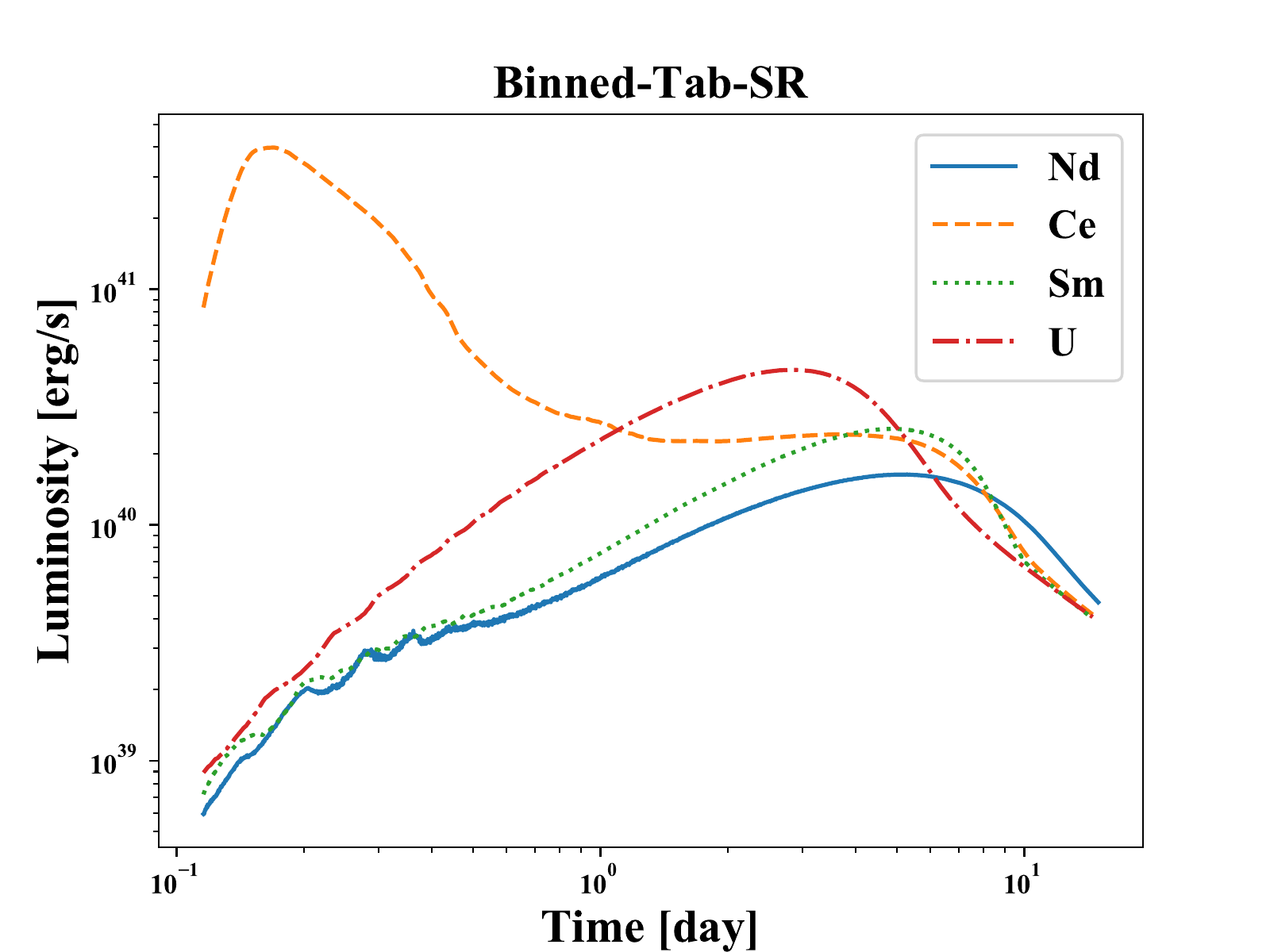}
    \label{fig:p6a}}
  {\includegraphics[clip=true,angle=0,width=1.0\columnwidth]{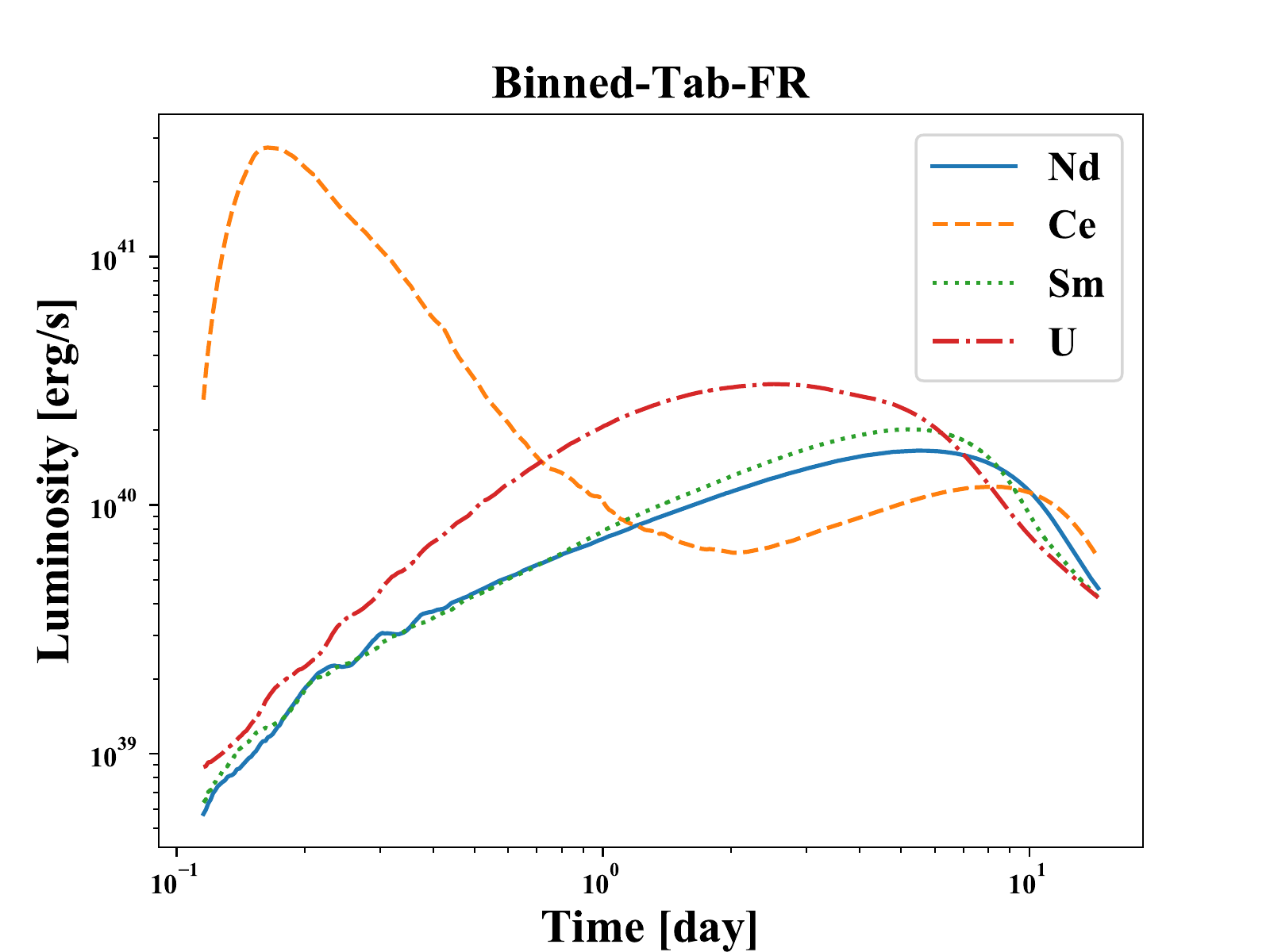}
    \label{fig:p6b}}
  \caption{
    Bolometric luminosity for different elements:
    Nd (blue solid), Ce (orange dashed), Sm
    (green dotted), and U (red dot-dashed) using tabulated
    semi-relativistic (left panel) or
    fully relativistic (right panel) binned opacities.
  }
  \label{fig:p6}
\end{figure*}

\begin{figure*}
  \centering
  {\includegraphics[clip=true,angle=0,width=1.0\columnwidth]{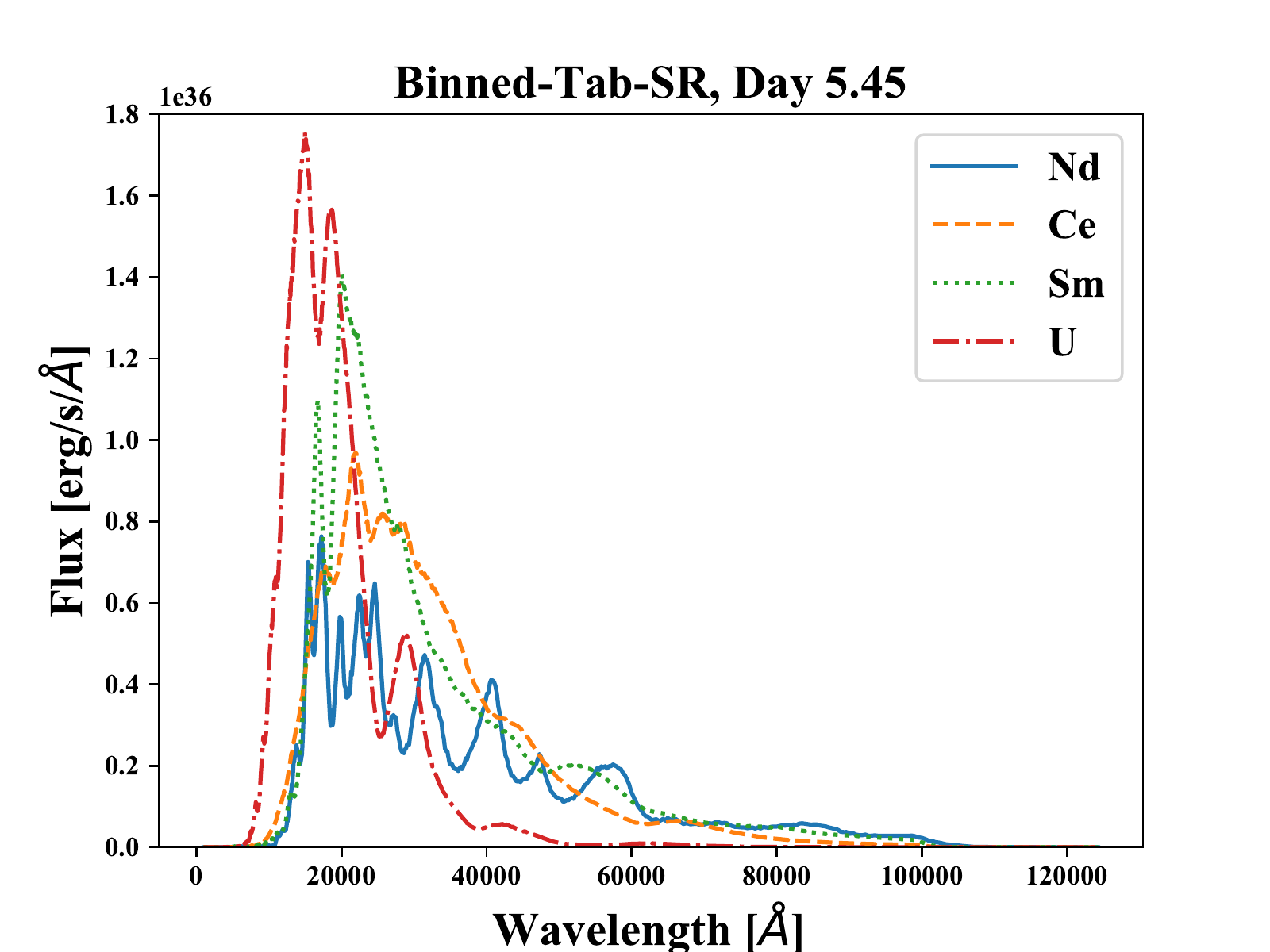}
    \label{fig:p7a}}
  {\includegraphics[clip=true,angle=0,width=1.0\columnwidth]{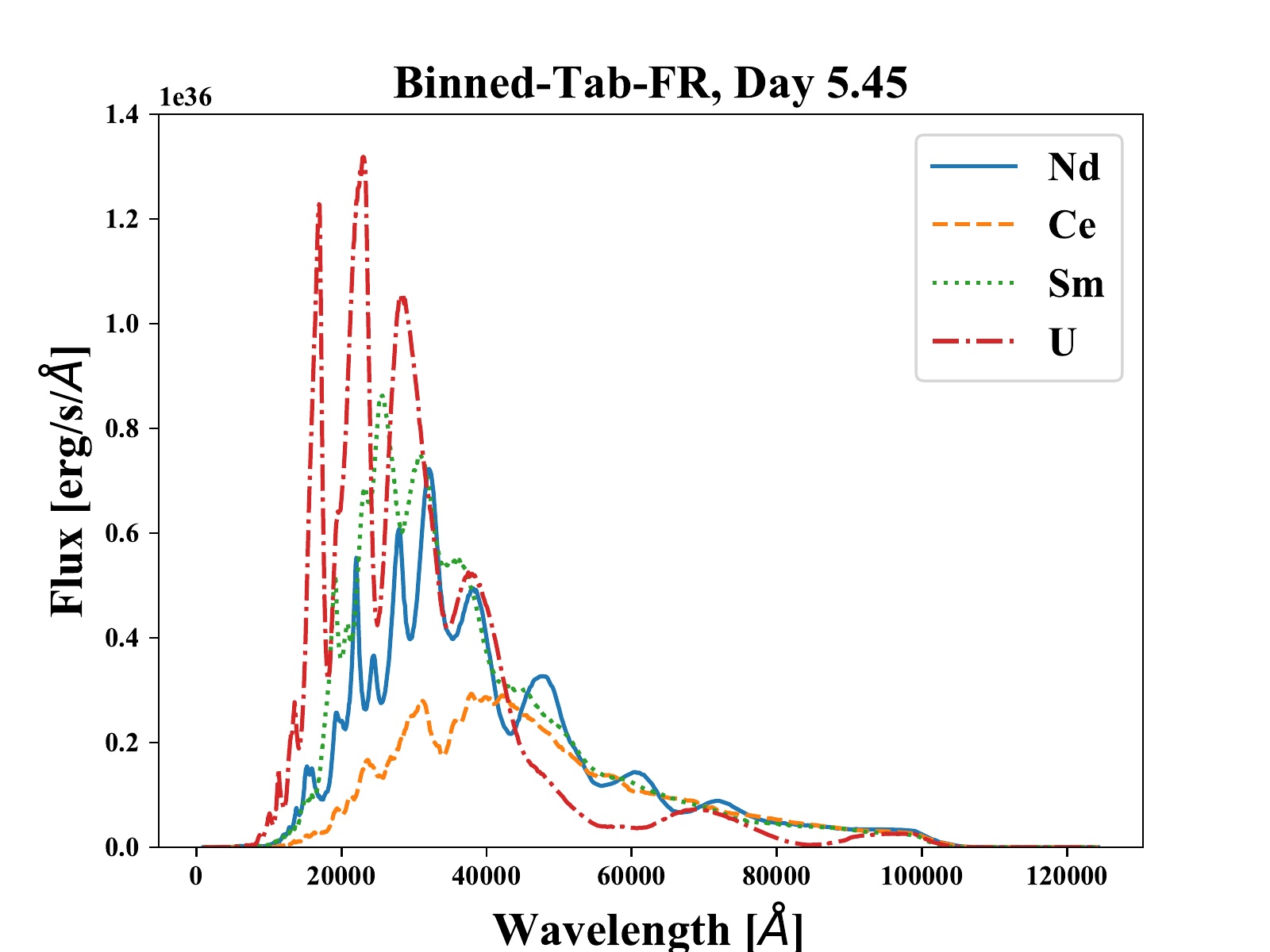}
    \label{fig:p7b}}
  \caption{
    Spectra at 5.45~days for different elements:
    Nd (blue solid), Ce (orange dashed), Sm
    (green dotted), and U (red dot-dashed) using tabulated
    semi-relativistic (left panel) or
    fully relativistic (right panel) binned opacities.
  }
  \label{fig:p7}
\end{figure*}

It is also worth noting that there is a common assumption among
all groups that the opacity contribution from the mixing of heavy elements
present in the ejecta can be reasonably represented by a surrogate,
single-element lanthanide (Nd is commonly assumed).
By considering different elements here,
we demonstrate that the choice of element can have a    
significant impact on the light curves and spectra as well.  
Figure~\ref{fig:p6} displays bolometric luminosity for line-binned opacity
tables of Nd along with Sm, Ce, and U.
Sm and Ce differ from Nd in atomic number by only +2 and -2, respectively,
but produce significantly brighter light curves,
by about a factor of two at 5.45~days post-merger.
Moreover, Ce has a bright early transient for both SR and FR opacity,
presumably due to its relatively low opacity that results from a smaller
number of $4f$ electrons. Nd and U are homologues, i.e. they appear
in the same column of the periodic table due to the similar
electron structure (with the principal quantum number of their valence shells
differing by one), but U displays a peak luminosity that is
about five times brighter and appears earlier by about three days.
The corresponding spectra for these four elements are presented
in Figure~\ref{fig:p7}. The peak emission varies in value by about a factor
of 2.5 for the SR calculations and a factor of 4.3 for the FR calculations,
and its wavelength location ranges from $\sim$10,000--30,000~\AA.
This type of sensitivity indicates that using a more complete set of lanthanide
elements to simulate kilonovae, rather than a single representative
element, may be appropriate.

Finally, we provide some basic timing information to demonstrate the
efficiency of the tabular, line-binned opacity method compared to the
corresponding inline approach, when simulating kilonovae emission.
For the case of $f_c = 10^{-6}$, SR atomic data and pure Nd,
the total time, in seconds, spent in the opacity calculations per MPI rank is:
0.1162 for tabular line-binned
(``Binned-Tab''), 167.001 for inline line-binned
(``Binned-Inl'') and 284.538 for expansion-opacity (``Expansion-Inl'').
(For reference, in our simulations the total mean time spent in the
radiation transport logic, for $\sim10^6$ particles per MPI rank, is 1060.51~s.)
Thus, the tabular approach is approximately three orders of magnitude faster
than the inline approach due to the amount of time required to obtain
the atomic level populations and to construct the line-binned opacities
for every required temperature-density.
Also of note is the fact that the inline expansion-opacity
calculation requires a factor of 1.7 more time than the corresponding
line-binned calculation.
The additional cost in the expansion-opacity method is due to the evaluation of the exponentials for each line.


\section{Summary}

We have proposed an area-preserving (line-binned),
tabular approach for calculating opacities
relevant to the modeling of kilonova light curves and spectra.
The line-binned approach is superior to the previous line-smeared approach
because the former is guaranteed to preserve the integral of the opacity.
The
area-preserving
approach differs from the more widely employed
expansion-opacity method in how the
line contribution to the opacity is calculated. However, our alternative
approach was demonstrated to produce similar emission results compared
to both the expansion-opacity method and to the more accurate, direct
implementation of Sobolev line transfer for a simplified problem
(see Figure~\ref{fig:test}).
In fact, light curves produced by
the approximate line-binned and expansion-opacity
methods were found to be lower than the direct Sobolev treatment by
a maximum of 8\% and 3\%, respectively. As expected, our line-binned
light curve is less than or equal to the expansion-opacity light curve,
with a maximum difference of 5\% occurring at peak for the simplified test case.
The spectral features show greater variability in the level of agreement
between the three opacity implementations, with our line-binned approach
agreeing well with the direct line treatment at late times.

We considered a more complete (``full'') problem
to study the errors associated with choices for the oscillator
strength cutoff value, $f_c$, which determines the number of lines that
are used in a particular simulation
(see Figures~\ref{fig:p2},~\ref{fig:p3}),
as well as sensitivities to the use of variant FR and SR atomic models
(see Figures~\ref{fig:p4},~\ref{fig:p5}).
We found that light curves are typically converged at a value of
$f_c = 10^{-4}$. Increasing the cutoff value by one order of magnitude,
$f_c = 10^{-3}$, results in a 50\% error in the peak of the curve and
a factor of three or so in the early-time behavior.
For the converged light curves produced with this full problem,
a larger difference of about 30\% was obtained in the peak luminosity
when comparing expansion-opacity and line-binned simulations.
Obtaining convergence for all major spectral features typically required
a lower tolerance of $f_c = 10^{-6}$.
As for SR versus FR models, the light curves were found to be relatively
insensitive to this option, but certain spectral features displayed
visible differences, and the FR calculation produced emission that
is shifted toward slightly higher (redder) wavelengths.

The choice of element comprising the ejecta was also investigated
and found to be a significant factor in predictions
of the brightness and shape of the bolometric luminosity light curve,
as well as the corresponding spectra, for single-element simulations.
(see Figures~\ref{fig:p6} and \ref{fig:p7}).
The peak luminosity varied by factors of 2--5 when considering
lanthanides and an actinide with several $f$-shell valence electrons
in the ground state of their neutral ion stages. The magnitude of
the peak spectral emission differed by similar amounts, i.e.
factors of 2.5--4.3.
A study of the relative importance of various elements in multi-element
light-curve and spectral simulations is reserved for future work.

From a practical, computational perspective,
the tabular line-binned approach presented here is more efficient
than the traditional expansion-opacity method
because the opacities are independent of the particular type
of expansion and can therefore be precomputed on a grid of temperatures
and densities.  It was found that the tabular approach
requires three orders of magnitude less time than
the expansion-opacity method to generate the relevant opacities for
typical kilonova simulations.
Tabular opacities for all 14 lanthanide elements, as well as for a
representative actinide element, uranium, have been generated using
the line-binned approach. These data are currently available, by request,
from the corresponding author and we plan to make them publicly
available via a new database developed in collaboration with NIST.

\section*{Acknowledgements}

Helpful discussions with H.A.~Scott, J.I.~Castor and S.P.~Owocki
are gratefully acknowledged.
This work was supported by the US Department of Energy through the Los Alamos
National Laboratory. Los Alamos National Laboratory is operated
by Triad National Security, LLC, for the National Nuclear Security
Administration of US Department of Energy (Contract No.~89233218CNA000001).





\bibliographystyle{mnras}
\bibliography{master}

\begin{thebibliography}{}
\makeatletter
\relax
\def\mn@urlcharsother{\let\do\@makeother \do\$\do\&\do\#\do\^\do\_\do\%\do\~}
\def\mn@doi{\begingroup\mn@urlcharsother \@ifnextchar [ {\mn@doi@}
  {\mn@doi@[]}}
\def\mn@doi@[#1]#2{\def\@tempa{#1}\ifx\@tempa\@empty \href
  {http://dx.doi.org/#2} {doi:#2}\else \href {http://dx.doi.org/#2} {#1}\fi
  \endgroup}
\def\mn@eprint#1#2{\mn@eprint@#1:#2::\@nil}
\def\mn@eprint@arXiv#1{\href {http://arxiv.org/abs/#1} {{\tt arXiv:#1}}}
\def\mn@eprint@dblp#1{\href {http://dblp.uni-trier.de/rec/bibtex/#1.xml}
  {dblp:#1}}
\def\mn@eprint@#1:#2:#3:#4\@nil{\def\@tempa {#1}\def\@tempb {#2}\def\@tempc
  {#3}\ifx \@tempc \@empty \let \@tempc \@tempb \let \@tempb \@tempa \fi \ifx
  \@tempb \@empty \def\@tempb {arXiv}\fi \@ifundefined
  {mn@eprint@\@tempb}{\@tempb:\@tempc}{\expandafter \expandafter \csname
  mn@eprint@\@tempb\endcsname \expandafter{\@tempc}}}

\bibitem[\protect\citeauthoryear{{Abbott} et~al.,}{{Abbott}
  et~al.}{2017a}]{abbott17h}
{Abbott} B.~P.,  et~al., 2017a, \mn@doi [Physical Review Letters]
  {10.1103/PhysRevLett.119.161101}, \href
  {http://adsabs.harvard.edu/abs/2017PhRvL.119p1101A} {119, 161101}

\bibitem[\protect\citeauthoryear{{Abbott} et~al.,}{{Abbott}
  et~al.}{2017b}]{abbott17a}
{Abbott} B.~P.,  et~al., 2017b, \mn@doi [\apjl] {10.3847/2041-8213/aa91c9},
  \href {http://adsabs.harvard.edu/abs/2017ApJ...848L..12A} {848, L12}

\bibitem[\protect\citeauthoryear{Abdallah, {R.E.H.~Clark}  \&
  {R.D.~Cowan}}{Abdallah et~al.}{1988}]{cats_man}
Abdallah J.,  {R.E.H.~Clark}  {R.D.~Cowan} 1988, {Los Alamos Manual LA-11436-M,
  Vol. I}

\bibitem[\protect\citeauthoryear{{Abdikamalov}, {Burrows}, {Ott},
  {L{\"o}ffler}, {O'Connor}, {Dolence}  \& {Schnetter}}{{Abdikamalov}
  et~al.}{2012}]{abdikamalov12}
{Abdikamalov} E.,  {Burrows} A.,  {Ott} C.~D.,  {L{\"o}ffler} F.,  {O'Connor}
  E.,  {Dolence} J.~C.,   {Schnetter} E.,  2012, \mn@doi [\apj]
  {10.1088/0004-637X/755/2/111}, \href
  {http://adsabs.harvard.edu/abs/2012ApJ...755..111A} {755, 111}

\bibitem[\protect\citeauthoryear{{Barnes} \& {Kasen}}{{Barnes} \&
  {Kasen}}{2013}]{barnes13}
{Barnes} J.,  {Kasen} D.,  2013, \mn@doi [\apj] {10.1088/0004-637X/775/1/18},
  \href {http://adsabs.harvard.edu/abs/2013ApJ...775...18B} {775, 18}

\bibitem[\protect\citeauthoryear{{Barnes}, {Kasen}, {Wu}  \&
  {Mart{\'{\i}}nez-Pinedo}}{{Barnes} et~al.}{2016}]{barnes16}
{Barnes} J.,  {Kasen} D.,  {Wu} M.-R.,   {Mart{\'{\i}}nez-Pinedo} G.,  2016,
  \mn@doi [\apj] {10.3847/0004-637X/829/2/110}, \href
  {http://adsabs.harvard.edu/abs/2016ApJ...829..110B} {829, 110}

\bibitem[\protect\citeauthoryear{{Castor}}{{Castor}}{1974}]{castor74}
{Castor} J.~L.,  1974, \mn@doi [\mnras] {10.1093/mnras/169.2.279}, \href
  {http://adsabs.harvard.edu/abs/1974MNRAS.169..279C} {169, 279}

\bibitem[\protect\citeauthoryear{{Castor}}{{Castor}}{2004}]{castor04}
{Castor} J.~I.,  2004, {Radiation Hydrodynamics}.
Cambridge, UK: Cambridge University Press

\bibitem[\protect\citeauthoryear{{Colgan} et~al.,}{{Colgan}
  et~al.}{2016}]{colgan_oplib}
{Colgan} J.,  et~al., 2016, \mn@doi [\apj] {10.3847/0004-637X/817/2/116}, \href
  {http://adsabs.harvard.edu/abs/2016ApJ...817..116C} {817, 116}

\bibitem[\protect\citeauthoryear{{C{\^o}t{\'e}} et~al.,}{{C{\^o}t{\'e}}
  et~al.}{2017}]{cote18}
{C{\^o}t{\'e}} B.,  et~al., 2017, \apj, \href
  {http://adsabs.harvard.edu/abs/2017arXiv171005875C} {}

\bibitem[\protect\citeauthoryear{{Cowan}}{{Cowan}}{1981}]{cowan}
{Cowan} R.~D.,  1981, {The theory of atomic structure and spectra}.
Berkeley: University of California Press

\bibitem[\protect\citeauthoryear{{Cowperthwaite} et~al.,}{{Cowperthwaite}
  et~al.}{2017}]{cowperthwaite17}
{Cowperthwaite} P.~S.,  et~al., 2017, \mn@doi [\apjl]
  {10.3847/2041-8213/aa8fc7}, \href
  {http://adsabs.harvard.edu/abs/2017ApJ...848L..17C} {848, L17}

\bibitem[\protect\citeauthoryear{Fleck \& Cummings}{Fleck \&
  Cummings}{1971}]{fleck71}
Fleck J.,  Cummings J.,  1971, Journal of Computational Physics, 8, 313

\bibitem[\protect\citeauthoryear{{Fontes}, {Fryer}, {Hungerford}, {Hakel},
  {Colgan}, {Kilcrease}  \& {Sherrill}}{{Fontes} et~al.}{2015a}]{fontes15}
{Fontes} C.~J.,  {Fryer} C.~L.,  {Hungerford} A.~L.,  {Hakel} P.,  {Colgan} J.,
   {Kilcrease} D.~P.,   {Sherrill} M.~E.,  2015a, \mn@doi [High Energy Density
  Physics] {10.1016/j.hedp.2015.06.002}, \href
  {http://adsabs.harvard.edu/abs/2015HEDP...16...53F} {16, 53}

\bibitem[\protect\citeauthoryear{{Fontes} et~al.,}{{Fontes}
  et~al.}{2015b}]{LANL_suite}
{Fontes} C.~J.,  et~al., 2015b, \mn@doi [Journal of Physics B Atomic Molecular
  Physics] {10.1088/0953-4075/48/14/144014}, \href
  {http://adsabs.harvard.edu/abs/2015JPhB...48n4014F} {48, 144014}

\bibitem[\protect\citeauthoryear{Fontes, Colgan  \& Abdallah~Jr}{Fontes
  et~al.}{2016}]{fontes_cr16}
Fontes C.~J.,  Colgan J.,   Abdallah~Jr J.,  2016, in {Ralchenko} Y.,  ed., ,
  Modern Methods in Collisional-Radiative Modeling of Plasmas.
New York: Springer, p.~17

\bibitem[\protect\citeauthoryear{{Fontes}, {Fryer}, {Hungerford}, {Wollaeger},
  {Rosswog}  \& {Berger}}{{Fontes} et~al.}{2017}]{fontes17}
{Fontes} C.~J.,  {Fryer} C.~L.,  {Hungerford} A.~L.,  {Wollaeger} R.~T.,
  {Rosswog} S.,   {Berger} E.,  2017, preprint, \href
  {http://adsabs.harvard.edu/abs/2017arXiv170202990F} {} (\mn@eprint {arXiv}
  {1702.02990})

\bibitem[\protect\citeauthoryear{{Frey}, {Even}, {Whalen}, {Fryer},
  {Hungerford}, {Fontes}  \& {Colgan}}{{Frey} et~al.}{2013}]{frey13}
{Frey} L.~H.,  {Even} W.,  {Whalen} D.~J.,  {Fryer} C.~L.,  {Hungerford} A.~L.,
   {Fontes} C.~J.,   {Colgan} J.,  2013, \mn@doi [\apjs]
  {10.1088/0067-0049/204/2/16}, \href
  {http://adsabs.harvard.edu/abs/2013ApJS..204...16F} {204, 16}

\bibitem[\protect\citeauthoryear{{Grossman}, {Korobkin}, {Rosswog}  \&
  {Piran}}{{Grossman} et~al.}{2014}]{grossman14}
{Grossman} D.,  {Korobkin} O.,  {Rosswog} S.,   {Piran} T.,  2014, \mn@doi
  [\mnras] {10.1093/mnras/stt2503}, \href
  {http://adsabs.harvard.edu/abs/2014MNRAS.439..757G} {439, 757}

\bibitem[\protect\citeauthoryear{{Hakel} \& {Kilcrease}}{{Hakel} \&
  {Kilcrease}}{2004}]{atomic2}
{Hakel} P.,  {Kilcrease} D.~P.,  2004, in {Cohen} J.~S.,  {Kilcrease} D.~P.,
  {Mazevet} S.,  eds,  American Institute of Physics Conference Series Vol.
  730, American Institute of Physics Conference Series. p.~190,
  \mn@doi{10.1063/1.1824870}

\bibitem[\protect\citeauthoryear{{Hakel} et~al.,}{{Hakel}
  et~al.}{2006}]{hakel06}
{Hakel} P.,  et~al., 2006, \mn@doi [\jqsrt] {10.1016/j.jqsrt.2005.04.007},
  \href {http://adsabs.harvard.edu/abs/2006JQSRT..99..265H} {99, 265}

\bibitem[\protect\citeauthoryear{{Huebner} \& {Barfield}}{{Huebner} \&
  {Barfield}}{2014}]{huebner}
{Huebner} W.~F.,  {Barfield} W.~D.,  2014, {Opacity}.
 Astrophysics and Space Science Library Vol. 402, New York: Springer,
  \mn@doi{10.1007/978-1-4614-8797-5}

\bibitem[\protect\citeauthoryear{{Just}, {Bauswein}, {Pulpillo}, {Goriely}  \&
  {Janka}}{{Just} et~al.}{2015}]{just15}
{Just} O.,  {Bauswein} A.,  {Pulpillo} R.~A.,  {Goriely} S.,   {Janka} H.-T.,
  2015, \mn@doi [\mnras] {10.1093/mnras/stv009}, \href
  {http://adsabs.harvard.edu/abs/2015MNRAS.448..541J} {448, 541}

\bibitem[\protect\citeauthoryear{{Karp}, {Lasher}, {Chan}  \&
  {Salpeter}}{{Karp} et~al.}{1977}]{karp77}
{Karp} A.~H.,  {Lasher} G.,  {Chan} K.~L.,   {Salpeter} E.~E.,  1977, \mn@doi
  [\apj] {10.1086/155241}, \href
  {http://adsabs.harvard.edu/abs/1977ApJ...214..161K} {214, 161}

\bibitem[\protect\citeauthoryear{{Kasen}, {Thomas}  \& {Nugent}}{{Kasen}
  et~al.}{2006}]{kasen06_2}
{Kasen} D.,  {Thomas} R.~C.,   {Nugent} P.,  2006, \mn@doi [\apj]
  {10.1086/506190}, \href {http://adsabs.harvard.edu/abs/2006ApJ...651..366K}
  {651, 366}

\bibitem[\protect\citeauthoryear{{Kasen}, {Badnell}  \& {Barnes}}{{Kasen}
  et~al.}{2013}]{kasen13}
{Kasen} D.,  {Badnell} N.~R.,   {Barnes} J.,  2013, \mn@doi [\apj]
  {10.1088/0004-637X/774/1/25}, \href
  {http://adsabs.harvard.edu/abs/2013ApJ...774...25K} {774, 25}

\bibitem[\protect\citeauthoryear{{Kasen}, {Metzger}, {Barnes}, {Quataert}  \&
  {Ramirez-Ruiz}}{{Kasen} et~al.}{2017}]{kasen17}
{Kasen} D.,  {Metzger} B.,  {Barnes} J.,  {Quataert} E.,   {Ramirez-Ruiz} E.,
  2017, \mn@doi [\nat] {10.1038/nature24453}, \href
  {http://adsabs.harvard.edu/abs/2017Natur.551...80K} {551, 80}

\bibitem[\protect\citeauthoryear{{Kohn} \& {Sham}}{{Kohn} \&
  {Sham}}{1965}]{kohnsham}
{Kohn} W.,  {Sham} L.~J.,  1965, \mn@doi [Physical Review]
  {10.1103/PhysRev.140.A1133}, \href
  {http://adsabs.harvard.edu/abs/1965PhRv..140.1133K} {140, 1133}

\bibitem[\protect\citeauthoryear{{Korobkin}, {Rosswog}, {Arcones}  \&
  {Winteler}}{{Korobkin} et~al.}{2012}]{korobkin12}
{Korobkin} O.,  {Rosswog} S.,  {Arcones} A.,   {Winteler} C.,  2012, \mn@doi
  [\mnras] {10.1111/j.1365-2966.2012.21859.x}, \href
  {http://adsabs.harvard.edu/abs/2012MNRAS.426.1940K} {426, 1940}

\bibitem[\protect\citeauthoryear{Kramida, {Yu.~Ralchenko}, Reader  \& {and NIST
  ASD Team}}{Kramida et~al.}{2018}]{nist}
Kramida A.,  {Yu.~Ralchenko} Reader J.,   {and NIST ASD Team} 2018, {NIST
  Atomic Spectra Database (ver. 5.6.1), [Online]. Available:
  {\tt{https://physics.nist.gov/asd}} [2019, January 29]. National Institute of
  Standards and Technology, Gaithersburg, MD.}

\bibitem[\protect\citeauthoryear{{Lucy}}{{Lucy}}{2005}]{lucy05}
{Lucy} L.~B.,  2005, \mn@doi [\aap] {10.1051/0004-6361:20041656}, \href
  {http://adsabs.harvard.edu/abs/2005A\%26A...429...19L} {429, 19}

\bibitem[\protect\citeauthoryear{{Magee} et~al.,}{{Magee}
  et~al.}{2004}]{atomic1}
{Magee} N.~H.,  et~al., 2004, in {Cohen} J.~S.,  {Kilcrease} D.~P.,   {Mazevet}
  S.,  eds,  American Institute of Physics Conference Series Vol. 730, American
  Institute of Physics Conference Series. New York: AIP, p.~168,
  \mn@doi{10.1063/1.1824868}

\bibitem[\protect\citeauthoryear{{Metzger}}{{Metzger}}{2017}]{metzger17y}
{Metzger} B.~D.,  2017, \mn@doi [Living Reviews in Relativity]
  {10.1007/s41114-017-0006-z}, \href
  {http://adsabs.harvard.edu/abs/2017LRR....20....3M} {20, 3}

\bibitem[\protect\citeauthoryear{{Metzger} \& {Berger}}{{Metzger} \&
  {Berger}}{2012}]{metzger12f}
{Metzger} B.~D.,  {Berger} E.,  2012, \mn@doi [\apj]
  {10.1088/0004-637X/746/1/48}, \href
  {http://adsabs.harvard.edu/abs/2012ApJ...746...48M} {746, 48}

\bibitem[\protect\citeauthoryear{{Narayan}, {Paczynski}  \& {Piran}}{{Narayan}
  et~al.}{1992}]{1992ApJ...395L..83N}
{Narayan} R.,  {Paczynski} B.,   {Piran} T.,  1992, \mn@doi [\apjl]
  {10.1086/186493}, \href {http://adsabs.harvard.edu/abs/1992ApJ...395L..83N}
  {395, L83}

\bibitem[\protect\citeauthoryear{{Pian} et~al.,}{{Pian} et~al.}{2017}]{pian17}
{Pian} E.,  et~al., 2017, \mn@doi [\nat] {10.1038/nature24298}, \href
  {http://adsabs.harvard.edu/abs/2017Natur.551...67P} {551, 67}

\bibitem[\protect\citeauthoryear{{Pinto} \& {Eastman}}{{Pinto} \&
  {Eastman}}{2000}]{pintoeast00}
{Pinto} P.~A.,  {Eastman} R.~G.,  2000, \mn@doi [\apj] {10.1086/308380}, \href
  {http://adsabs.harvard.edu/abs/2000ApJ...530..757P} {530, 757}

\bibitem[\protect\citeauthoryear{{Rosswog}, {Liebend{\"o}rfer}, {Thielemann},
  {Davies}, {Benz}  \& {Piran}}{{Rosswog} et~al.}{1999}]{1999A&A...341..499R}
{Rosswog} S.,  {Liebend{\"o}rfer} M.,  {Thielemann} F.-K.,  {Davies} M.~B.,
  {Benz} W.,   {Piran} T.,  1999, \aap, \href
  {http://adsabs.harvard.edu/abs/1999A\%26A...341..499R} {341, 499}

\bibitem[\protect\citeauthoryear{{Rosswog}, {Korobkin}, {Arcones}, {Thielemann}
   \& {Piran}}{{Rosswog} et~al.}{2014}]{rosswog14}
{Rosswog} S.,  {Korobkin} O.,  {Arcones} A.,  {Thielemann} F.-K.,   {Piran} T.,
   2014, \mn@doi [\mnras] {10.1093/mnras/stt2502}, \href
  {http://adsabs.harvard.edu/abs/2014MNRAS.439..744R} {439, 744}

\bibitem[\protect\citeauthoryear{{Sampson}, {Zhang}  \& {Fontes}}{{Sampson}
  et~al.}{2009}]{sampson_physrep}
{Sampson} D.~H.,  {Zhang} H.~L.,   {Fontes} C.~J.,  2009, \mn@doi [\physrep]
  {10.1016/j.physrep.2009.04.002}, \href
  {http://adsabs.harvard.edu/abs/2009PhR...477..111S} {477, 111}

\bibitem[\protect\citeauthoryear{{Sobolev}}{{Sobolev}}{1960}]{sobolev60}
{Sobolev} V.~V.,  1960, {Moving envelopes of stars}.
Cambridge: Harvard University Press

\bibitem[\protect\citeauthoryear{{Symbalisty} \& {Schramm}}{{Symbalisty} \&
  {Schramm}}{1982}]{1982ApL....22..143S}
{Symbalisty} E.,  {Schramm} D.~N.,  1982, \aplett, \href
  {http://adsabs.harvard.edu/abs/1982ApL....22..143S} {22, 143}

\bibitem[\protect\citeauthoryear{{Tanaka} et~al.,}{{Tanaka}
  et~al.}{2018}]{tanaka18}
{Tanaka} M.,  et~al., 2018, \mn@doi [\apj] {10.3847/1538-4357/aaa0cb}, \href
  {http://adsabs.harvard.edu/abs/2018ApJ...852..109T} {852, 109}

\bibitem[\protect\citeauthoryear{{Tanvir} et~al.,}{{Tanvir}
  et~al.}{2017}]{tanvir17}
{Tanvir} N.~R.,  et~al., 2017, \mn@doi [\apjl] {10.3847/2041-8213/aa90b6},
  \href {http://adsabs.harvard.edu/abs/2017ApJ...848L..27T} {848, L27}

\bibitem[\protect\citeauthoryear{{Troja} et~al.,}{{Troja}
  et~al.}{2017}]{troja17}
{Troja} E.,  et~al., 2017, \mn@doi [\nat] {10.1038/nature24290}, \href
  {http://adsabs.harvard.edu/abs/2017Natur.551...71T} {551, 71}

\bibitem[\protect\citeauthoryear{{Wollaeger} \& {van Rossum}}{{Wollaeger} \&
  {van Rossum}}{2014}]{wollaeger14}
{Wollaeger} R.~T.,  {van Rossum} D.~R.,  2014, \mn@doi [\apjs]
  {10.1088/0067-0049/214/2/28}, \href
  {http://adsabs.harvard.edu/abs/2014ApJS..214...28W} {214, 28}

\bibitem[\protect\citeauthoryear{{Wollaeger} et~al.,}{{Wollaeger}
  et~al.}{2018}]{wollaeger18}
{Wollaeger} R.~T.,  et~al., 2018, \mn@doi [\mnras] {10.1093/mnras/sty1018},
  \href {http://adsabs.harvard.edu/abs/2018MNRAS.478.3298W} {478, 3298}

\makeatother
\end{thebibliography}




\appendix


\section{List of configurations used in this work}
\label{app:config_list}

Table~\ref{tab:configs} in this appendix contains a list of configurations
that were used in calculating the energy levels and oscillator strengths for the
14 lanthanide elements and sole actinide (uranium) element considered
in this work. Based on the relevant conditions of kilonova ejecta, only
the first four ion stages were calculated for each element.
The list of configurations was chosen to obtain a good representation
of the lowest lying energy levels that are necessary to: (a) obtain converged
atomic level populations via Saha-Boltzmann statistics and (b) calculate
converged opacities with respect to the number of bound-bound transitions
in the photon energy range of interest.
The choice of configurations was based on the energy-level entries in the NIST
database \citep{nist} as well as ab initio atomic structure calculations.
\begin{table*}
\centering
\caption{\rm A list of configurations, number of fine-structure levels,
and number of (electric dipole) absorption lines for the various ion stages
considered in this work. A completely filled Xe core is assumed
for the 14 lanthanide elements, while a filled Rn core is assumed for uranium.
For the first two ion stages of ytterbium, Yb {\sc i} and {\sc ii}, the
orbital angular momentum symbol $\ell$ represents the range of values
$\ell = s, p, d, f$ and $g$. }
\vspace*{0.5\baselineskip}
\begin{tabular}{lcrr}
\hline
Ion stage   &  Configurations &  \# of levels & \# of lines \\
\hline
La {\sc i} &
$5d^1 6s^2$,
$5d^2 6s^1$,
$5d^3$,
$5d^1 6s^1 6p^1$,
$4f^1 6s^2$,
$6s^2 6p^1$, & 366 & 16,163 \\
&
$5d^2 6p^1$,
$4f^1 5d^1 6s^1$,
$4f^1 6s^1 6p^1$,
$4f^1 5d^2$,
$4f^1 5d^1 6p^1$ \\

La {\sc ii} &
$5d^2$,
$5d^1 6s^1$,
$4f^1 6s^1$,
$6s^2$,
$4f^1 5d^1$, & 79 & 768 \\
&
$5d^1 6p^1$,
$6s^1 6p^1$,
$4f^1 6p^1$,
$4f^2$ \\

La {\sc iii} &
$5d^1$,
$4f^1$,
$6s^1$,
$6p^1$ & 7 & 8 \\

La {\sc iv} &
$5p^6$,
$5p^5 4f^1$,
$5p^5 6s^1$,
$5p^5 5d^1$,
$5p^5 6p^1$ & 39 & 211 \\
\hline

Ce {\sc i} &
$4f^2 6s^2$,
$4f^1 5d^1 6s^2$,
$4f^1 5d^2 6s^1$,
$4f^1 5d^1 6s^1 6p^1$,
$4f^2 5d^1 6s^1$, & 2,546 & 626,112 \\
&
$4f^2 6s^1 6p^1$,
$4f^1 5d^3$,
$4f^1 6s^2 6p^1$,
$4f^1 5d^2 6p^1$,
$4f^2 5d^2$,
$4f^2 5d^1 6p^1$ \\

Ce {\sc ii} &
$4f^2 6s^1$,
$4f^2 5d^1$,
$4f^2 6p^1$,
$4f^1 5d^2$,
$4f^1 6s^2$, & 519 & 28,887 \\
&
$4f^1 5d^1 6s^1$,
$4f^1 5d^1 6p^1$,
$4f^1 6s^1 6p^1$,
$5d^3$,
$4f^3$ \\

Ce {\sc iii} &
$4f^2 $,
$4f^1 6s^1$,
$4f^1 5d^1$,
$4f^1 6p^1$,
$5d^2$,
$5d^1 6s^1$ & 62 & 452 \\

Ce {\sc iv} &
$4f^1$,
$6s^1$,
$5d^1$,
$6p^1$ & 7 & 8 \\
\hline

Pr {\sc i} &
$4f^3 6s^2$,
$4f^2 5d^1 6s^2$,
$4f^2 5d^2 6s^1$,
$4f^3 5d^1 6s^1$, & 7,362 & 4,750,354 \\
&
$4f^3 5d^2$,
$4f^2 5d^1 6s^1 6p^1$,
$4f^3 5d^1 6p^1$,
$4f^3 6s^1 6p^1$ \\

Pr {\sc ii} &
$4f^3 6s^1$,
$4f^3 5d^1 $,
$4f^3 6p^1 $,
$4f^2 5d^2 $, & 2,145 & 412,027 \\
&
$4f^2 5d^1 6s^1$,
$4f^2 5d^1 6p^1$,
$4f^2 6s^1 6p^1$ \\

Pr {\sc iii} &
$4f^3$,
$4f^2 6s^1$,
$4f^2 5d^1$, & 361 & 12,394 \\
&
$4f^2 6p^1$,
$4f^1 5d^2$,
$4f^1 5d^1 6s^1$ \\

Pr {\sc iv} &
$4f^2$,
$4f^1 6s^1$,
$4f^1 5d^1$,
$4f^1 6p^1$ & 49 & 296 \\
\hline

Nd {\sc i} &
$4f^4 6s^2$,
$4f^3 5d^1 6s^2$,
$4f^4 5d^1 6s^1$,
$4f^4 5d^2$, & 18,104 & 25,224,451 \\
&
$4f^3 5d^1 6s^1 6p^1$,
$4f^4 5d^1 6p^1$,
$4f^4 6s^1 6p^1$ \\

Nd {\sc ii} &
$4f^4 6s^1$,
$4f^4 5d^1$,
$4f^4 6p^1$,
$4f^3 5d^2$, & 6,888 & 3,958,977 \\
&
$4f^3 5d^1 6s^1$,
$4f^3 5d^1 6p^1$,
$4f^3 6s^1 6p^1$ \\

Nd {\sc iii} &
$4f^4$,
$4f^3 6s^1$,
$4f^3 5d^1$,
$4f^3 6p^1$, & 1,650 & 233,822 \\
&
$4f^2 5d^2$,
$4f^2 5d^1 6s^1$,
$4f^1 5d^2 6s^1$ \\

Nd {\sc iv} &
$4f^3 $,
$4f^2 6s^1$,
$4f^2 5d^1$,
$4f^2 6p^1$ & 241 & 5,784 \\
\hline

Pm {\sc i} &
$4f^5 6s^2$,
$4f^4 5d^1 6s^2$,
$4f^5 5d^1 6s^1$,
$4f^5 5d^2$, & 37,504 & 102,137,397 \\
&
$4f^4 5d^1 6s^1 6p^1$,
$4f^5 5d^1 6p^1$,
$4f^5 6s^1 6p^1$ \\

Pm {\sc ii} &
$4f^5 6s^1$,
$4f^5 5d^1$,
$4f^5 6p^1$,
$4f^4 5d^2$, & 16,595 & 21,306,571 \\
&
$4f^4 5d^1 6s^1$,
$4f^4 5d^1 6p^1$,
$4f^4 6s^1 6p^1$ \\

Pm {\sc iii} &
$4f^5$,
$4f^4 6s^1$,
$4f^4 5d^1$,
$4f^4 6p^1$, & 5,274 & 2,262,145 \\
&
$4f^3 5d^2$,
$4f^3 5d^1 6s^1$,
$4f^2 5d^2 6s^1$ \\

Pm {\sc iv} &
$4f^4$,
$4f^3 6s^1$,
$4f^3 5d^1$,
$4f^3 6p^1$ & 817 & 57,765 \\
\hline

Sm {\sc i} &
$4f^6 6s^2$,
$4f^5 5d^1 6s^2$,
$4f^6 5d^1 6s^1$,
$4f^6 5d^2$, & 60,806 & 249,301,825 \\
&
$4f^5 5d^1 6s^1 6p^1$,
$4f^6 5d^1 6p^1$,
$4f^6 6s^1 6p^1$\\

Sm {\sc ii} &
$4f^6 6s^1$,
$4f^6 5d^1$,
$4f^6 6p^1$,
$4f^5 5d^2$, & 29,970 & 67,743,385 \\
&
$4f^5 5d^1 6s^1$,
$4f^5 5d^1 6p^1$,
$4f^5 6s^1 6p^1$ \\

Sm {\sc iii} &
$4f^6$,
$4f^5 6s^1$,
$4f^5 5d^1$,
$4f^5 6p^1$, & 13,170 & 13,318,114 \\
&
$4f^4 5d^2$,
$4f^4 5d^1 6s^1$,
$4f^3 5d^2 6s^1$ \\

Sm {\sc iv} &
$4f^5$,
$4f^4 6s^1$,
$4f^4 5d^1$,
$4f^4 6p^1$ & 1,994 & 320,633 \\
\hline

Eu {\sc i} &
$4f^7 6s^2$,
$4f^6 5d^1 6s^2$,
$4f^7 5d^1 6s^1$,
$4f^7 5d^2$,
$4f^6 5d^2 6s^1$,
$4f^6 5d^1 6s^1 6p^1$, & 110,887 & 902,515,995 \\
&
$4f^7 5d^1 6p^1$,
$4f^7 6s^1 6p^1$,
$4f^7 6s^1 6d^1$,
$4f^7 6s^1 7s^1$,
$4f^7 6s^1 7p^1$ \\

Eu {\sc ii} &
$4f^7 6s^1$,
$4f^7 5d^1$,
$4f^7 6p^1$,
$4f^6 5d^2$,
$4f^6 5d^1 6s^1$, & 46,213 & 152,795,763 \\
&
$4f^6 5d^1 6p^1$,
$4f^6 6s^1 6p^1$,
$4f^7 6d^1$,
$4f^7 7s^1$ \\

Eu {\sc iii} &
$4f^7$,
$4f^6 6s^1$,
$4f^6 5d^1$, & 17,058 & 13,844,004 \\
&
$4f^6 6p^1$,
$4f^5 5d^2$,
$4f^5 5d^1 6s^1$ \\

Eu {\sc iv} &
$4f^6$,
$4f^5 6s^1$,
$4f^5 5d^1$,
$4f^5 6p^1$ & 3,737 & 1,045,697 \\
\hline

Gd {\sc i} &
$4f^8 6s^2$,
$4f^7 5d^1 6s^2$,
$4f^7 5d^2 6s^1$,
$4f^7 5d^2 6p^1$,
$4f^7 6s^2 6p^1$,
$4f^7 6s^1 6p^2$, & 228,048 & 3,583,266,975 \\
&
$4f^7 5d^3$,
$4f^8 5d^1 6s^1$,
$4f^8 5d^2$,
$4f^7 5d^1 6s^1 6p^1$,
$4f^8 5d^1 6p^1$,
$4f^8 6s^1 6p^1$ \\

Gd {\sc ii} &
$4f^8 6s^1$,
$4f^8 5d^1$,
$4f^8 6p^1$,
$4f^7 5d^2$,
$4f^7 6s^2$, & 46,733 & 160,595,610 \\
&
$4f^7 5d^1 6s^1$,
$4f^7 5d^1 6p^1$,
$4f^7 6s^1 6p^1$ \\

Gd {\sc iii} &
$4f^8$,
$4f^7 6s^1$,
$4f^7 5d^1$,
$4f^7 6p^1$, & 39,324 & 113,346,746 \\
&
$4f^6 5d^2$,
$4f^6 5d^1 6s^1$,
$4f^5 5d^2 6s^1$ \\

Gd {\sc iv} &
$4f^7$,
$4f^6 6s^1$,
$4f^6 5d^1$,
$4f^6 6p^1$ & 5,323 & 2,073,701 \\
\hline
\end{tabular}
\label{tab:configs}
\end{table*}
\addtocounter{table}{-1}
\begin{table*}
\centering
\caption{\rm Continued\ldots}
\vspace*{0.5\baselineskip}
\begin{tabular}{lcrr}
\hline
Ion stage   &  Configurations &  \# of levels & \# of lines \\
\hline
Tb {\sc i} &
$4f^9 6s^2$,
$4f^8 5d^1 6s^2$,
$4f^8 5d^2 6s^1$,
$4f^8 6s^2 6p^1$,
$4f^9 5d^1 6s^1$, & 84,779 & 530,258,427 \\
&
$4f^9 5d^2$,
$4f^8 5d^1 6s^1 6p^1$,
$4f^9 5d^1 6p^1$,
$4f^9 6s^1 6p^1$ \\

Tb {\sc ii} &
$4f^9 6s^1$,
$4f^9 5d^1$,
$4f^9 6p^1$,
$4f^8 6s^2$,
$4f^8 5d^2$, & 40,502 & 120,186,078 \\
&
$4f^8 5d^1 6s^1$,
$4f^8 5d^1 6p^1$,
$4f^8 6s^1 6p^1$ \\

Tb {\sc iii} &
$4f^9$,
$4f^8 6s^1$,
$4f^8 5d^1$,
$4f^8 6p^1$, & 48,048 & 166,697,126 \\
&
$4f^7 5d^2$,
$4f^7 5d^1 6s^1$,
$4f^6 5d^2 6s^1$ \\

Tb {\sc iv} &
$4f^8$,
$4f^7 6s^1$,
$4f^7 5d^1$,
$4f^7 6p^1$ & 5,983 & 2,545,968 \\
\hline

Dy {\sc i} &
$4f^{10} 6s^2$,
$4f^9 5d^1 6s^2$,
$4f^9 5d^1 6s^1 6p^1$,
$4f^{10} 5d^1 6s^1$,
$4f^{10} 6s^1 6p^1$,
$4f^{10} 6s^1 6d^1$, & 55,116 & 229,181,735 \\
&
$4f^{10} 6s^1 7s^1$,
$4f^{10} 6s^1 7p^1$,
$4f^9 6s^2 6p^1$,
$4f^9 5d^2 6s^1$,
$4f^{10} 6p^2$,
$4f^{10} 5d^1 6p^1$ \\

Dy {\sc ii} &
$4f^{10} 6s^1$,
$4f^{10} 5d^1$,
$4f^{10} 6p^1$,
$4f^9 5d^2$,
$4f^9 6s^2$, & 26,968 & 55,381,943 \\
&
$4f^9 5d^1 6s^1$,
$4f^9 5d^1 6p^1$,
$4f^9 6s^1 6p^1$ \\

Dy {\sc iii} &
$4f^{10}$,
$4f^9 6s^1$,
$4f^9 5d^1$, & 20,834 & 12,685,641 \\
&
$4f^9 6p^1$,
$4f^8 5d^2$,
$4f^8 5d^1 6s^1$ \\

Dy {\sc iv} &
$4f^9$,
$4f^8 6s^1$,
$4f^8 5d^1$,
$4f^8 6p^1$ & 5,194 & 1,943,961 \\
\hline

Ho {\sc i} &
$4f^{11} 6s^2$,
$4f^{10} 5d^1 6s^2$,
$4f^{10} 5d^1 6s^1 6p^1$,
$4f^{11} 5d^1 6s^1$,
$4f^{11} 6s^1 6p^1$,
$4f^{11} 6s^1 6d^1$, & 26,759 & 57,149,392 \\
&
$4f^{11} 6s^1 7s^1$,
$4f^{11} 6s^1 7p^1$,
$4f^{10} 6s^2 6p^1$,
$4f^{10} 5d^2 6s^1$,
$4f^{11} 6p^2$,
$4f^{11} 5d^1 6p^1$ \\

Ho {\sc ii} &
$4f^{11} 6s^1$,
$4f^{11} 5d^1$,
$4f^{11} 6p^1$,
$4f^{10} 5d^2$,
$4f^{10} 6s^2$, & 13,970 & 15,323,035 \\
&
$4f^{10} 5d^1 6s^1$,
$4f^{10} 5d^1 6p^1$,
$4f^{10} 6s^1 6p^1$ \\
 
Ho {\sc iii} &
$4f^{11}$,
$4f^{10} 6s^1$,
$4f^{10} 5d^1$,
$4f^{10} 6p^1$ & 1,837 & 259,812 \\

Ho {\sc iv} &
$4f^{10}$,
$4f^9 6s^1$,
$4f^9 5d^1$,
$4f^9 6p^1$ & 3,549 & 915,339 \\
\hline

Er {\sc i} &
$4f^{12} 6s^2$,
$4f^{11} 5d^1 6s^2$,
$4f^{11} 5d^1 6s^1 6p^1$,
$4f^{12} 5d^1 6s^1$,
$4f^{12} 6s^1 6p^1$,
$4f^{12} 6s^1 6d^1$, & 9,904 & 8,223,793 \\
&
$4f^{12} 6s^1 7s^1$,
$4f^{12} 6s^1 7p^1$,
$4f^{11} 6s^2 6p^1$,
$4f^{11} 5d^2 6s^1$,
$4f^{12} 6p^2$,
$4f^{12} 5d^1 6p^1$ \\

Er {\sc ii} &
$4f^{12} 6s^1$,
$4f^{12} 5d^1$,
$4f^{12} 6p^1$,
$4f^{11} 5d^2$,
$4f^{11} 6s^2$, & 5,333 & 2,432,666 \\
&
$4f^{11} 5d^1 6s^1$,
$4f^{11} 5d^1 6p^1$,
$4f^{11} 6s^1 6p^1$ \\

Er {\sc iii} &
$4f^{12}$,
$4f^{11} 6s^1$,
$4f^{11} 5d^1$,
$4f^{11} 6p^1$ & 723 & 42,671 \\

Er {\sc iv} &
$4f^{11}$,
$4f^{10} 6s^1$,
$4f^{10} 5d^1$,
$4f^{10} 6p^1$ & 1,837 & 259,812 \\
\hline

Tm {\sc i} &
$4f^{13} 6s^2$,
$4f^{12} 5d^1 6s^2$,
$4f^{12} 5d^1 6s^1 6p^1$,
$4f^{13} 5d^1 6s^1$,
$4f^{13} 6s^1 6p^1$, & 1,684 & 155,754 \\
&
$4f^{13} 6s^1 6d^1$,
$4f^{13} 6s^1 7s^1$,
$4f^{13} 6s^1 7p^1$,
$4f^{12} 6s^2 6p^1$,
$4f^{13} 6p^2$,
$4f^{13} 5d^1 6p^1$ \\

Tm {\sc ii} &
$4f^{13} 6s^1$,
$4f^{13} 5d^1$,
$4f^{13} 6p^1$,
$4f^{12} 5d^2$,
$4f^{12} 6s^2$, & 1,484 & 205,258 \\
&
$4f^{12} 5d^1 6s^1$,
$4f^{12} 5d^1 6p^1$,
$4f^{12} 6s^1 6p^1$ \\

Tm {\sc iii} &
$4f^{13}$,
$4f^{12} 6s^1$,
$4f^{12} 5d^1$,
$4f^{12} 6p^1$ & 202 & 3,797 \\

Tm {\sc iv} &
$4f^{12}$,
$4f^{11} 6s^1$,
$4f^{11} 5d^1$,
$4f^{11} 6p^1$ & 723 & 42,671 \\
\hline

Yb {\sc i} &
$4f^{14} 6s^1 6\ell^1$,
$4f^{14} 6s^1 7\ell^1$,
$4f^{14} 6s^1 8\ell^1$,
$4f^{14} 6s^1 9\ell^1$,
$4f^{14} 6s^1 10\ell^1$,
$4f^{14} 6s^1 11\ell^1$, & 1,117 & 111,828 \\
&
$4f^{14} 5d^1 6s^1$,
$4f^{14} 5d^1 6p^1$,
$4f^{14} 6s^1 5f^1$,
$4f^{14} 6s^1 5g^1$,
$4f^{14} 6p^2$,
$4f^{14} 5d^2$,
$4f^{13} 5d^1 6s^2$, \\
&
$4f^{13} 5d^2 6s^1$,
$4f^{13} 5d^1 6s^1 6p^1$,
$4f^{13} 6s^2 6p^1$,
$4f^{13} 6s^1 6p^2$,
$4f^{13} 6s^2 6d^1$,
$4f^{13} 5d^2 6p^1$ \\

Yb {\sc ii} &
$4f^{14} 6\ell^1$,
$4f^{14} 7\ell^1$,
$4f^{14} 8\ell^1$,
$4f^{14} 5d^1$,
$4f^{14} 5f^1$,
$4f^{14} 5g^1$, & 292 & 9,673 \\
&
$4f^{13} 5d^2$,
$4f^{13} 6s^2$,
$4f^{13} 5d^1 6s^1$,
$4f^{13} 5d^1 6p^1$,
$4f^{13} 6s^1 6p^1$ \\

Yb {\sc iii} &
$4f^{14}$,
$4f^{13} 6s^1$,
$4f^{13} 5d^1$,
$4f^{13} 6p^1$,
$4f^{13} 6d^1$,
$4f^{13} 6f^1$, & 171 & 3,192 \\
&
$4f^{13} 5f^1$,
$4f^{13} 7s^1$,
$4f^{13} 7p^1$,
$4f^{13} 7d^1$,
$4f^{13} 7f^1$ \\

Yb {\sc iv} &
$4f^{13}$,
$4f^{12} 6s^1$,
$4f^{12} 5d^1$,
$4f^{12} 6p^1$ & 202 & 3,797 \\
\hline

U {\sc i} &
$5f^4 7s^2$,
$5f^3 6d^1 7s^2$,
$5f^4 6d^1 7s^1$, & 16,882 & 20,948,831 \\
&
$5f^4 6d^2$,
$5f^3 6d^1 7s^1 7p^1$,
$5f^4 6d^1 7p^1$\\

U {\sc ii} &
$5f^3 7s^2$,
$5f^4 7s^1$,
$5f^4 6d^1$,
$5f^4 7p^1$, & 6,929 & 4,016,742 \\
&
$5f^3 6d^2$,
$5f^3 6d^1 7s^1$,
$5f^3 6d^1 7p^1$,
$5f^3 7s^1 7p^1$\\

U {\sc iii} &
$5f^4$,
$5f^3 7s^1$,
$5f^3 6d^1$,
$5f^3 7p^1$, & 1,650 & 233,822 \\
&
$5f^2 6d^2$,
$5f^2 6d^1 7s^1$,
$5f^1 6d^2 7s^1$\\

U {\sc iv} &
$5f^3$,
$5f^2 7s^1$,
$5f^2 6d^1$,
$5f^2 7p^1$ & 241 & 5,784 \\
\hline
\end{tabular}
\end{table*}

\section{Sobolev implementation in the {\tt SuperNu} code}
\label{app:dir_sob}

In this appendix, we discuss the implementation of a direct
Sobolev treatment in the {\tt SuperNu} code.
To do so, we write down the equations and describe details
of the algorithm.

The Sobolev method needs to be made compatible with
Implicit Monte Carlo (IMC, \citet{fleck71}).
Consequently, it is instructive to revisit the basic equations
implemented in {\tt SuperNu}, which like the {\tt SEDONA} code, assumes
the homologous approximation \citep{kasen06_2},
\begin{equation}
  \label{eq1}
  \vec{v} = \frac{\vec{r}}{t} \;\;,
\end{equation}
where $\vec{v}$ is velocity and $t$ is time.
Incorporating \eqnref{eq1} into the $O(v/c)$ transport
equation gives (see, for instance, \citet{castor04})
\begin{equation}
  \label{eq2}
  \frac{1}{c}\frac{D\psi}{D t}
  + \hat{\Omega}\cdot\nabla\psi
  + \frac{1}{ct}\lambda
  \frac{\partial\psi}{\partial\lambda}
  + \frac{3}{ct}\psi
  = j - \rho\kappa\psi \;,
\end{equation}
where $\hat{\Omega}$ is the direction unit vector, $\lambda$ is
wavelength, $\psi$ is angular intensity, $j$ is emissivity, and
$\kappa$ is opacity.
All quantities, except the spatial coordinate and time, are
evaluated in the comoving frame.
The corresponding IMC formulation requires an equation for the
internal energy (or temperature of the matter).
Assuming LTE, after performing a linear temporal expansion in $T^4$ of
the Lagrangian internal energy equation \citep{fleck71},
the IMC equations in {\tt SuperNu} are
\begin{subequations}
  \label{eq3}
  \begin{gather}
    \frac{1}{c}\frac{D\psi}{D t}
    + \hat{\Omega}\cdot\nabla\psi
    + \frac{1}{ct_n}\lambda
    \frac{\partial\psi}{\partial\lambda}
    + \frac{3}{ct_n}\psi =
    \nonumber\\
    \qquad\qquad
    f_n\rho_n\kappa_{n}B_n
    + \frac{b_n\kappa_n}{4\pi\kappa_{P,n}}(1-f_n)
    \nonumber\\
    \qquad\qquad
    \times
    \int_{4\pi}d\Omega'\int_0^{\infty}d\lambda'\rho_n\kappa_n'\psi'
    - f_n\rho_n\kappa_n\psi \;, \label{eq3a}\\
    \rho_nc_{v,n}\frac{DT}{Dt}
    = \rho_nf_n\int_{4\pi}d\Omega\int_0^{\infty}d\lambda
    (\kappa_n\psi - \kappa_nB_n) \;, \label{eq3b} \\
    f_n = \frac{1}{1 + c\Delta t_n\kappa_{P,n}4acT_n^3/c_{v,n}} \;,
    \label{eq3c}
  \end{gather}
\end{subequations}
where quantities subscripted with $n$ are evaluated at the beginning
of time step $n$, $f_n$ is the Fleck factor, $B$ is the Planck function
$b$ is the normalized Planck function, $\kappa_P$ is the Planck opacity,
and $c_v$ is heat capacity.
For simplicity, we have neglected contributions from $r$-process decay and
Thomson scattering (though generally speaking,
these are included in the simulations).
It is worth noting that the Fleck factor introduces an ``effective scattering''
term into the transport equation (the second term on the right-hand side of
\eqnref{eq3a}).
Monte Carlo particles undergoing an effective scattering event are
thermally redistributed in their wavelengths (the $b_n\kappa_n/\kappa_{P,n}$
coefficient is the redistribution kernel).

In {\tt SuperNu}, the opacity is typically discretized into groups in
wavelength; these groups are defined in the comoving frame (local to each
fluid parcel).
However, the wavelength for Monte Carlo particles is left continuous.
Consequently, the first two equations in \eqnref{eq3} become
\begin{subequations}
  \label{eq4}
  \begin{gather}
    \frac{1}{c}\frac{D\psi}{D t}
    + \hat{\Omega}\cdot\nabla\psi
    + \frac{1}{ct_n}\lambda
    \frac{\partial\psi}{\partial\lambda}
    + \frac{3}{ct_n}\psi =
    \nonumber\\
    \qquad\qquad
    f_n\rho_n\kappa_{g,n}B_{g,n}
    + \frac{b_{g,n}\kappa_{g,n}}{4\pi\kappa_{P,n}}(1-f_n)
    \nonumber\\
    \qquad\qquad
    \times
    \int_{4\pi}d\Omega'\sum_{g'}\rho_n\kappa_{g',n}'\psi'
    - f_n\rho_n\kappa_{g,n}\psi \;, \label{eq4a}\\
    \rho_nc_{v,n}\frac{DT}{Dt}
    = \rho_nf_n\int_{4\pi}d\Omega \sum_g
    \left(\int_gd\lambda\kappa_{g,n}\psi
    - \kappa_{g,n}B_{g,n}\Delta\lambda_g\right) \;, \label{eq4b}
  \end{gather}
\end{subequations}
where subscript $g$ denotes a group-averaged quantity,
$b_g = \int_gd\lambda \;b / \Delta\lambda_g$, and the Planck opacity is
evaluated via a sum over groups, $\kappa_P = \sum_g\kappa_g b_g$.
Since the particle wavelength is left continuous, the particles Doppler
shift continuously across groups as they stream through space.
This particle behavior corresponds to having left the third term on the
left-hand side of \eqnref{eq4a} to remain continuous in wavelength.

Alternatively, we may keep some of the opacity continuous in wavelength.
To obtain the Sobolev formulation of the IMC equations, we simply substitute
\begin{equation}
  \label{eq5}
  \kappa_n = \tilde{\kappa}_{g,n} + \kappa_n^{b-b}
\end{equation}
for $\kappa_{g,n}$ in Equations~(\ref{eq4}),
where $\tilde{\kappa}_{g,n}$ are all
continuum contributions (discretized into groups) and $\kappa_n^{b-b}$
is a monochromatic superposition of Dirac delta distributions corresponding
to the line contributions.
For our Sobolev test problem, we neglect $\tilde{\kappa}_{g,n}$.

For fully discrete opacity, we also discretize the bound-bound contribution,
$\kappa_n^{b-b}\rightarrow\kappa_{g,n}^{b-b}$.
Using the formulae from the paper,
\begin{equation}
  \label{eq6}
  \kappa_{g,n}^{b-b} =
  \frac{1}{\rho_nct_n}\frac{1}{\Delta\lambda_g}\sum_{i\in\Delta\lambda_g}
  \begin{cases}
    \lambda_i\tau_i, \text{for binned} \;,\\
    \lambda_i(1-e^{-\tau_i}), \text{for expansion} \;,
  \end{cases}
\end{equation}
where $i$ is a line index and $\tau_i$ is the corresponding Sobolev optical
depth.
It is straightforward to see that the expansion opacity formula reduces to the
binned opacity formula for $\tau_i\ll 1$.
It is also straightforward to see that the line-binned formula
is in fact a direct
integral-average of the monochromatic opacity over wavelength under the
assumption that the line profiles are Dirac delta distributions:
\begin{eqnarray}
  \label{eq7}
  \kappa_{g,n}^{b-b} &=&
  \frac{1}{\rho_nct_n}\frac{1}{\Delta\lambda_g}
  \sum_{i\in\Delta\lambda_g}\lambda_i\tau_i
    \nonumber\\
  &=&
  \frac{1}{\rho_nct_n}\frac{1}{\Delta\lambda_g}
  \sum_{i\in\Delta\lambda_g}t_n\frac{\pi e^2}{m_ec}N_i|f_i|\lambda_i^2
    \nonumber\\
  &=&
  \frac{1}{\rho_n}\frac{1}{\Delta\lambda_g}
  \sum_{i\in\Delta\lambda_g}\frac{\pi e^2}{m_ec^2}N_i|f_i|\lambda_i^2
  \;,
\end{eqnarray}
where $N_i$ is the number density of the initial level for line $i$ and $f_i$ is
the oscillator strength.

\subsection{Monte Carlo Algorithm with Discrete Opacity}

The algorithm for the system of equations with fully discretized opacity
(Equations~(\ref{eq4})) is as follows:
\begin{enumerate}
\item Store particles from time step $n-1$, sample source particles in
  time step $n$:
  \begin{enumerate}
  \item integrate first term on right-hand side
    of \eqnref{eq4a} in $\Omega$,
    sum over groups, and integrate over the volume of a spatial cell, then
    partition the energy across the prescribed number of source particles
    for the cell,
  \item sample each particle's position in the cell, in time step $n$,
    and direction ($\hat{\Omega}$),
  \item sample each particle's wavelength ($\lambda$) by first sampling
    the group from a cumulative distribution function (CDF) generated by
    $b_{g,n}\kappa_{g,n}/\kappa_{P,n}$, then uniformly within the sampled group.
  \end{enumerate}
\item Transport the updated set of particles.
  For each particle:
  \begin{enumerate}
  \item Calculate the set of velocity distances \citep{kasen06_2}
    to stream to each possible next particle event:
    \begin{subequations}
      \label{eq8}
      \begin{gather}
        d_{\rm es} = -\frac{\ln(\xi)}{t_n(1-f_n)\rho_n\kappa_{g,n}} \;, \\
        d_{\rm red} = c(\lambda_{g+1/2}-\lambda)/\lambda \;, \\
        d_b = |\vec{r}_b(\hat{\Omega}) - \vec{r}| \;, \\
        d_{\rm end} = c(t_{n+1}-t)/t_n \;,
      \end{gather}
    \end{subequations}
    where $d_{\rm es}$, $d_{\rm red}$, $d_b$, and $d_{\rm end}$ are distances to
    effective scattering, redshift out of the group edge, the cell boundary
    along $\hat{\Omega}$, and to the end of time step $n$.
    The quantity $\xi\in[0,1]$ is a uniformly sampled random number.
  \item Select the event from the minimum distance $d$, and update the
    particle's properties via streaming:
    \begin{subequations}
      \label{eq9}
      \begin{gather}
        t' = t + \frac{t_nd}{c} \;, \\
        \vec{r}' = \vec{r} + d\hat{\Omega} \;, \\
        E' = Ee^{-(1-f_n)\rho_n\kappa_{g,n}t_nd} \;, \\
        \lambda' = \lambda\left(1 + \frac{d}{c}\right) \;,
      \end{gather}
    \end{subequations}
    where, for simplicity, the update to the spatial position has been
    written in Cartesian coordinates.
  \item Add $E(1-e^{-(1-f_n)\rho_n\kappa_{g,n}t_nd})$ to the tally of absorbed
    energy in the time step.
  \item If the particle effectively scatters, sample $\hat{\Omega}$ isotropically
    and resample $\lambda$ thermally from $b_{g,n}\kappa_{g,n}/\kappa_{P,n}$.
  \item If the particle redshifts out of the group, set $g$ to $g+1$ for the
    next round of distance calculations.
  \item If the particle crosses the cell boundary, use the adjacent cell's
    properties for the next round of distance calculations.
    If the edge of the simulation grid is reached, tally the particle as
    escaping flux for the observables (light curves and spectra).
  \item If the particle reaches the end of the time step, stop transporting
    and store it for $n+1$.
  \end{enumerate}
\item Account for redshift on particle energy weights using operator-split
  equation \citep{abdikamalov12},
  \begin{equation}
    \label{eq10}
    \frac{\partial\psi}{\partial t} + \frac{\psi}{t_n} = 0 \;,
  \end{equation}
  which has solution $E''=E'e^{-\Delta t_n/t_n}$ for each particle energy
  weight.
\item With the energy absorption tallied during the transport of the particles,
  update the internal energy of the ejecta with \eqnref{eq4b}
  (the absorbed
  energy tally is represented by the term with $\psi$ on the right-hand side).
\end{enumerate}

The above algorithm applies to both expansion and line-binned opacity
treatements.
The only difference between the approaches to treating line-binned and expansion
opacity is in how the bound-bound contribution of $\kappa_{n,g}$ is calculated
(in \eqnref{eq6}).
Again, \eqnref{eq6} shows that the expansion and
line-binned opacity methods approach one another when $\tau_i\ll 1$.

\subsection{Monte Carlo Algorithm with Sobolev Line Treatment}

The Sobolev algorithm is nearly the same, but with modifications to steps
1 and 2 resulting from incorporating \eqnref{eq5}
into Equations~(\ref{eq4}).
This algorithm is intended to be as close to that of \citet{kasen06_2}
as possible, within the context of IMC and {\tt SuperNu}.
The Sobolev algorithm is as follows:
\begin{enumerate}
\item Store particles from time step $n-1$, sample source particles in
  time step $n$:
  \begin{enumerate}
  \item integrate first term on the right-hand side
    of \eqnref{eq4a} in $\Omega$,
    sum over groups, and integrate over the volume of a spatial cell, then
    partition the energy across the prescribed number of source particles
    for the cell,
  \item sample each particle's position in the cell, in time step $n$,
    and direction ($\hat{\Omega}$),
  \item sample each particle's wavelength ($\lambda$) by first sampling
    the group from a CDF generated by
    $b_{g,n}\kappa_{g,n}/\kappa_{P,n}$, then
    \begin{enumerate}
    \item uniformly within the sampled group if
      \begin{equation}
        \label{eq11}
        \xi < \frac{\tilde{\kappa}_{g,n}}{\kappa_{g,n}} \;\;,
      \end{equation}
      where $\xi\in[0,1]$ is a uniformly sampled random number, or
    \item at a line center $\lambda_i$, using a CDF
      constructed from $1-e^{-\tau_i}$ as the probability for line $i$.
    \end{enumerate}
  \end{enumerate}
  The expansion opacity formula (see \eqnref{eq6}) is used for
  $\kappa_{g,n}^{b-b} = \kappa_{g,n} - \tilde{\kappa}_{g,n}$.
\item For each particle,
  \begin{enumerate}
  \item Calculate the set of velocity distances \citep{kasen06_2}
    to stream to each possible next particle event:
    \begin{subequations}
      \label{eq12}
      \begin{gather}
        d_{\rm es} = -\frac{\ln(\xi)}{t_n(1-f_n)\rho_n\tilde{\kappa}_{g,n}} \;, \\
        d_{\rm red} = c(\lambda_{g+1/2}-\lambda)/\lambda \;, \\
        d_b = |\vec{r}_b(\hat{\Omega}) - \vec{r}| \;, \\
        d_{\rm Sob} = c(\lambda_i-\lambda)/\lambda \;, \\
        d_{\rm end} = c(t_{n+1}-t)/t_n \;,
      \end{gather}
    \end{subequations}
    where $d_{\rm es}$, $d_{\rm red}$, $d_b$, $d_{\rm Sob}$, and $d_{\rm end}$ are
    distances
    to effective scattering due to continuum opacity, redshift out of the group edge,
    the cell boundary along $\hat{\Omega}$, to resonance with line $i$, and to the
    end of time step $n$.
    The quantity $\xi\in[0,1]$ is a uniformly sampled random number.
  \item Select the event from the minimum distance $d$, and update the particle's
    properties via streaming:
    \begin{subequations}
      \label{eq13}
      \begin{gather}
        t' = t + \frac{t_nd}{c} \;, \\
        \vec{r}' = \vec{r} + d\hat{\Omega} \;, \\
        E' = Ee^{-(1-f_n)\rho_n\tilde{\kappa}_{g,n}t_nd} \;, \\
        \lambda' = \lambda\left(1 + \frac{d}{c}\right) \;,
      \end{gather}
    \end{subequations}
    where, for simplicity, the update to the spatial position has been
    written in Cartesian coordinates.
  \item Add $E(1-e^{-(1-f_n)\rho_n\tilde{\kappa}_{g,n}t_nd})$ to the tally of
    absorbed energy in the time step.
  \item If the particle effectively scatters, sample $\hat{\Omega}$
    isotropically
    and resample $\lambda$ thermally from $b_{g,n}\kappa_{g,n}/\kappa_{P,n}$.
  \item If the particle redshifts out of the group, set $g$ to $g+1$ for the
    next round of distance calculations.
  \item If the particle crosses the cell boundary, use the adjacent cell's
    properties for the next round of distance calculations.
    If the edge of the simulation grid is reached, tally the particle as
    escaping flux for the observables (light curves and spectra).
  \item If the particle comes into resonance with line $i$, sample if interaction
    with the line occurs,
    \begin{equation}
      \label{eq14}
      \xi < 1 - e^{-(1-f_n)\tau_i} \;,
    \end{equation}
    where $\xi\in[0,1]$ is again a uniformly sampled random number.
    Note that we have introduced an ``effective line scattering'', to be
    consistent with the IMC formulation.
    \begin{enumerate}
    \item If an interaction occurs, sample $\hat{\Omega}$ isotropically
      and resample $\lambda$ thermally from $b_{g,n}\kappa_{g,n}/\kappa_{P,n}$,
      where the line contribution to the discrete opacity is calculated with
      the expansion-opacity formula (following \citet{kasen06_2}).
      Within the span of the newly sampled group, $g'$, sample whether the emission
      is from the continuum,
      \begin{equation}
        \label{eq15}
        \xi' < \frac{\tilde{\kappa}_{g',n}}{\kappa_{g',n}} \;\;,
      \end{equation}
      where $\xi'$ is another uniform random variable.
      If so, uniformly sample $\lambda$ in $g'$. if not, sample $\lambda$ at one
      of the line centers in $g'$ with a CDF
      constructed from $1-e^{-\tau_{i'}}$ as the probability for each new line $i'$.
    \item If $\xi \geq 1 - e^{-(1-f_n)\tau_i}$, lower the particle energy,
      \begin{equation}
        \label{eq16}
        E' = Ee^{-f_n\tau_i} \;\;,
      \end{equation}
      and add $E(1-e^{-f_n\tau_i})$ to the absorption tally, but keep the particle's
      direction $\hat{\Omega}$ and wavelength $\lambda = \lambda_i$ unchanged.
    \end{enumerate}
  \item If the particle reaches the end of the time step, stop transporting
    and store it for $n+1$.
  \end{enumerate}
\item Account for redshift on particle energy weights using operator-split
  equation \citep{abdikamalov12},
  \begin{equation}
    \label{eq17}
    \frac{\partial\psi}{\partial t} + \frac{\psi}{t_n} = 0 \;\;,
  \end{equation}
  which has solution $E''=E'e^{-\Delta t_n/t_n}$ for each particle energy weight.
\item With the energy absorption tallied during the transport of the particles,
  update the internal energy of the ejecta with \eqnref{eq4b} with
  \eqnref{eq5} substituted for $\kappa_{g,n}$ in the first term
  on the right-hand side (this term is solved for by the tally of absorbed
  energy).
\end{enumerate}

\subsection{Motivation for Sobolev Emission Method}

Assuming $\psi$ represents the intensity of photons that have streamed to line
$i$, the absorption rate in the Sobolev formulation is
\begin{equation}
  \label{eq18}
  \mathcal{A}_i = \psi(1-e^{-\tau_i}) \;,
\end{equation}
which is in units of erg/s/cm$^2$/Hz/sr (as is $\psi$).
In a Monte Carlo framework,
\eqnref{eq18} corresponds to the sampling procedure in
\eqnref{eq14} (with Equations~(\ref{eq14}) and (\ref{eq16}),
we have employed a variance reduction technique, implicit capture,
which preserves the \eqnref{eq18} as the expectation value).
To obtain an absorption rate density, \eqnref{eq18}
can be multiplied by a spatial Dirac delta distribution over the particle
path length,
\begin{equation}
  \label{eq20}
  \mathbb{A}_i = \mathcal{A}_i\delta(x - x_i) =
               \psi(1-e^{-\tau_i})\delta(x - x_i) \;,
\end{equation}
where $x$ is the photon path length and $x_i$ is the path length at which the
photon may interact with line $i$.
\eqnref{eq20} is in units of erg/s/cm$^3$/Hz/sr,
consistent with the emissivity, $j$, in \eqnref{eq2}.
Using the identity,
\begin{equation}
  \label{eq21}
  \delta(x-x_i)dx = \delta(\lambda-\lambda_i)d\lambda \;\;,
\end{equation}
where $dx$ and $d\lambda$ are infinitesimal path length and wavelength, and using
\begin{equation}
  \label{eq22}
  \frac{d\lambda}{dx} = \frac{\lambda}{ct} \;,
\end{equation}
implied from \eqnref{eq2}, the wavelength can be substituted for path length in
the absorption rate density for line $i$,
\begin{equation}
  \label{eq23}
  \mathbb{A}_i = \psi(1-e^{-\tau_i})\delta(\lambda - \lambda_i)\frac{\lambda}{ct} \;.
\end{equation}

In full equilibrium,
\begin{equation}
  \psi = B \;.
\end{equation}
Requiring detailed balance, the emission rate density (or emissivity) for line $i$ is
\begin{equation}
  \label{eq24}
  j_i = \mathbb{A}_i = B(1-e^{-\tau_i})\delta(\lambda - \lambda_i)\frac{\lambda}{ct} \;\;.
\end{equation}
In the implementation, we have discretized the Planck function over group,
but this is a subdominant error with our 1024 group structure.
The total emissivity is
\begin{equation}
  \label{eq25}
  j = \sum_i j_i = \sum_i B(1-e^{-\tau_i})\delta(\lambda - \lambda_i)
    \frac{\lambda}{ct} \;,
\end{equation}
which, with the discretization of the Planck function over groups,
can be refactored as
\begin{equation}
  \label{eq26}
  j = \sum_g B_g \sum_{i\in\Delta\lambda_g}(1-e^{-\tau_i})
      \delta(\lambda - \lambda_i)\frac{\lambda}{ct} \;.
\end{equation}
\eqnref{eq26} can be sampled exactly/analytically in two steps:
\begin{enumerate}
\item sample group wavelength interval $\lambda_g$ with a CDF generated by integrating $j$ over each group,
  \begin{equation}
    \label{eq27}
    \int_gd\lambda\; j =
      B_g\sum_{i\in\Delta\lambda_g}(1-e^{-\tau_i})\frac{\lambda_i}{ct} \;,
  \end{equation}
  which, when divided by $\int d\lambda \;j = \sum_{g'}\int_{g'}d\lambda\; j$,
  is the probability of being emitted in group $g$,
\item sample line $i\in\Delta\lambda_g$ with a CDF generated by integrating $j$
over intervals about each of the lines,
  \begin{equation}
    \label{eq28}
    \int_id\lambda\; j = B_g(1-e^{-\tau_i})\frac{\lambda_i}{ct} \;,
  \end{equation}
  which, when divided by the right-hand side of \eqnref{eq27},
  gives the probability of getting emitted at line $i$ of group $g$.
\end{enumerate}

\section{Description of simplified test case}
\label{app:test_case}

In this appendix, we describe the ejecta and initial conditions of the
simplified test problem that is employed in Section~\ref{subsub:comp_simp}
to compare the line-binned and expansion-opacity discretizations
to our direct Sobolev implementation.
The equations for the state of the ejecta are~\citep{wollaeger18}
\begin{subequations}
  \label{eq1_app}
  \begin{gather}
    v(r,t) = \frac{r}{t} \;\;, \\
    \rho(r,t) = \rho_0\left(\frac{t}{t_0}\right)^{-3}
    \left(1 - \frac{r^2}{(v_{\max}t)^2}\right)^3 \;, \\
    T(r,t) = T_0\left(\frac{t}{t_0}\right)^{-1}
    \left(1 - \frac{r^2}{(v_{\max}t)^2}\right) \;,
  \end{gather}
\end{subequations}
where $r$ is the radius, $v$ is velocity, $\rho$ is density, $T$
is radiation temperature, $\rho_0$ is the initial maximum density, $t_0$
is the initial time, $v_{\max}$ is maximum ejecta velocity, and $T_0$
is the initial maximum temperature.
We set $v_{\max}=0.25c$ and $t_0=4$ days.
For an ejecta mass of $1.4\times10^{-2}$ M$_{\odot}$, this gives
$\rho_0=2.51\times10^{-15}$ g/cm$^3$.

The initial innermost temperature is $T_0 = 5.70\times10^3$ K.
The initial matter temperature is set to the initial radiation temperature.

For the source, we use~\citep{korobkin12}
\begin{equation}
  \label{eq2_app}
  \dot{\varepsilon} = \dot{\varepsilon}_0\left(\frac{t}{t_0}\right)^{-1.3} \;,
\end{equation}
where $\dot{\varepsilon}$ and $\dot{\varepsilon}_0$ are the specific
heating rate and initial specific heating rate.
For $t_0=4$ days, $\dot{\varepsilon}_0=8.2\times10^8$ erg/s/g.

The composition of the ejecta is assumed to be 100\% Nd.
We neglect bound-free, free-free, and Thomson scattering contributions
to the opacity.
We also neglect lines with $f_c < 10^{-3}$.

The numerical resolutions are as follows:
\begin{itemize}
\item 64 uniform spatial cells from 0 to $0.25c$,
\item 180 logarithmic time steps from day 4 to day 16,
\item 1024 logarithmic wavlength groups from $10^3$ to $1.28\times10^5$ \AA,
\item $2^{18}=262,144$ source particles generated per time step.
\end{itemize}
The mass per spatial cell is obtained by analytically integrating
the density profile over volume and the temperature per cell is obtained
by evaluating the temperature profile at radial cell centers.

We do not apply the observer-time correction to the simplified problem,
which is~\citep{lucy05}
\begin{equation}
  \label{eq3_app}
  t_{\rm obs} = t\left(1 - \frac{\mu v_{\max}}{c}\right) \;,
\end{equation}
where $\mu$ is the dot product of the particle direction with the radial
unit vector.
Consequently, the time $t$ for each particle is not adjusted to $t_{\rm obs}$
when it is tallied as escaping flux.
This choice permits the {\tt SuperNu} code to print out flux for more time
steps, and should not affect the comparison between methods.




\bsp    
\label{lastpage}
\end{document}